\documentclass[%
 reprint,
superscriptaddress,
 amsmath,amssymb,
 aps,
prd,
]{revtex4-1}
\usepackage{amsmath}
\usepackage{amssymb}
\usepackage{amsthm}
\usepackage{array}
\usepackage{float}
\usepackage{graphicx}
\usepackage{subfigure}
\usepackage{color}
\usepackage{hyperref}

\newcounter{ichi}
\newcounter{ni}
\newcounter{san}
\newcounter{yon}
\def\be{\begin{equation}}
\def\ee{\end{equation}}
\def\ba{\begin{eqnarray}}
\def\ea{\end{eqnarray}}

\definecolor{ao}{rgb}{0.0, 0.5, 0.0}

\preprint{APS/123-QED}
\begin{document}
\title{High-energy neutrino emission subsequent to gravitational wave radiation\\
from supermassive black hole mergers}
%from relativistic jets \peter{in} supermassive black hole mergers}
%\thanks{}
\author{Chengchao Yuan}\email{cxy52@psu.edu}
\affiliation{Department of Physics; Department of Astronomy \& Astrophysics; Center for Multimessenger Astrophysics, Institute for Gravitation and the Cosmos, The Pennsylvania State University, University Park, PA 16802, USA}

\author{Kohta Murase}
\affiliation{Department of Physics; Department of Astronomy \& Astrophysics; Center for Multimessenger Astrophysics, Institute for Gravitation and the Cosmos, The Pennsylvania State University, University Park, PA 16802, USA}
\affiliation{Center for Gravitational Physics, Yukawa Institute for Theoretical Physics, Kyoto University, Kyoto, Kyoto 606-8502, Japan}

\author{Shigeo S. Kimura}
\affiliation{Department of Physics; Department of Astronomy \& Astrophysics; Center for Multimessenger Astrophysics, Institute for Gravitation and the Cosmos, The Pennsylvania State University, University Park, PA 16802, USA}
\affiliation{Frontier Research Institute for Interdisciplinary Sciences; Astronomical Institute, Tohoku University, Sendai 980-8578, Japan}

\author{P\'eter M\'esz\'aros}
\affiliation{Department of Physics; Department of Astronomy \& Astrophysics; Center for Multimessenger Astrophysics, Institute for Gravitation and the Cosmos, The Pennsylvania State University, University Park, PA 16802, USA}

\date{\today}
\begin{abstract}
Supermassive black hole (SMBH) coalescences are {ubiquitous} in the history of the Universe and often exhibit strong accretion activities and powerful jets. These SMBH mergers are also promising candidates for future gravitational wave detectors such as {Laser Space Inteferometric Antenna (LISA)}. In this work, we consider neutrino counterpart emission originating from the jet-induced shocks. The physical picture is that relativistic jets launched after the merger will push forward inside the premerger disk wind material, and then they subsequently {get} collimated, leading to the formation of internal shocks, collimation shocks, forward shocks and reverse shocks. 
Cosmic rays can be accelerated in these sites and neutrinos are expected via the photomeson production process. 
We formulate the jet structures and relevant interactions therein, and then evaluate neutrino emission from each shock site. We find that month-to-year high-energy neutrino emission from the postmerger jet after the gravitational wave event is detectable by IceCube-Gen2 within approximately five to ten years of operation in optimistic cases where the cosmic-ray loading is sufficiently high and a mildly super-Eddington accretion is achieved.
We also estimate the contribution of SMBH mergers to the diffuse neutrino intensity, and find that a significant fraction of the observed very high-energy ($E_\nu\gtrsim1$ PeV) IceCube neutrinos could originate from them in the optimistic cases. 
In the future, such neutrino counterparts together with gravitational wave observations can be used in a multimessenger approach to elucidate in greater detail the evolution and the physical mechanism of SMBH mergers.
\end{abstract}

\maketitle

%%%%%%%%%%%%%%%%% sec 1
\section{\label{sec:intro}Introduction}
The coincident detection of gravitational waves (GWs) and the broadband electromagnetic (EM) counterpart from {the binary neutron star (NS) merger event GW 170817} \cite{abbott2017gw170817,abbott2017multi} heralds a new era of multimessenger astronomy. Since the initial discovery of GWs from binary black hole (BH) mergers by {the advanced Laser Interferometric Gravitational Wave Observatory (LIGO)} \cite{abbott2016observation,abbott2017gw170814}, 
{intense efforts have been dedicated to searching for the possible} associated neutrino emissions from binary NS/BH mergers (see a review~\cite{Murase:2019tjj} and Refs.~\cite{albert2017search,2019ApJ...870..134A,kimura2017high,kimura2018transejecta,fang2017high,decoene2020high}). 
The joint analysis of different messengers {would shed significantly} more light on the physical conditions of compact objects, as well as on the origin of their high-energy emissions. One vivid example that manifests the power of including high-energy neutrino observations as an additional messenger is the detection of the IceCube-170922A neutrino coincident with the flaring blazar TXS 0506+056~\cite{telescope2018multimessenger}. 
The combined analyses of EM and neutrino emissions from TXS 0506+056 provided stringent constraints on the blazar’s particle acceleration processes and the flare models~\cite{keivani2018multimessenger,Murase:2018iyl,ansoldi2018blazar,padovani2018dissecting,cerruti2019leptohadronic,gao2019modelling,reimer2019cascading,rodrigues2019leptohadronic,Petropoulou:2019zqp}. 

High-energy neutrino astrophysics began in 2012--2013 by the discovery of the cosmic high-energy neutrino background~\cite{aartsen2013first,icecube2013evidence}.
Despite the fact that the diffuse neutrino background has been studied for several years~\cite{aartsen2014observation,aartsen2015combined,aartsen2016observation,aartsen2020characteristics}, its origin still remains {unknown}, having given rise to a number of theoretical models aimed at explaining the observations (see, e.g., Refs. \cite{ahlers2018opening,Meszaros:2019xej} for reviews). 
Candidate source classes include bright jetted AGN \cite{murase2014diffuse,dermer2014photopion,padovani2015simplified,petropoulou2015photohadronic,yuan2020complementarity}, hidden cores of AGN \cite{stecker1991high,kimura2015neutrino,murase2016hidden,murase2019hidden}, galaxy clusters and groups~\cite{murase2008cosmic,murase2013testing,fang2018linking}, and starburst galaxies~\cite{loeb2006cumulative,murase2013testing} that contain supernovae and hypernovae as cosmic-ray (CR) accelerators~\cite{senno2015extragalactic} or AGN winds or galaxy mergers~\cite{Liu:2017bjr,kashiyama2014galaxy,yuan2018cumulative}. 
All the above models require CR acceleration up to 10--100 PeV to explain PeV neutrinos, because the typical neutrino energy produced by $pp$ or $p\gamma$ interactions is $E_\nu\sim(0.03-0.05)E_p$~\cite{murase2013testing}, where $E_p$ and $E_\nu$ are energies of protons and neutrinos, respectively. 
The same CR interactions also produce neutral 
pions that decay into high-energy gamma rays, which quickly interact with much lower-energy diffuse interstellar photons, degrading the gamma rays down to energies below $\sim$ TeV, which can be compared to the diffuse GeV-TeV gamma-rays background observed by \emph{Fermi} \cite{ackermann2015spectrum,ackermann2016resolving}. An important constraint that all such models must satisfy is that the resulting secondary diffuse gamma-ray flux must not exceed the diffuse isotropic gamma-ray background~\cite{murase2013testing,murase2016hidden}. 
The various models mentioned above satisfy, with varying degrees of the success, the observed neutrino and gamma-ray spectral energy densities, but there is uncertainty concerning the occurrence rate of the posited sources at various redshifts, due to our incomplete observational knowledge about the behavior of the corresponding luminosity functions at high redshifts. 

Recent observations have provided increasing evidence that a large fraction of nearby galaxies harbor supermassive black holes (SMBHs). One influential {scenario} for the formation of these SMBHs is that they, like the galaxies, have grown {their} mass through hierarchical mergers (e.g., Ref.~\cite{richstone1998supermassive}). 
SMBH mergers are ubiquitous across the history of the Universe especially at high redshifts where the minor galaxy mergers are more frequent. When galaxies merge, the SMBHs residing in each galaxy may sink to the center of the new merged galaxy and subsequently form a SMBH binary~\cite{begelman1980massive,kormendy2013coevolution}. 
The SMBHs gradually approach each other as the gravitational radiation takes away the angular momentum, which eventually leads to their coalescence, accompanied by a GW burst. The GW burst from the final stage of coalescing can be detected by future missions such as the \emph{Laser Interferometer Space Antenna} (LISA)~\cite{amaro2017laser}, providing through this channel valuable and prompt information about the merger rates, SMBH masses and redshift. In addition, SMBH mergers are usually associated with mass accretion activities and relativistic jets, which may lead to detectable EM and neutrino emission. For example, SMBH mergers may trigger AGN activities \cite{barnes1996transformations}. In this picture, the merger of SMBHs will become an important target for future multimessenger astronomy (e.g., Ref. ~\cite{milosavljevic2005afterglow}). 

In this paper, we present a concrete model for high-energy neutrino emission from four possible sites in the relativistic jet of SMBH mergers, namely, the collimation shock (CS), internal shock (IS), forward shock (FS) and reverse shock (RS). In \S~\ref{sec:phys-condition} we discuss the physical conditions in the jet and the gaseous envelope surrounding the merging SMBHs. In \S~\ref{sec:timescales} we discuss the various relevant dynamic and particle interaction timescales. In \S~\ref{sec:results} we calculate the neutrino emission from each site and investigate the neutrino detection rates for IceCube and its successor, IceCube-Gen2. We also integrated over redshift for parametrized merger rates compatible with our current knowledge and show that our model can contribute a significant portion to the diffuse neutrino background without violating the gamma-ray constraints. We summarize and discuss the implications of our results in \S~\ref{sec:summary}. 

Throughout the paper, we use the conventional notation $Q_x=Q/10^x$ and quantities are written in CGS units, unless otherwise specified. The integration over redshift is carried out in the $\Lambda$CDM universe with $H_0=71\rm\ km\ s^{-1}\ Mpc^{-1}$, $\Omega_{\rm m}=0.3$ and $\Omega_\Lambda=0.7.$

%%%%%%%%%%%%%%%%% sec 2
\section{\label{sec:phys-condition} Physical conditions of the premerger circumnuclear environment and the jet}
\begin{figure*}\centering
    \includegraphics[width=1.0\textwidth]{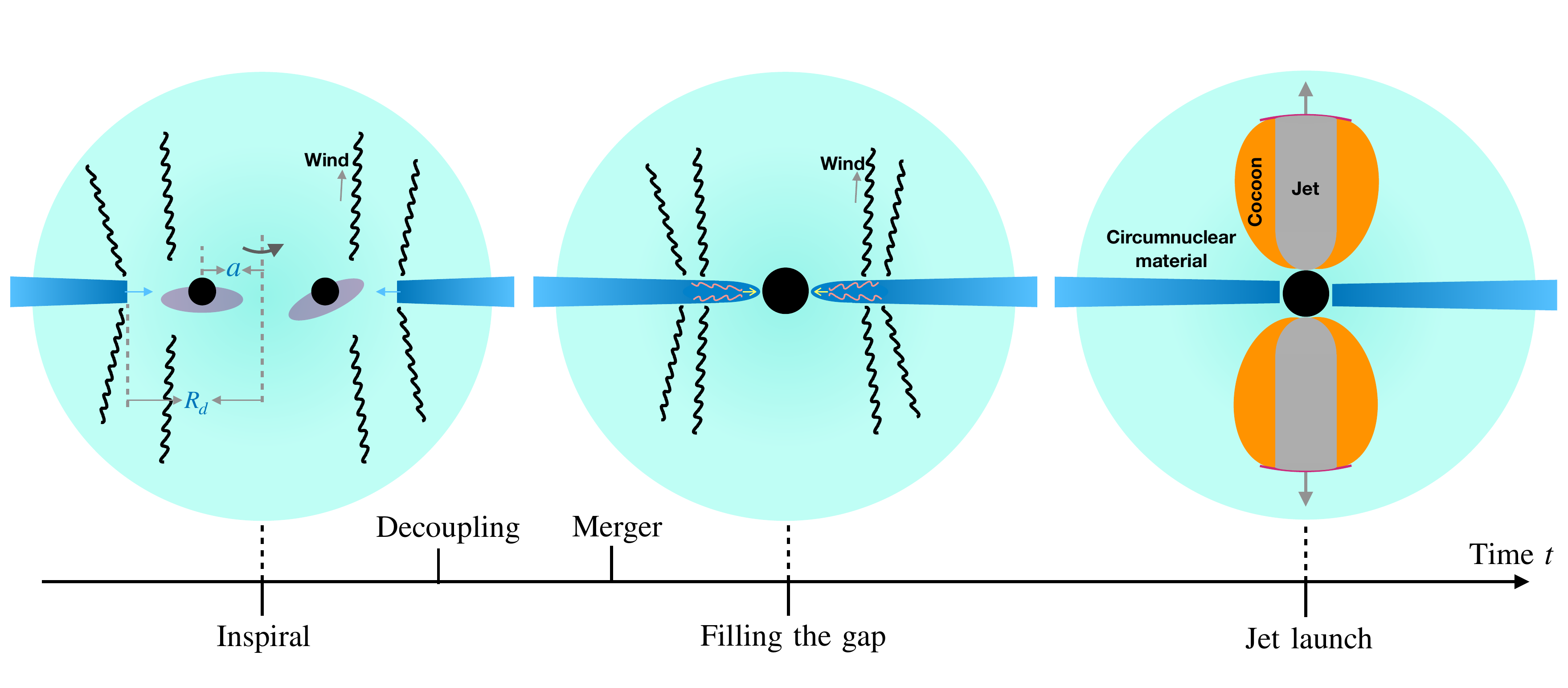}
    \caption{Schematic description of the merger of SMBHs with minidisks. The black wavy lines in the first and second panels illustrate the disk wind that forms the premerger circumnuclear material. The second {panel} shows the evolution of the circumbinary disk after the merger, while the third panel shows the postmerger jet-cocoon system. The stages of the evolution are marked on the time arrow below the figures.}
    \label{fig:schematic1}
\end{figure*}

The premerger circumnuclear material is thought to form from disk winds driven by the inspiralling binary SMBHs, in which a postmerger jet {is launched}, powered by the rotational energy of the remnant of the merger. It consists of two components originating respectively from the winds from the circumbinary disks around the binary system and from the minidisks surrounding each SMBH. Differently from the relativistic jet, the bulk velocity of the
winds is nonrelativistic and the mass outflow carried by the wind {spreads out quasi-spherically} above and below the disks \cite{narayan2012grmhd,yuan2012numerical,skadowski2013energy}. 
Although many jet and wind models have been proposed, currently there is no unambiguous way to demarcate the wind and the jet temporally. In this work, from the practical standpoint, we conjecture that the accretion by the binary system before the merger dominates the circumnuclear material, while the jet is launched after the merger and subsequently it propagates inside the existing premerger disk wind. This viewpoint is supported by {numerical} models of disk winds and relativistic jets. One of the most promising theoretical models to power relativistic jets is the Blandford-Znajek (BZ) mechanism \cite{blandford1977electromagnetic}, which posits that the jet is primarily driven by the rotational energy of the central SMBH, while it is widely accepted that the accretion outflows dominantly produce the nonrelativistic winds. In this case, it is reasonable to assume that the launch of the jets occurs after the binary SMBH coalescence, as a more massive SMBH is formed, and the wind bubble arises from the inspiral epoch during which the powerful tidal torque powers the strong winds. The schematic picture in Fig.~\ref{fig:schematic1} illustrates the evolution of the system.

As the jet penetrates deeply into the premerger disk wind, it sweeps up the gaseous material, leading to a high-pressure region which forces the encountered gas to flow sideaway to form a cocoon \cite{ramirez2002events,zhang2003relativistic,matzner2003supernova,bromberg2011propagation,mizuta2013opening,nakar2016observable,lazzati2017off} {(see also Refs.~\cite{matsumoto2018delayed,10.1093/mnras/stz3044,hamidani2020jet} for the jet propagation in expanding mediums).}
In this process, a forward shock and a reverse shock are also formed due to {the interaction between the jet and the premerger disk wind}. The shocks together with the shocked material are generally referred to as the jet head. A collimation shock will appear if the cocoon pressure is high enough to bend the jet boundary toward the axis of the jet, which as a consequence, collimates the jet. Moreover, the velocity fluctuation in the plasma inside the jet may produce internal shocks~\cite{rees1994unsteady}. 

For the purpose of conciseness, we use the abbreviations CS, IS, FS and RS to represent the collimation shock, internal shock, forward shock and revers shock in the following text, respectively. We show that all four of these sites can be CR accelerators, and we discuss the neutrino emissions from each site.
In~\ref{subsec:2.1}, we describe the premerger physical processes in details and derive a quantitative estimation of the premerger circumnuclear environment, while the jet structure and the shock properties are discussed in~\ref{subsec:2.2}.

\subsection{\label{subsec:2.1}Premerger circumnuclear environment}
{The} existence of circumbinary and minidisks may have a profound impact on the evolution of the binary system especially in the early inspiral stage where angular momentum losses due to gravitational radiation are subdominant compared with that from the circumbinary disk~\cite{armitage2002accretion,escala2005role,dotti2007supermassive}. {There are significant uncertainties in formulating a rigorous model of the disk-binary interactions throughout the merger, and this is} beyond the scope of this work. 
Here, we consider {three major factors that can dominate the disk and binary evolution in the late inspiral phase, namely, the viscosity, the tidal torques on the disks, and the gravitational radiation of the binary system, and use these to formulate a simplified treatment for deriving} the density profile of the premerger circumnuclear material. 
{This treatment can be justified because the previously launched disk wind material will be overtaken by the fast wind from the late inspiral stage, which} implies that we only need to model the disk-binary interactions in a short time interval immediately before the merger occurs.

Considering a circumbinary disk of inner radius $R_d$ around a SMBH binary of total mass $M_{\rm BH}$, the viscosity time for the disk is (e.g., Ref.~\cite{pringle1981accretion})
\be
t_{\rm vis}=\frac{1}{\alpha\Omega_K}\left(\frac{R_d}{H}\right)^2\simeq 0.31{\ \rm yr}\ {M^{-1/2}_{{\rm BH},6}}R_{d,14}^{3/2}\alpha_{-1}^{-1}(h/0.3)^{-2},
\label{eq:vis_time}
\ee
where $\alpha\sim0.1$ is the viscosity parameter, $H$ is the disk scale height, $\Omega_K=\sqrt{GM_{\rm BH}/R_d^3}$ is the Kepler rotation angular velocity, $M_{\rm BH}=10^6~M_{{\rm BH},6}M_\odot$ is the total mass of the binary SMBHs, and the dimensionless parameter $h$ is defined by $h=H/R_d$. {In this study, we consider high mass accretion rates, and assume optically thick circumbinary disks with $h\approx0.3$.}
%for a standard thin disk model where the {disk is mainly supported by the centrifugal force.}
For illustrative purposes we take the SMBH mass to be $M_{\rm BH}=10^6M_\odot$ as in \cite{DAscoli2018} and assume the mass ratio of the two SMBHs is $\zeta=1$. Initially, before the merger, the binary system has a large semi-major axis $a$, implying that the influence of the GWs for the disk is inferior to that of the viscosity, e.g., $t_{\rm GW}\gg t_{\rm vis}$. Here, the timescale of the GW inspiral is (e.g., Ref.~\cite{shapiro2008black}) 
\begin{equation}
t_{\rm GW}=\frac{5}{64}\frac{c^5a^4}{G^3M_{\rm BH}^3}\frac{(1+\zeta)^2}{\zeta}\simeq1.0\times10^4{\rm\ yr\ }M^{-3}_{\rm BH,6}a_{14}^4,
\label{eq:gw_time}
\end{equation}
As the two SMBHs gradually approach each other, the effects of the GWs become increasingly important. However, the circumbinary disk is still able to respond promptly to the slowly shrinking binary system until $t_{\rm GW}=t_{\rm vis}$. In this phase, the ratio of $R_d$ and $a$ remains roughly constant, e.g., $R_d\sim 2a$, as a result of the balance of the internal viscosity torque and the tidal torque exerted by the binary system. 
Later on, when the semi-major axis shortens down to or below a certain length, the binary system starts to evolve much faster and the gas in the circumbinary disk cannot react fast enough since GWs take away an increasingly large amount of energy from the binary system. 
The critical radius is referred to as the decoupling radius. Equating $t_{\rm vis}$ with $t_{\rm GW}$ we obtain the decoupling radius as
\be
R_{d,\rm dec}\simeq4.8\times10^{12}{\ \rm cm\ }M_{{\rm BH},6}\alpha_{-1}^{-2/5}(h/0.3)^{-4/5}.
\label{eq:decoupling}
\ee

The accretion activity also produces disk winds that blow away a fraction of the accreted mass, resulting in a premerger circumnuclear material above {and} below the circumbinary disk. 
In this study, we assume that the accretion rate is mildly larger than the Eddington rate, as $\dot M_{\rm BH}=\dot m\dot M_{\rm Edd}\equiv10\dot m L_{\rm Edd}/c^2\sim0.2(\dot m/10)\ \rm M_\odot\ yr^{-1}$. Given the accretion rate, we parameterize the mass outflow rate as $\dot M_w=\eta_w\dot
M_{\rm BH}$. 
After the disk becomes decoupled, $R_d$ remains roughly constant until merger occurs. The time interval between the disk decoupling and the merger, $t_{\rm m}$, can be estimated using Eq.~(\ref{eq:gw_time}) in combination with $t_{\rm GW}=a/|da/dt|$. 
After the merger, the gap between the disk and the newly formed SMBH cannot be preserved and the gas starts to fill the cavity in the {viscosity timescale (e.g., Ref.~\cite{farris2015binary}).} 
Our estimate suggests that both {$t_{\rm m}\sim8\times10^{-4}~{\rm \ yr}M_{\rm BH,6}\alpha_{-1}^{-8/5}(h/0.3)^{-16/5}$ and $t_{\rm vis}\sim3\times10^{-3}{\rm \ yr}~M_{\rm BH,6}\alpha_{-1}^{-8/5}(h/0.3)^{-16/5}$ at decoupling are approximately of the order of $10^{-3}\ \rm yr$, which is much shorter than the timescales to be considered later for the neutrino production.} {In such a short time duration, the wind formed at decoupling can reach only up to $\sim10^{13}-10^{14}$~cm, but one may extrapolate the density profile to a farther radius by incorporating different disk winds into one smooth profile.} 
Therefore, we neglect the modifications to the disk wind due to these two short term processes and we use the density profile at the decoupling to derive the jet structure. Moreover, we assume that the jet driven by the BZ mechanism is launched immediately after the cavity is occupied by gas. The evolution of the binary system is shown in the schematic pictures in Fig.~\ref{fig:schematic1}. Given the wind mass outflow rate $\dot M_w$ and the decoupling radius $R_{d,\rm dec}$, we have the density distribution of the premerger circumnuclear material
\be
\varrho_w(r)=\frac{\eta_w\dot M_{\rm BH}(1+\chi)}{4\pi r^2}\sqrt{\frac{R_{d,\rm dec}}{2GM_{\rm BH}}},
\label{eq:density_r}
\ee
where the enhancement factor $\chi\approx\frac{\dot M_{\rm mini}}{\dot M_w}\frac{\sqrt{2GM_{\rm BH}/R_{d,\rm dc}}}{v_{\rm mini}}$ takes into account the contribution of minidisks. In this expression, $\dot M_{\rm mini}$ represents the rate of accretion to the binary system from the minidisks, while {$v_{\rm mini}\approx\sqrt{2GM_{\rm BH}/(a/2)}$ is the typical escape velocity from the minidisks. 
We expect $R_d\sim 2a$, which implies that $v_{\rm mini}$ is about twice as much as the wind velocity of the circumbinary disk, i.e. $v_{\rm mini}\approx 2\sqrt{2GM_{\rm BH}/R_{d,\rm dc}}$. On the other hand, we expect a lower mass accretion rate onto the minidisk, e.g., $\dot M_{\rm mini} < \dot M_{\rm BH}$, as a result of the suppression due to the binary tidal torque}. In this case, we conclude that the factor $\chi$ is close to unity. The parameter $\eta_w$ depends strongly on $\dot m$ and on the disk magnetic field. For the standard and normal evolution (SANE) model the magnetic field is weak and $\eta_w$ ranges from ~$10^{-1}$ to ~$10^{-4}$ for super- and sub-Eddington accretions~\cite{Jiang2019a,Jiang2019,Ohsuga2009}, respectively. However, more powerful outflows could be produced in the magnetically arrested disk (MAD) model. In this case, $\eta_w$ can reach $10^{-2}-10^{-1}$ \cite{firstm87}. Here, we assume a fiducial value, $\eta_w\sim10^{-2}$ {and we will discuss the impact of a higher $\eta_w$, e.g., $\eta_w=0.1$, later.}

\subsection{\label{subsec:2.2}Postmerger jet structure and CR acceleration}
\begin{figure}
    \centering
    \includegraphics[width=0.5\textwidth]{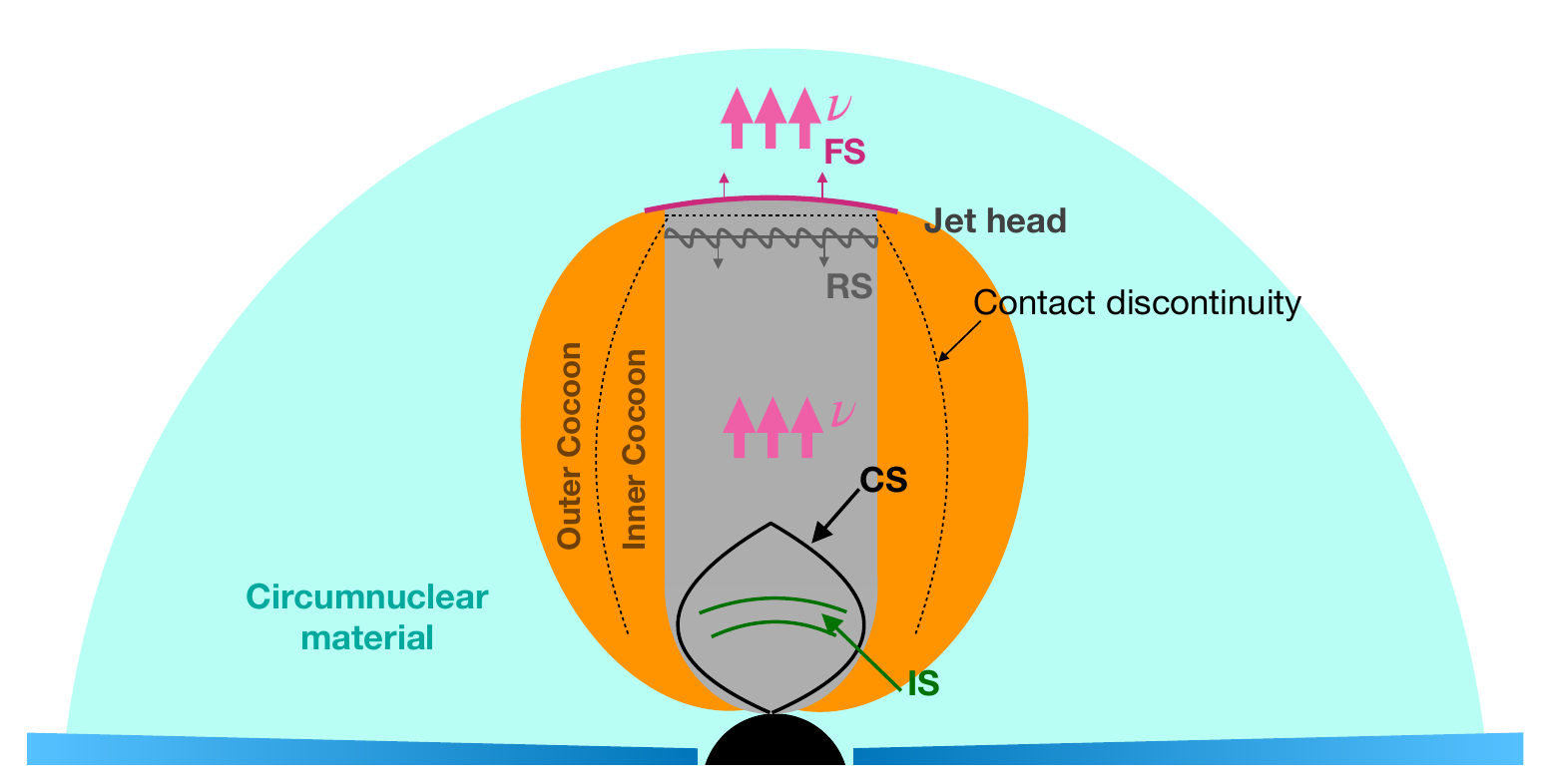}
    \caption{Schematic description of the structure of the collimated jet, where CS, IS, FS and RS stand for collimation shock, internal shock, forward shock and reverse shock. The contact discontinuity is illustrated as the dashed line.}
    \label{fig:schematic2}
\end{figure}
The central engines of strong, highly relativistic jets are generally assumed to be related to magnetized accretion flows and rotation of compact objects.
According to general relativistic magnetohydrodynamics (GRMHD) simulations, the threading magnetic flux $\Phi_B=\pi B_HR_g^2$ can reach the maximum saturation value~\cite{tchekhovskoy2011efficient}, $\Phi_B\sim50~\dot M_{\rm BH}^{1/2}R_gc^{1/2}$, for a given accretion rate $\dot M_{\rm BH}$ and horizon radius $R_g=GM_{\rm BH}/c^2$. Here, $B_H$ is the magnetic field that threads the SMBH horizon and we assume that the accretion rate remains unchanged before and after the merger, e.g., $\dot M_{\rm BH}\sim
0.2~\ M_\odot~{\rm yr^{-1}}~(\dot m/10)$, where the parameter $\dot m$ is defined as the ratio of $\dot M_{\rm BH}$ and the Eddington value $\dot M_{\rm Edd}\equiv 10L_{\rm Edd}/c^2$. {In the case of the magnetically arrested accretion, we estimate the jet kinetic luminosity to be
\begin{equation}
L_{k,j}\approx\eta_j\dot M_{\rm BH}c^2\simeq3.4\times10^{46}~(\dot m/10)(\eta_j/3)\ \rm erg\ s^{-1},
\label{eq:L_kj}
\end{equation}
where $\eta_j$ is the efficiency with which the accretion system converts accretion energy into jet energy~\cite{tchekhovskoy2011efficient}. Since this parameter is degenerate with $\dot m$, we assume $\eta_j=3$ in the following text.}

\begin{figure*}\centering
    \includegraphics[width=0.49\textwidth]{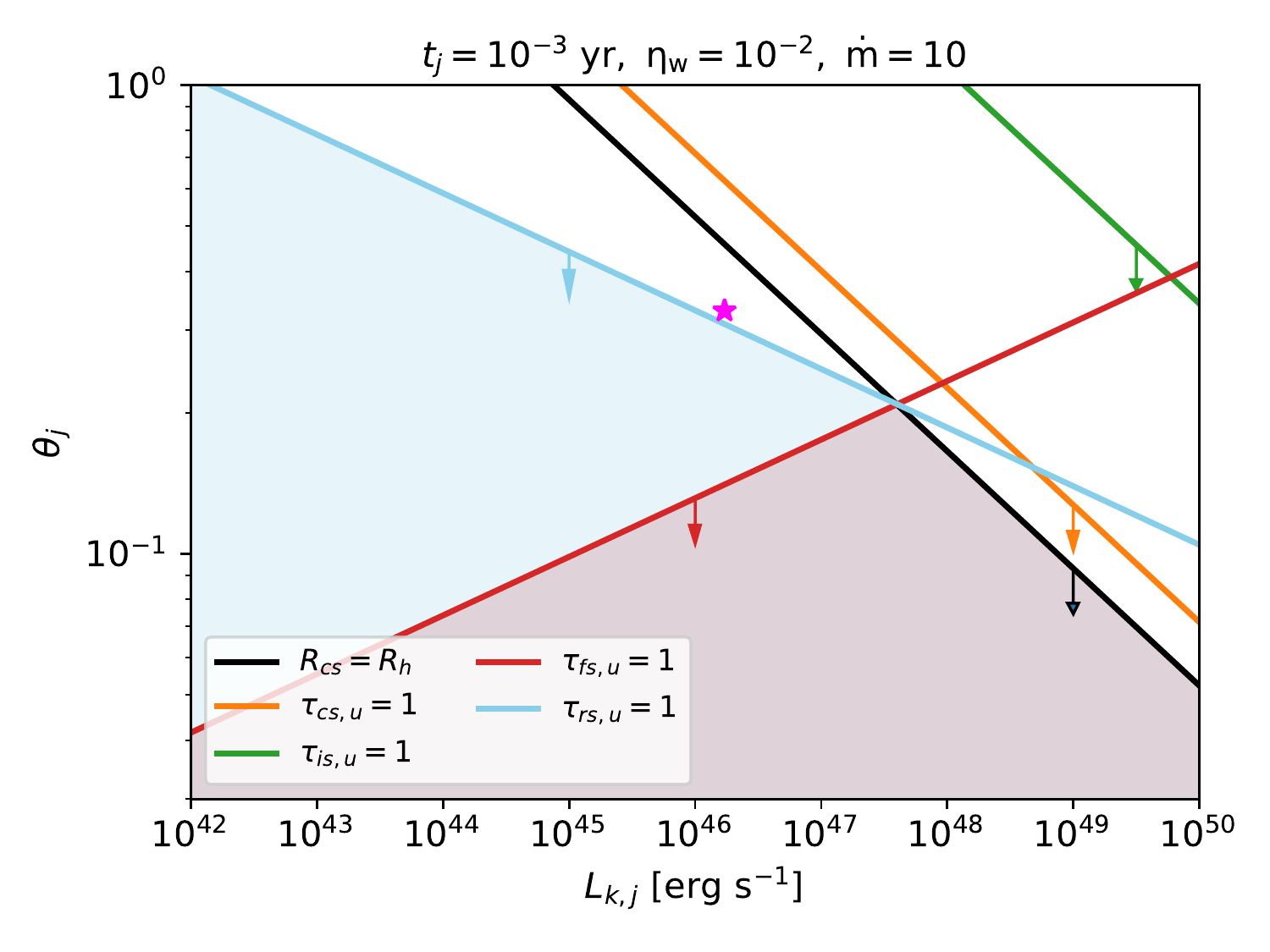}
    \includegraphics[width=0.49\textwidth]{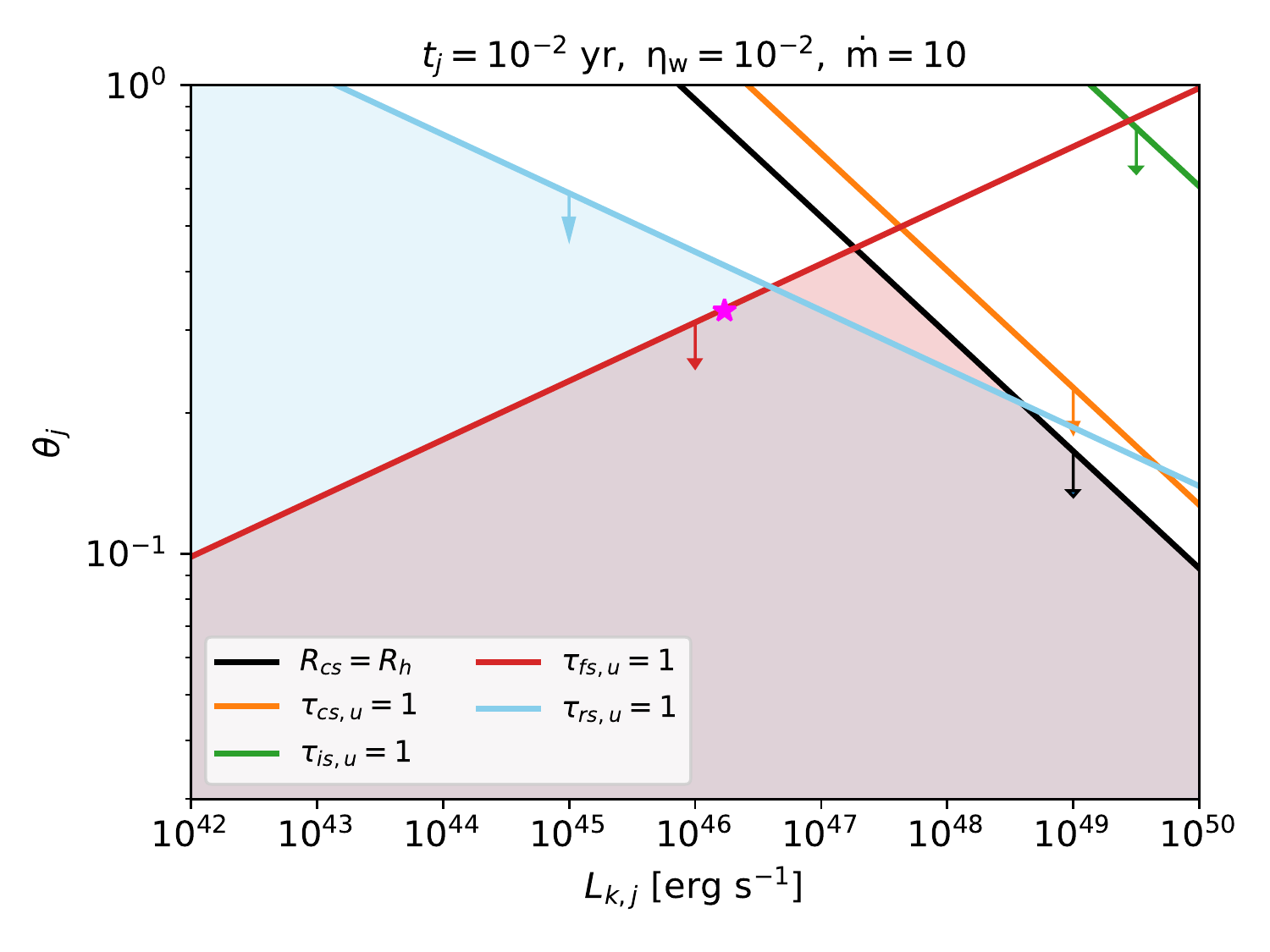}
    \caption{Radiation constraints, $\tau_{i,u}<1$, on $\theta_j-L_{k,j}$ plane at $t_j=10^{-3}\ \rm yr$ (left panel) and $t_j=10^{-2}\rm\ yr$ (right panel) for $i=$CS (orange lines), IS (green lines), FS (red lines) and RS (blue lines). The magenta stars show the parameters that are used, $\theta_j^{-1}=3$ and $L_{k,j}\simeq3.4\times10^{46}\ \rm erg\ s^{-1}$. The black solid line in each panel corresponds to the jet collimation condition, $R_{\rm cs}\lesssim R_h$. The blue and red areas illustrate the FS and RS constraints respectively, whereas the overlapped areas represent the joint
    constraints.}
    \label{fig:constraints}
\end{figure*}

Once the jet kinetic luminosity is specified, the shock structure is determined by the ambient gas density distribution and the Lorentz factor of the unshocked material, $\Gamma_j$. We now discuss the conditions under which the jets are collimated and for which CRs can be efficiently accelerated in each of the shock regions including the CS, IS, RS and FS. The jet is typically collimated for a sufficiently high cocoon density. Considering a jet of opening angle $\theta_j$, jet kinetic luminosity $L_{k,j}$ and isotropic equivalent kinetic luminosity
$L_{k,\rm iso}\approx2L_{k,j}/\theta_j^2$, the jet head position for the collimated jet is estimated to be ( {e.g., Refs.~ \cite{bromberg2011propagation,harrison2018numerically}}),
\be
R_{h}\approx\Xi^{1/5}L_{k,j}^{1/5}\hat\varrho_w^{-1/5}\theta_j^{-4/5}t_{j}^{3/5}
\label{eq:jet_head}
\ee
where $\Xi=16/\pi$ is a constant, $t_{j}$ is the jet propagation time reckoned from the launch of the jet and $\hat\varrho_w=(1/R_h)\int_{2R_{g}}^{R_h}\varrho_w(r)dr$ is the average density over the cocoon volume assuming that the cocoon's shape is cylindrical. Combining Eq.~(\ref{eq:jet_head}) with the definition of $\hat\varrho_w$, we are able to solve $R_h$ and $\hat\varrho_w$. According to the jet-cocoon model, the collimation shock forms at 
\be
R_{cs}\approx(2\pi)^{-1/2}\Xi^{-1/5}c^{-1/2}\hat\varrho_w^{-3/10}\theta_j^{-1/5}t_j^{2/5}L_{k,j}^{3/10}.
\label{eq:R_cs}
\ee
One precondition for these equations is that the jet should be collimated, which requires $R_{\rm cs}\lesssim R_h$. 
From the black lines of Fig.~\ref{fig:constraints}, we find that the jets  with the typical parameters $\theta_j\approx1/\Gamma_{\rm cj}\simeq0.33$ and $L_{k,j}\simeq3.4\times10^{46}\rm\ erg\ s^{-1}$ satisfies this requirement {if $t_j\gtrsim10^{-3}\rm\ yr$}, where $\Gamma_{\rm cj}\approx1/\theta_j\simeq3$ is the Lorentz factor of the downstream material of the collimation shock.

In the precollimation region, we assume the Lorentz factor of the unshocked material to be comparable to that of blazars, e.g., $\Gamma_j\sim10$, which is typically lower than the case of GRBs. Internal shocks usually arise in this region as a result of velocity fluctuations inside the outflow, resulting in faster and slower gas shells. Numerical simulations indicate that the fast material shells with Lorentz factor $\Gamma_r$ will catch up with the slower ones with $\Gamma_s$ nearly at the position of the collimation shock (e.g., Ref.~\cite{gottlieb2019high}). 
Hence, we may approximate the radius of the internal shocks to be
\begin{equation}
    \label{eq:R_is}
    R_{\rm is}\approx \min\left[R_{\rm cs},\ 2\Gamma_j^2 c  t_{\rm var}\right],
\end{equation}
where $t_{\rm var}\simeq10^5\ \rm s$ is the variability time.
%be \peter{approxximately the same as the collimation radius}, e.g., $R_{is}\sim %R_{cs}$. 

Fig.~\ref{fig:schematic2} schematically describes the structure of the jet-cocoon system as well as the shocks inside the jet. We consider CR acceleration and neutrino production in four different shock sites, including the CS, IS, FS and RS, as the jet propagates. 
One necessary condition for efficient CR acceleration through the shock acceleration mechanism is that the shock should have a sufficiently strong jump between the upstream and downstream material. Therefore, a collisionless shock {mediated} by plasma instabilities would be necessary rather than a {radiation-mediated} shock where velocity discontinuities are smeared out \cite{budnik2010relativistic,nakar2012relativistic}. Motivated by this, we
obtain one necessary constraint on the upstream of the shock for particle acceleration (see Refs.~\cite{murase2013tev,kimura2018transejecta} for details)
\be
\tau_u=n_u\sigma_Tl_u\lesssim{\rm min}[1,\Pi(\Gamma_{\rm sh})]
%1\ {\rm or\ }\tau_u=n_u\sigma_Tl_u\lesssim\frac{0.1\Gamma_{\rm  sh}}{1+2\ln\Gamma_{\rm sh}^2}
\label{eq:CR_constraints}
\ee
where $\tau_u$ is the upstream optical depth, $n_u$ is the comoving number density of upstream material, $\sigma_T$ is the Thomson cross section, $l_u$ is the length scale of the upstream fluid, $\Gamma_{\rm sh}$ stands for the relative Lorentz factor between the shock downstream and upstream, $\Pi(\Gamma_{\rm rel})$ is the function that depends on details of the pair enrichment. %In this equation, the second part mainly applies to 
Although the pairs are important for ultrarelativistic shocks, we impose $\tau_u<1$ for conservative estimates. However, our results are not much affected by this assumption, because the neutrino production continues to occur when the system becomes optically thin. The Lorentz factors for the shocks that are considered lie in the range $1<\Gamma_{\rm sh}\lesssim 5$, therefore we focus on the first constraint in Eq.~(\ref{eq:CR_constraints}) for our mildly relativistic shocks. As for the collimation shocks, combining the number density of the upstream $n_{{\rm cs},u}\approx L_{k,\rm iso}/(4\pi\Gamma_j^2R_{{\rm cs}}^2m_pc^3)$ with the comoving length of upstream fluid $l_{{\rm cs},u}\sim R_{cs}/\Gamma_j$, we have for the optical depth
\be
\tau_{{\rm cs},u}\approx n_{{\rm cs},u}l_{{\rm cs},u}\sigma_T\approx\frac{L_{k,\rm iso}\sigma_T}{4\pi\Gamma_j^3R_{\rm cs}m_pc^3}.
\label{eq:cs_optical_depth}
\ee
In the precollimated region, particles are mainly accelerated by internal shocks. The downstream of the internal shock can be regarded as the upstream of the collimation shock, and one may use $n_{{\rm is},u}\sim n_{{\rm is},d}/\Gamma_{\rm rel-is}\sim n_{{\rm cs},u}/\Gamma_{\rm rel-is}$ (ignoring coefficients), where $\Gamma_{\rm rel-is}\approx\Gamma_r/2\Gamma_j$ is the relative Lorentz factor between the upstream and the downstream of internal shocks. Here, we assume $\Gamma_{\rm rel-is}\approx5$ and obtain
\be
\tau_{{\rm is},u}\approx\frac{L_{k,\rm iso}\sigma_T}{{ 4}\pi R_{\rm is}m_pc^3\Gamma_j^3\Gamma_{\rm rel-is}^2}
\label{eq:is_optical_depth}
\ee
where the relationship $l_{{\rm is},u}\sim R_{\rm is}/\Gamma_j/\Gamma_{\rm rel-is}$ is used because the upstream unshocked flows are moving with a higher Lorentz factor $\Gamma_r$. 

In the jet head, the gas is rapidly decelerated to subrelativistic speeds, implying that the Lorentz factor is close to unity, e.g.,, $\Gamma_h\gtrsim1$. {Nevertheless}, the shock still satisfies the criteria for strong shocks. 
{The ambient gas enters the jet head through the forward shock and forms the outer cocoon, whereas the the shocked material from the jet constitutes the inner cocoon.}
The dashed lines in Fig.~\ref{fig:schematic2} show the contact discontinuity that separates the outer and inner cocoon components. In this case, we estimate that the head shock upstream number density is $n_{{\rm fs},u}=n_{\rm ext}=\varrho_w(R_h)/m_p$, where $n_{\rm ext}$ is the number density of the exterior premerger circumstellar material at $R_h$. 
With this we can write down the optical depth as
\be
\tau_{{\rm fs},u}\approx \frac{\varrho_w(R_h)\sigma_TR_h }{m_p}.
\label{eq:fs_optical_depth}
\ee
This simplified treatment is computationally convenient, albeit with the caveat that it is optimistic when computing the maximum energy of CRs accelerated by the FS. Similarly, repeating this procedure, we can get the corresponding quantities for the reverse shock, $n_{{\rm rs},u}=n_{{\rm cs},d}\sim n_{{\rm cs},u}\Gamma_{\rm rel-cs}$ and $l_{{\rm rs},u}\sim R_h/\Gamma_{\rm cj}$, where $\Gamma_{\rm rel-cs}\approx\Gamma_j/2\Gamma_{\rm cj}$ is the relative Lorentz factor. Substituting these quantities into Eq.~ (\ref{eq:CR_constraints}) yields
\be
\tau_{{\rm rs},u}\approx\frac{L_{k,\rm iso}\sigma_TR_h}{{ 4}\pi R_{\rm cs}^2m_pc^3\Gamma_j^3\Gamma_{\rm rel-cs}^{-2}}.
\label{eq:rs_optical_depth}
\ee

Fig.~\ref{fig:constraints} shows the radiation-mediated shock constraints at $t_j=10^{-3}\ \rm yr$ (left panel) and $t_j=10^{-2}\ \rm yr$ (right panel). The magenta star corresponds to the parameter set that is used in this work. The conditions for the jet collimation are shown by the black solid lines. From this figure, we find that the jet typically gets collimated in a short time $\sim10^{-3}\rm \ yr$ after the jet is launched. 
When the jet is collimated, the upstreams of the CS and IS are optically thin, implying that CRs may be efficiently accelerated at these two sites. However, the forward shock and reverse shock could still be radiation dominated for $t_j\lesssim10^{-3}\rm\ yr$, and subsequently become optically thin as the exterior gas envelop gets less denser. Therefore, there is a time $t_{*}$ at which the optical depth becomes unity, e.g., $\tau_{{\rm fs},u}(t_*)=1$, and $\tau_{{\rm fs},u}$ continues decreasing after that time. Since in the time interval $t_j\lesssim t_{*}$ the CR acceleration and the neutrino production are suppressed, we introduce a Heaviside function $H(t_j-t_*)$ in the expression for the CR and neutrino spectra to ensure that CRs are only accelerated after the onset time $t_*$.

%%%%%%%%%%%%%%%% sec 3t
\section{\label{sec:timescales} Interaction Timescales}
\subsection{\label{subsec:3.1}Nonthermal target photon fields}
In the following, we focus on the cases where the shock is collisionless and radiation unmediated. In astrophysical environments, neutrinos are produced through the decay of pions created by CRs via $pp$ and/or $p\gamma$ interactions.
Since the collimated jet is optically thin, we focus on nonthermal photons produced by the accelerated electrons and treat each site as an independent neutrino emitter, where the subtle interactions between particles from different regions are not considered. Here, we take a semianalytical approach to model the synchrotron and synchrotron-self-Compton (SSC) components of the target photon fields. 

We assume a power-law injection spectrum of electrons in terms of the Lorentz factor $dN_e/d\gamma_e\propto\gamma_e^{-p}$ for $\gamma_{e,\rm min}<\gamma_e<\gamma_{e,\rm max}$, where $p$ is the spectral index, $\gamma_{e,\rm min}$ and $\gamma_{e,\rm max}$ are the maximum and minimum electron Lorentz factors. Defining $\epsilon_e$ as the fraction of internal energy that is transferred to electrons and assuming the shocked gas {mainly consists of hydrogen, rather than $e^+e^-$ pairs,} one has $\gamma_{e,\rm min}=\epsilon_e\zeta_e\Gamma_{\rm rel}(m_p/m_e)$, where the parameter $\zeta_e$ has the typical value in the range $0.3-0.4$ (e.g., Refs.~\cite{sari2001synchrotron,murase2011implications}), and $\Gamma_{\rm rel}$ is the relative Lorentz factor between the upstream and the downstream, e.g., $\Gamma_{\rm rel-cs}$ for electrons from the collimation shock. The maximum electron Lorentz factor from the collimation shock acceleration can be obtained by equating the acceleration time $t_{e,\rm acc}\approx{\gamma_em_ec}/({eB_{{\rm cs},d}})$ with the
radiation cooling time $t_{e,c}\approx6\pi m_ec/[{\gamma_e\sigma_TB_{{\rm cs},d}^2(1+\tilde Y)}]$, where {$B_{{\rm cs},d}\approx(32\pi\epsilon_B\Gamma_{\rm rel-cs}^2n_{{\rm cs},u}m_pc^2)^{1/2}$ is the downstream magnetic field, $\epsilon_B\simeq0.01$ is the amplification factor that describes the fraction of the internal energy of unshocked materials converted to the magnetic field}, $\tilde Y$ is the Compton parameter and 
{given in Ref.}~\cite{panaitescu2000analytic}. 
Explicitly, we write the maximum Lorentz factor as 
\be
\gamma_{e,\rm max}=\left[\frac{18\pi e}{\sigma_T{ B_{{\rm cs},d}}(1+\tilde Y)}\right]^{1/2}.
\label{eq:elec_e_max}
\ee
Another important quantity that characterizes the shape of the {radiation}
%synchrotron 
spectrum is the cooling Lorentz factor, 
\begin{equation}
\gamma_{e,c}=\frac{6\pi m_ec}{t_{e,c}\sigma_TB_{{\rm cs},d}^2(1+\tilde Y)},
\label{eq:elec_e_cooling}
\end{equation}
above which electrons lose most of their energy {by radiation.}
%to the radiation fields. 
In this expression, $t_{e,c}\approx \min[t_j,t_{\rm cs,dyn}]$ is the radiation cooling time scale, where $t_{\rm cs,dyn}\approx R_{h}/(\Gamma_{\rm cj}c)$ is the dynamical time of the collimation shock.

\begin{figure}
    \centering
    \includegraphics[width=0.49\textwidth]{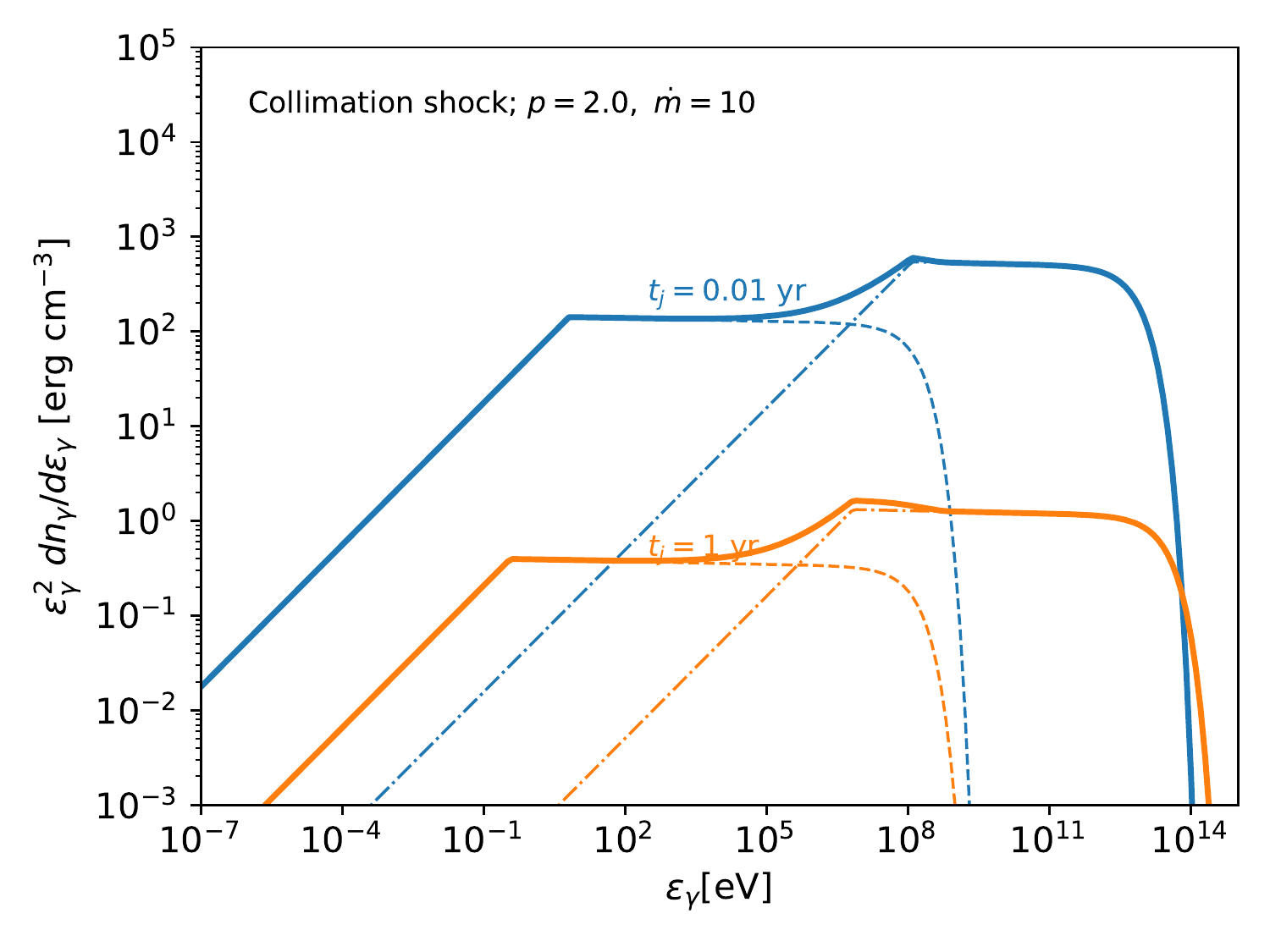}
    \caption{Collimation shock photon density distribution in the jet comoving frame at $t_j=0.01$ yr (blue lines) and $t_j=1$ yr (orange lines) for the super-Eddington accretion rate $\dot m=10$. The synchrotron and SSC components are shown as dashed and dash-dotted lines, respectively. The parameters, $\epsilon_e=0.1$, $\epsilon_B=0.01$, $\dot m=10$ and $\Gamma_{\rm cj}=\theta_j^{-1}=3$ are used.}
    \label{fig:photon_cs}
\end{figure}

Using $\gamma_{e,\rm min}$, $\gamma_{e,c}$ and $\gamma_{e,\rm max}$, the typical, cooling and maximum synchrotron emission energies in the jet comoving frame are respectively given by 
\be\begin{split}
    &\varepsilon_{\gamma,m}=\frac{3}{4\pi}\gamma_{e,\rm min}^2\frac{eB_{{\rm cs},d}}{m_ec^2},\\
    &\varepsilon_{\gamma,c}=\frac{3}{4\pi}\gamma_{e,\rm c}^2\frac{eB_{{\rm cs},d}}{m_ec^2},\\
    &\varepsilon_{\gamma,M}=\frac{3}{4\pi}\gamma_{e,\rm max}^2\frac{eB_{{\rm cs},d}}{m_ec^2}.\\
\end{split}
\ee
{If $\gamma_{e,\rm min}>\gamma_{e, c}$, the electrons are in the fast cooling regime} and we obtain the energy spectrum of the synchrotron radiation  (e.g. \cite{sari2001synchrotron,zhang2001high,murase2011implications})
\begin{equation}\begin{split}
\varepsilon_{\gamma}^2\frac{dn_{\gamma}^{\rm syn}}{d\varepsilon_\gamma}=&\frac{L_{\gamma}^{\rm syn}}{4\pi R_{\rm cs}^2 \Gamma_{\rm cj}^2 c \mathcal C_\gamma^{\rm syn}}\\
&\times
\begin{cases}
\left(\frac{\varepsilon_\gamma}{\varepsilon_{\gamma,c}}\right)^{\frac{4}{3}},\ &\varepsilon_\gamma<\varepsilon_{\gamma,c}\\
\left(\frac{\varepsilon_\gamma}{\varepsilon_{\gamma,c}}\right)^{\frac{1}{2}},\ &\varepsilon_{\gamma,c}<\varepsilon_\gamma<\varepsilon_{\gamma,m}\\
\left(\frac{\varepsilon_{\gamma,m}}{\varepsilon_{\gamma,c}}\right)^{\frac{1}{2}}\left(\frac{\varepsilon_\gamma}{\varepsilon_{\gamma,m}}\right)^{\frac{2-p}{2}},\ &\varepsilon_{\gamma,m}<\varepsilon_\gamma<\varepsilon_{\gamma,M}\\
\end{cases}
\end{split}
\label{eq:syn_spec}
\end{equation}
where {$L_{\gamma}^{\rm syn}={\epsilon_eL_{k,\rm iso}}/({1+\tilde Y})$, and $\mathcal C_\gamma^{\rm syn}$ is the normalization coefficient that ensures $\int \varepsilon_\gamma(dn_\gamma^{\rm syn}/d\varepsilon_\gamma) d\varepsilon_\gamma={L_{\gamma}^{\rm syn}}/[{4\pi R_{\rm cs}^2 \Gamma_{\rm cj}^2 c}]$}, and $\epsilon_e/(1+\tilde Y)$ represents the fraction of jet kinetic energy transferred to synchrotron radiation. In this work we assume $\epsilon_e=0.1.$ As for SSC, {we neglect the Klein-Nishina effect, since the highest energy photons do not contribute significant $p\gamma$ interactions}. The SSC spectrum in the Thomson regime is then given by
\be
\begin{split}
\varepsilon_\gamma^2 \frac{dn_{\gamma}^{\rm ssc}}{d\varepsilon_\gamma}=&
\frac{L_{\gamma}^{\rm ssc}}{4\pi R_{\rm cs}^2 \Gamma_{\rm cj}^2 c \mathcal C_\gamma^{\rm ssc}}\\
&\times
\begin{cases}
    \left(\frac{\varepsilon_\gamma}{\varepsilon_{\gamma,c}^{\rm ssc}}\right)^{\frac{4}{3}},&\ \varepsilon_\gamma<\varepsilon_{\gamma,c}^{\rm ssc}\\
    \left(\frac{\varepsilon_\gamma}{\varepsilon_{\gamma,c}^{\rm ssc}}\right)^{\frac{1}{2}},&\ \varepsilon_{\gamma,c}^{\rm ssc}<\varepsilon_\gamma<\varepsilon_{\gamma,m}^{\rm ssc}\\
        \left(\frac{\varepsilon_{\gamma,m}^{\rm ssc}}{\varepsilon_{\gamma,c}^{\rm ssc}}\right)^{\frac{1}{2}}\left(\frac{\varepsilon_\gamma}{\varepsilon_{\gamma,m}^{\rm ssc}}\right)^{\frac{2-p}{2}},&\ \varepsilon_{\gamma,m}^{\rm ssc}<\varepsilon_\gamma<\varepsilon_{\gamma,M}^{\rm ssc}\\
\end{cases}
\end{split}
\label{eq:ssc_spec}
\ee
where $L_{\gamma}^{\rm ssc}\approx\tilde Y  L_{\gamma}^{\rm syn}$ and the break energies are defined as $\varepsilon_{\gamma,m}^{\rm ssc}=2\gamma_{e,\rm min}^2\varepsilon_{\gamma,m}$, $\varepsilon_{\gamma,c}^{\rm ssc}=2\gamma_{e,c}^2\varepsilon_{\gamma,c}$ and $\varepsilon_{\gamma,M}^{\rm ssc}=\gamma_{e,\rm max}m_ec^2$. {Likewise, the normalization factor $\mathcal C_{\gamma}^{\rm ssc}$ is determined by $\int \varepsilon_\gamma(dn_\gamma^{\rm ssc}/d\varepsilon_\gamma) d\varepsilon_\gamma={L_{\gamma}^{\rm ssc}}/[4\pi R_{\rm cs}^2 \Gamma_{\rm cj}^2 c]$}. 
In the early stage of the jet propagation, the electrons are commonly in the fast cooling regime,
{and the equation controlling the distribution of nonthermal photons is}
\begin{equation}
    \varepsilon_\gamma\frac{dn_\gamma}{d\varepsilon_\gamma}=\varepsilon_\gamma\frac{dn_\gamma^{\rm syn}}{d\varepsilon_\gamma}+\varepsilon_\gamma\frac{dn_\gamma^{\rm ssc}}{d\varepsilon_\gamma}.
\end{equation}
The cooling of the electrons tends to be less efficient when the magnetic field decreases as jet expands, and the energy spectra for slow cooling electrons should be used if the order of $\gamma_{e,c}$ and $\gamma_{e,m}$ is reversed, i.e., $\gamma_{e,c}>\gamma_{e,\rm min}$. In this case, the synchrotron and SSC spectra should be rewritten by swapping $\varepsilon_{\gamma,m}$ and $\varepsilon_{\gamma,c}$ in Eq.~(\ref{eq:syn_spec}), and swapping $\varepsilon_{\gamma,m}^{\rm ssc}$ and
$\varepsilon_{\gamma,c}^{\rm ssc}$ in Eq.~(\ref{eq:ssc_spec}), respectively. {We also need to replace the index 1/2 by $(3-p)/2$ in both equations. Considering that only electrons with $\gamma_e$ greater than $\gamma_{e,c}$ can convert their kinetic energies to electromagentic emission, we introduce one extra parameter 
\be
\eta_e=\frac{\int_{\gamma_{e,c}}^{\gamma_{e,\rm max}}\gamma_e(dN_e/d\gamma_e)d\gamma_e}{\int_{\gamma_{e,\rm min}}^{\gamma_{e,\rm max}}\gamma_e(dN_e/d\gamma_e)d\gamma_e}\lesssim1,\ \gamma_m<\gamma_c<\gamma_M
\label{eq:fast_cooling_frac}
\ee
into the photon density for the slow cooling case. We adopt the spectral index $p=2.0$ for electrons. Fig. \ref{fig:photon_cs} shows the distribution of photon densities in the jet comoving frame for collimation shocks at $t_j=0.01$ yr (blue lines) and $t_j=1$ yr (orange lines) for the super-Eddington accretion rate $\dot m=10$. 
}

Similarly we can derive the photon distribution in other shocks given the dynamic times for IS, FS and RS, e.g.,
{$t_{\rm is, dyn}\approx R_{\rm is}/(\Gamma_jc),\ t_{\rm fw,\rm dyn}\approx R_h/({\beta_hc})\approx t_{\rm rs,\rm dyn}$}, 
where $\beta_hc=c\sqrt{1-1/\Gamma_h^2}$ is jet head speed and 
\be
\Gamma_h=\min\left[\Gamma_{\rm cj},\sqrt{1+\tilde L^{1/2}}\right]
\ee
is the jet head Lorentz factor. In this expression, we follow \cite{bromberg2011propagation} to define $\tilde L$
\be
\tilde L=\Xi^{2/5}L_{k,j}^{2/5}\hat\varrho_w^{-2/5}c^{-2}\theta_j^{-8/5}t_j^{-4/5}.
\ee
Since the jet head decelerates while sweeping up the exterior circumnuclear material and ends up being sub-relativistic ($\Gamma_h\gtrsim1.0$), we use the jet head velocity rather than the Lorentz transformation to compute $t_{fs,\rm dyn}$. {The photon spectra for the IS, FS and RS look similar to Fig. \ref{fig:photon_cs}, so for the purpose of conciseness, we merely show $dn_\gamma/d\varepsilon_\gamma$ for the CS case.}

%EDD

\subsection{\label{subsec:3.2}Timescales for the CRs and pions}
\begin{figure*}
    \centering
    \includegraphics[width=0.50\textwidth]{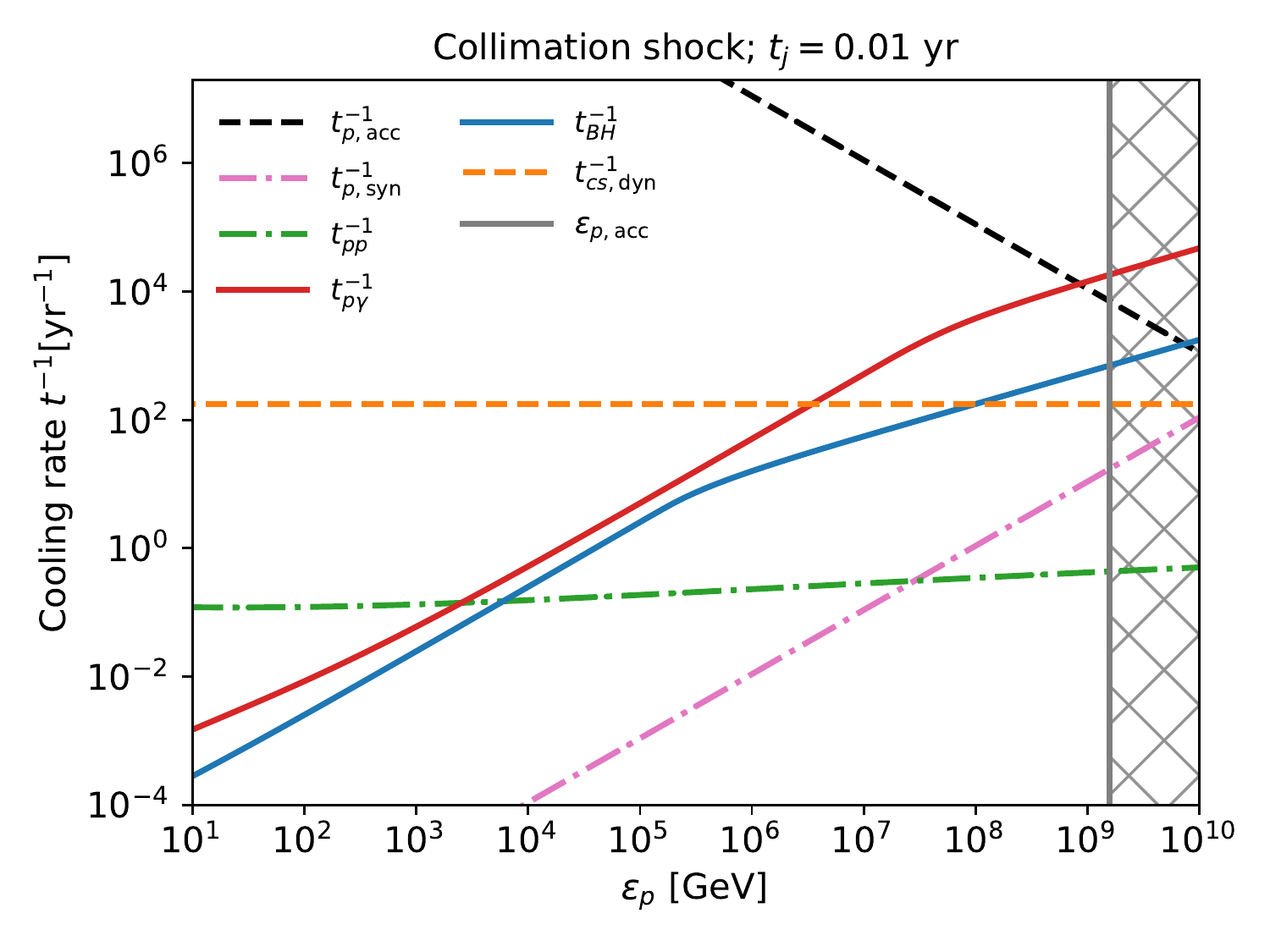}\hfil
    \includegraphics[width=0.50\textwidth]{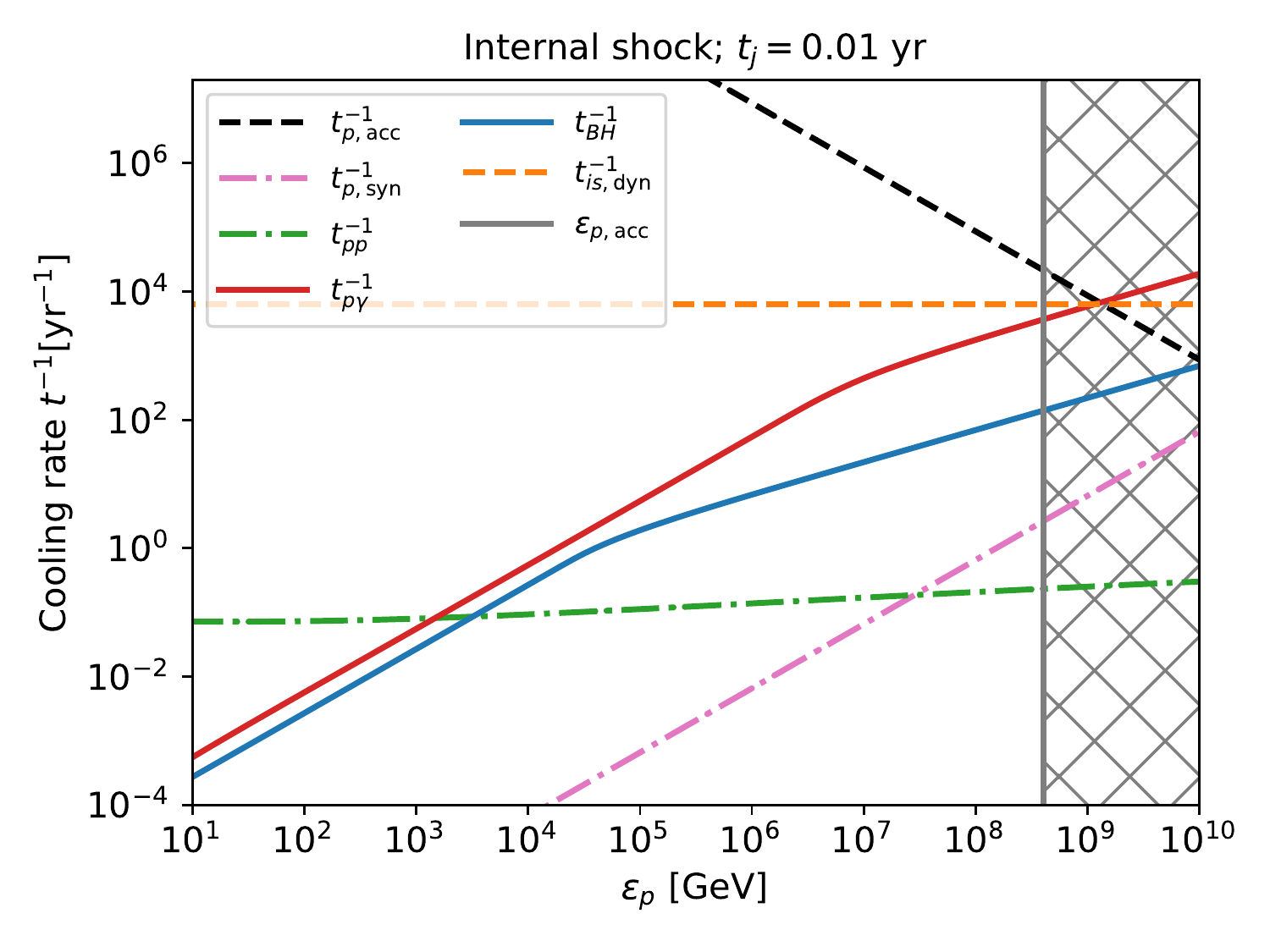}\par\medskip
    \centering
    \includegraphics[width=0.50\textwidth]{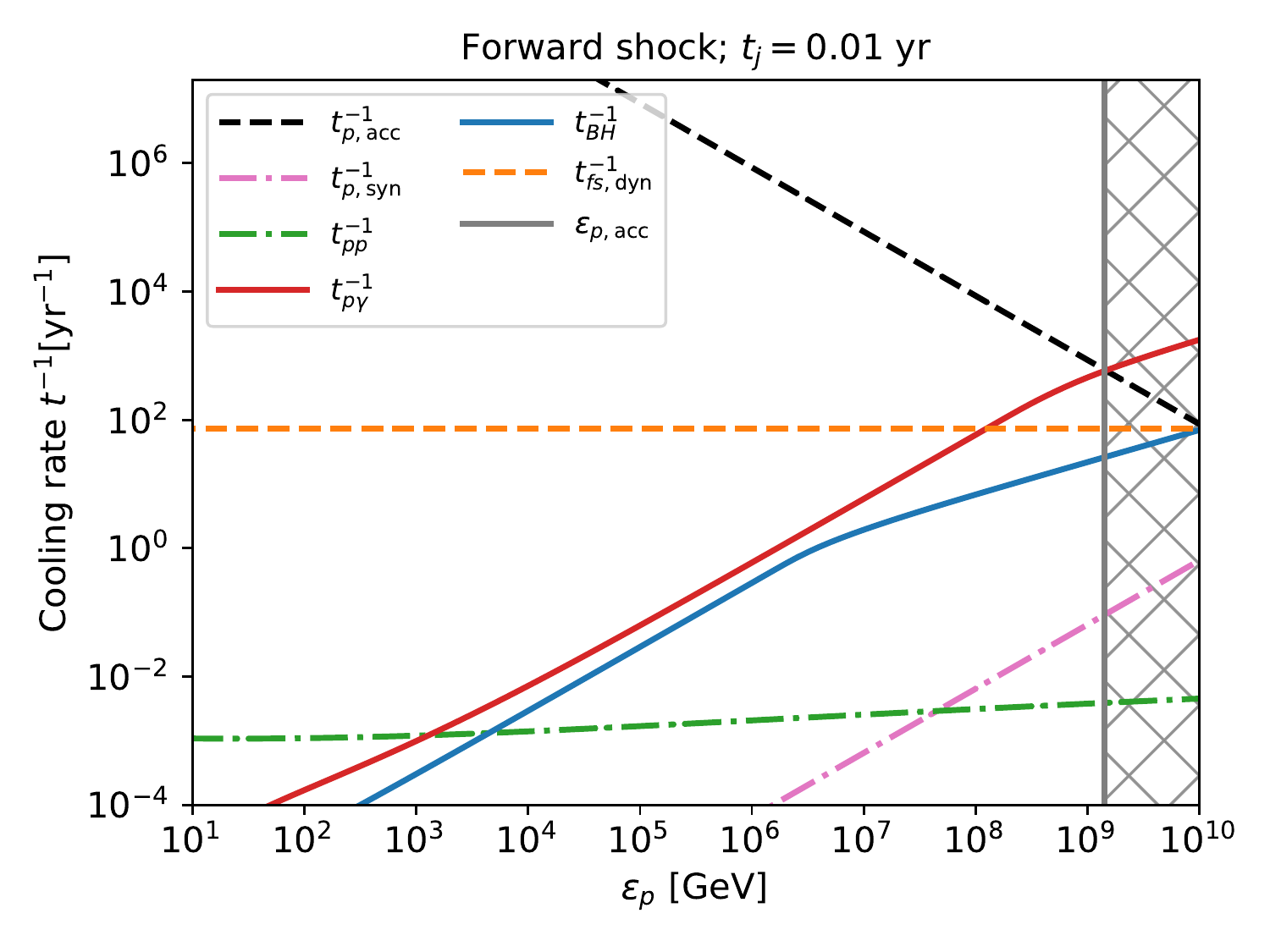}\hfil
    \includegraphics[width=0.50\textwidth]{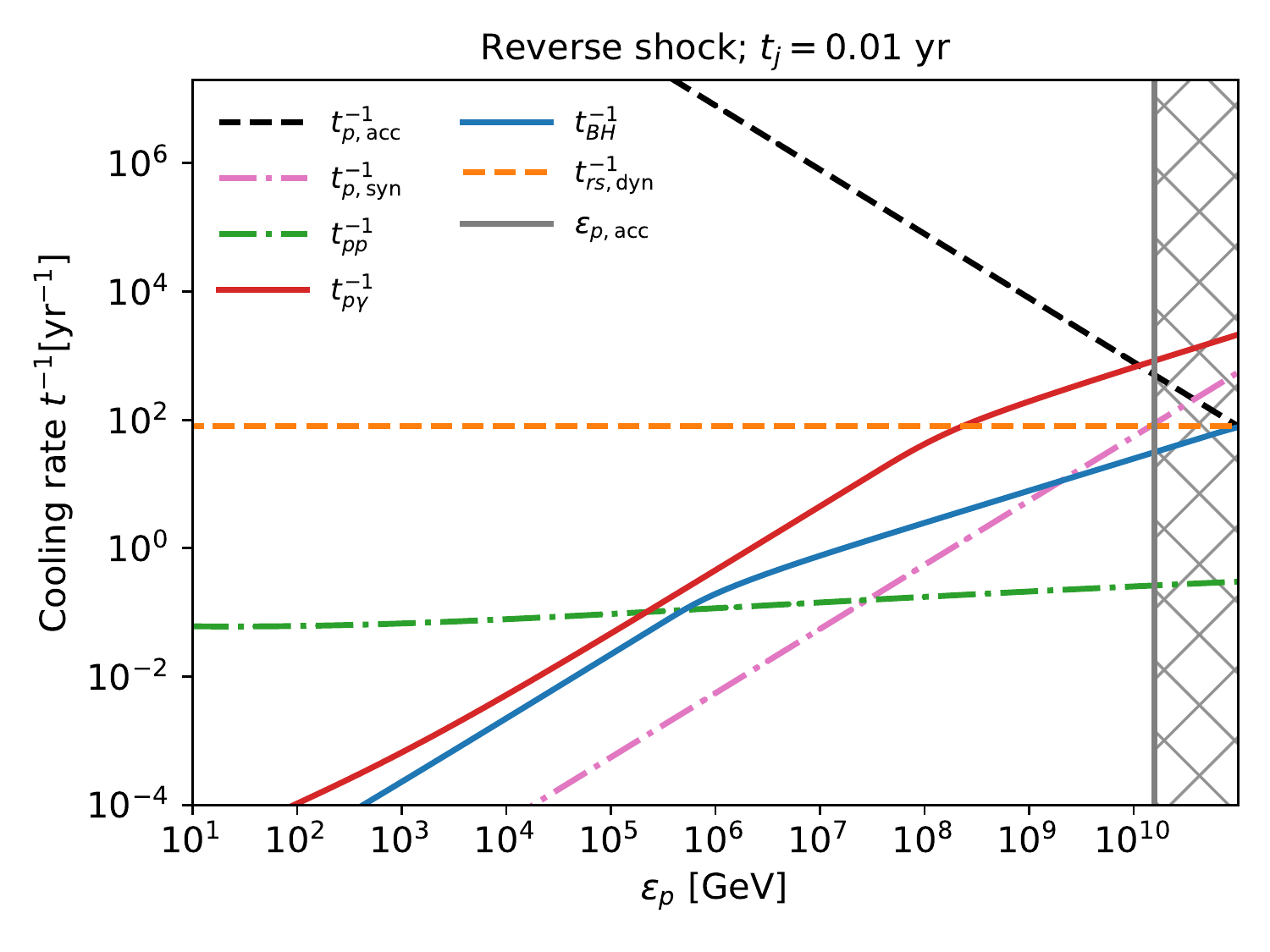}
    \caption{Snapshots of cooling, acceleration and dynamic timescales for CS (left up), IS (right up), FS (left down) and RS (right down) at $t_j=10^{-2}\ \rm yr$. The vertical line represents the maximum proton energy from acceleration, $\varepsilon_{p,\rm acc}$, whereas the hatches imply the unreachable proton energies. The parameters, $\epsilon_e=0.1$, $\epsilon_B=0.01$, $\dot m=10$, $\Gamma_j=10$ and $\Gamma_{\rm cj}=\theta_j^{-1}=3$ are used.}
    \label{fig:cooling_rates}
\end{figure*}

To calculate the neutrino emission, we need to estimate the cooling and acceleration timescales of the protons. Here we consider the CS case as an example, and it is straightforward to rewrite the relevant equations to cover the IS, FS and RS scenarios. For the CS case, the acceleration time for protons with an energy $\varepsilon_p$ is estimated to be $t_{p,\rm acc}\approx
\varepsilon_p/(eB_{{\rm cs},d}c)$. While propagating in the jet, the high-energy protons are subject to photomeson ($p\gamma$) interactions, the Bethe-Heitler (BH) process, proton-proton ($pp$) inelastic collisions and synchrotron radiation. The energy loss rate due to $p\gamma$ interactions is
\be
t_{p\gamma}^{-1}=\frac{c}{2\gamma_p^2}\int_{\bar \varepsilon_{\rm th}}^{\infty}d\bar \varepsilon_\gamma\sigma_{p\gamma}\kappa_{p\gamma}\bar\varepsilon_\gamma\int_{\frac{\bar\varepsilon_\gamma}{2\gamma_p}}^{\infty}d\varepsilon_\gamma\varepsilon_\gamma^{-2}\frac{dn_\gamma}{d\varepsilon_\gamma},
\label{eq:p_gamma}
\ee
where $\gamma_p=\varepsilon_p/(m_pc^2)$ is the proton Lorentz factor, $\bar\varepsilon_{\rm th}\simeq145\ \rm MeV$ is the threshold energy for $p\gamma$ meson production, and $\bar\varepsilon_\gamma$ is the photon energy in the proton rest frame. In this equation, $\sigma_{p\gamma}$ and $\kappa_{p\gamma}$ represent the $p\gamma$ cross section and inelasticity, respectively. We use the results of Ref.~\cite{murase2006high} for $\sigma_{p\gamma}$ and $\kappa_{p\gamma}$. Similarly we use Eq.~(\ref{eq:p_gamma}) to evaluate the BH cooling rate, $t_{\rm BH}^{-1}$, by replacing $\sigma_{p\gamma}$ and $\kappa_{p\gamma}$ with $\sigma_{\rm BH}$ and $\kappa_{\rm BH}$ whose fitting formulae are given by \cite{stepney1983numerical} and \cite{chodorowski1992reaction}, respectively. The time scale of $pp$ interactions can be written as $t_{pp}^{-1}\approx n_{{\rm cs},d}\sigma_{pp}\kappa_{pp}c$, where $\kappa_{pp}\approx 0.5$ is the
inelasticity and $\sigma_{pp}$ is the cross section for inelastic $pp$ collisions. As for the synchrotron radiation, the cooling timescale for protons is estimated to be $t_{p,\rm syn}=6\pi m_p^4c^3/(m_e^2\sigma_TB_{{\rm cs},d}^2\varepsilon_p)$.
%For the completeness purpose, we also take the cooling rate due to proton SSC radiation, $t_{p,\rm ssc}^{-1}$, into account, following the formalism presented in \cite{murase2011implications}.
Assuming $\epsilon_e=0.1$ and $\epsilon_B=0.01$, Fig. \ref{fig:cooling_rates} shows the cooling rates, acceleration and dynamical timescales for CS, IS, FS and RS scenarios at the jet time $t_j=10^{-2}\ \rm yr$. {The vertical lines represent the maximum proton energy by $Fermi$ acceleration, $\varepsilon_{p,\rm acc}\approx \frac{3}{20}eB_{i,d}t_{i,\rm dyn}c$.} From Fig. \ref{fig:cooling_rates}, we also find that the $pp$ interactions are
subdominant in comparison with photomeson ($p\gamma$) process. Given the timescales for protons, we are able to derive the energy-dependent neutrino production efficiencies from $p\gamma$ and $pp$ interactions respectively
\be\begin{split}
    &f_{p\gamma-{\rm cs}}=\frac{t_{p\gamma}^{-1}}{t_{p,c}^{-1}+t_{\rm cs,\rm dyn}^{-1}},\\
    &f_{pp-{\rm cs}}=\frac{t_{pp}^{-1}}{t_{p,c}^{-1}+t_{\rm cs,\rm dyn}^{-1}},
\end{split}
\label{eq:f_pgamma}
\ee
where $t_{p,c}^{-1}\equiv t_{p\gamma}^{-1}+t_{\rm BH}^{-1}+t_{pp}^{-1}+t_{p,\rm syn}^{-1}$ is the total cooling rate and the dynamic time $t_{\rm cs,\rm dyn}$ is included to constrain the timescale of interactions. If $t_{\rm cs,\rm dyn}^{-1}$ is high, protons tend to leave this site very fast before sufficiently participating in the interactions listed above. Likewise, we can obtain the neutrino production efficiencies for the IS, FS and RS.
As expected, in Fig. \ref{fig:cooling_rates} we find that $p\gamma$ interactions dominate the neutrino production, instead of the $pp$ collisions. The reason is that the jet is neither dense enough nor has a sufficiently large size to allow efficient $pp$ interactions.

The secondary pions produced from $p\gamma$ and $pp$ interactions may also lose energy through synchrotron and hadronic processes, e.g., $\pi p$ collisions. The pion synchrotron cooling timescale is $t_{\pi,\rm syn}=(m_\pi^4/m_p^4)t_{p,\rm syn}$, where $m_\pi\approx139.57\rm\ MeV$ is the mass of charged pions. Approximately, the hadronic cooling time scale can be written as $t_{\pi p}\approx n_{{\rm cs},d}\sigma_{\pi p}\kappa_{\pi p}c$, where $\sigma_{\pi p}\sim5\times10^{-26}\ \rm cm^2$ and
$\kappa_{\pi p}\sim 0.8$ are used in our calculation. Using the rest life time charged pions, $t_{\pi}\simeq8.2\times10^{-16}\ \rm yr$, the charged pion decay rate {is estimated to be} $t_{\pi,\rm dec}^{-1}\approx1/(\gamma_\pi t_{\pi})$. For a PeV pion, the decay rate is approximately $1.7\times10^8\ \rm yr^{-1}$, which is much larger than the reciprocal of the dynamic time ($t_{\rm cs,\rm dyn}^{-1}$) and the cooling rate ($t_{\pi,\rm syn}^{-1}$), implying that the pion decay efficiency is nearly unity, e.g., 
\be
f_{\pi,\rm sup-cs}\approx1-\exp\left(-\frac{t_{\pi,\rm dec}^{-1}}{t_{\rm cs,\rm dyn}^{-1}+t_{\pi,\rm syn}^{-1}}\right)\sim1.
\label{eq:f_pion}
\ee
We see that this is true in the other sites as well, and the relation $f_{\pi,\rm sup}\sim1$ will be used in the following text. 
{For neutrinos from secondary muon decay, we introduce another suppression factor besides $f_{\pi,\rm sup}$, e.g., $f_{\mu,\rm sup}=1-\exp({-t_{\mu,\rm dec}^{-1}/{t_{\mu,c}^{-1}}})$. For a 100 PeV muon, the decay rate is $t_{\mu,\rm dec}^{-1}\approx1/(\gamma_{\mu}t_{\mu})\simeq1.5\times10^4{\ \rm yr^{-1}}$, where $t_{\mu}$ is the muon lifetime. We conclude the ratio $t_{\mu,\rm dec}^{-1}/t_{\mu,c}^{-1}\approx(m_\mu^4/m_p^4)t_{p,\rm syn}/(\gamma_\mu t_{\mu})\simeq38\times(\varepsilon_\mu/100{\rm\ PeV})^{-2}(B_d/10\rm G)^{-2}$, depending on the shock site and jet time $t_j$. In the energy range studied in this paper and considering that the neutrino emission can last from years to decades (which will be shown later), the approximation of $f_{\mu,\rm sup}\approx1$ is valid. 
Ultrahigh-energy neutrinos (with $\gtrsim1$~EeV) from the muon decay can be suppressed by $f_{\mu,\rm sup}$ in the very early stage (e.g., $t_j<10^{-2}\rm yr$), which could change the observed flavor ratio.}

\begin{figure*}[]
    \centering\includegraphics[width=0.49\textwidth]{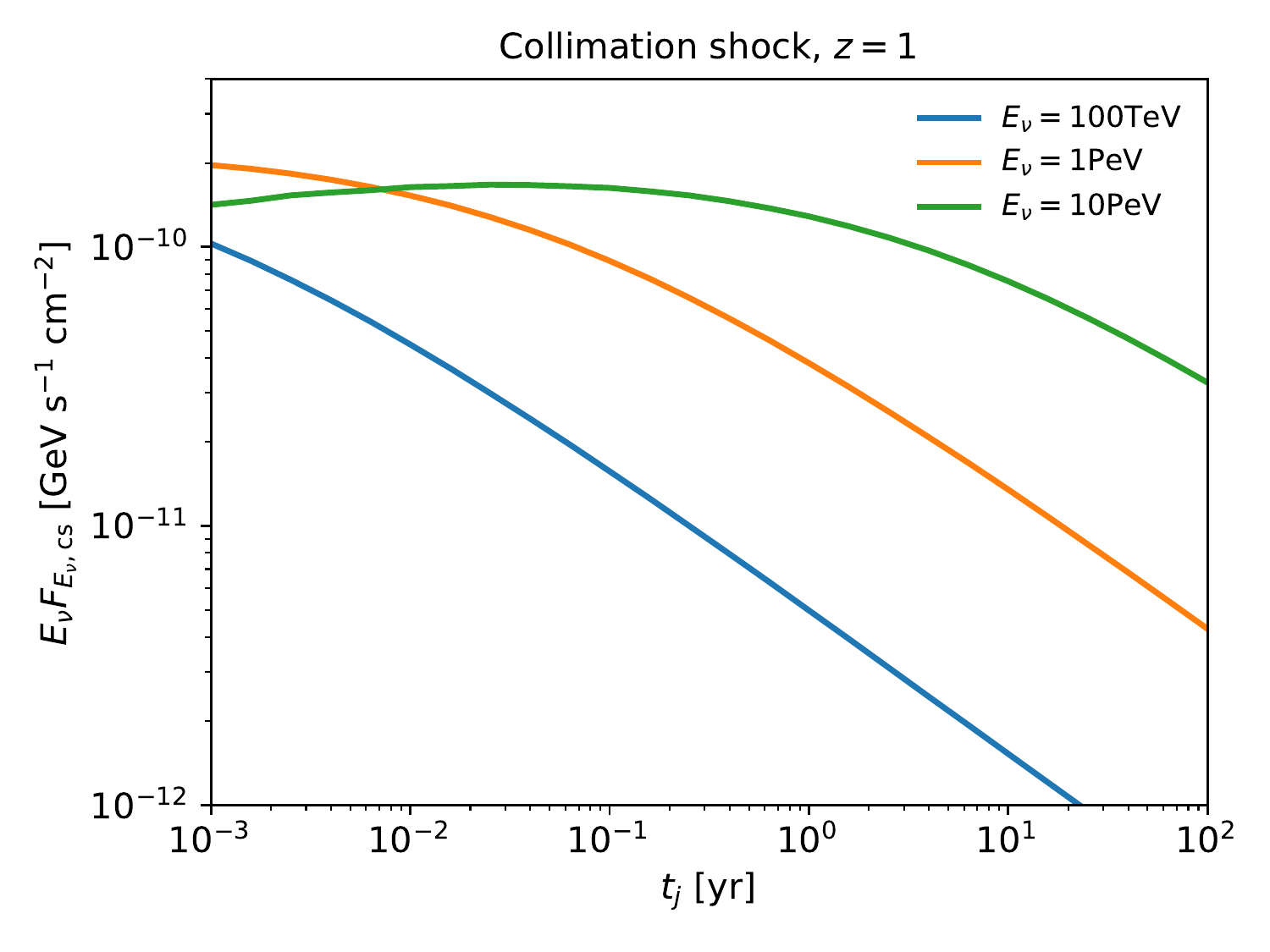}
    \centering\includegraphics[width=0.49\textwidth]{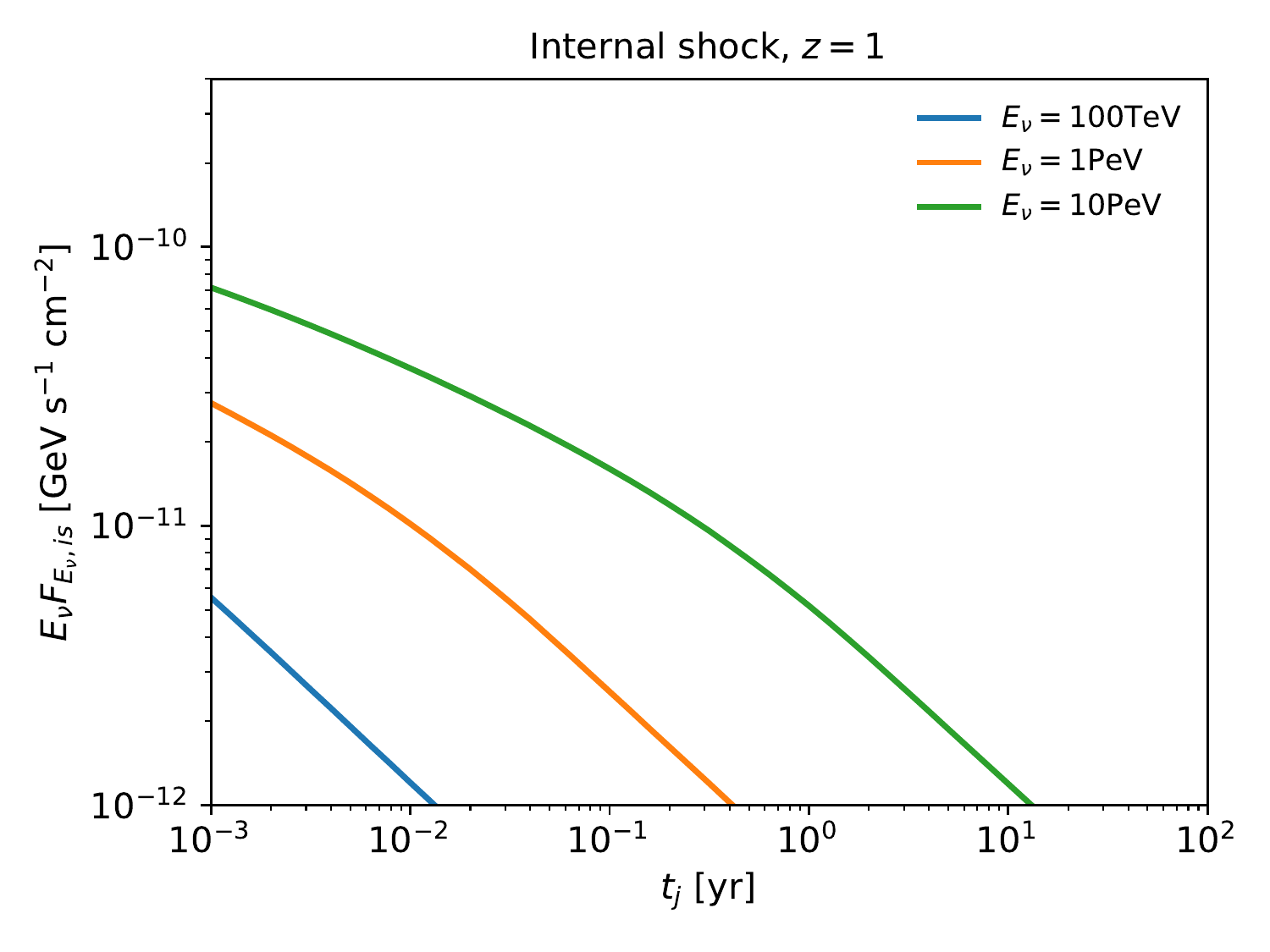}
    \centering\includegraphics[width=0.49\textwidth]{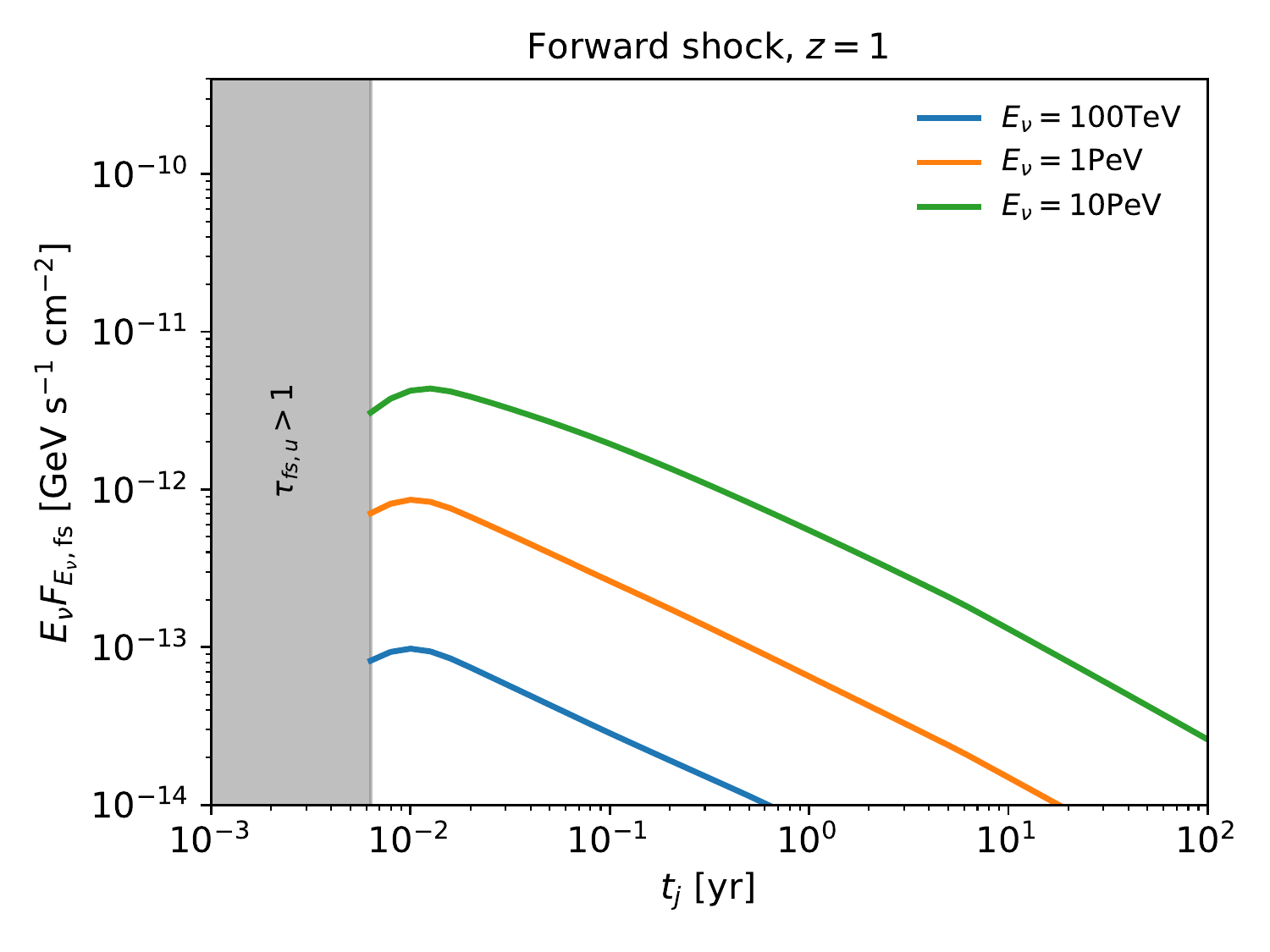}
    \centering\includegraphics[width=0.49\textwidth]{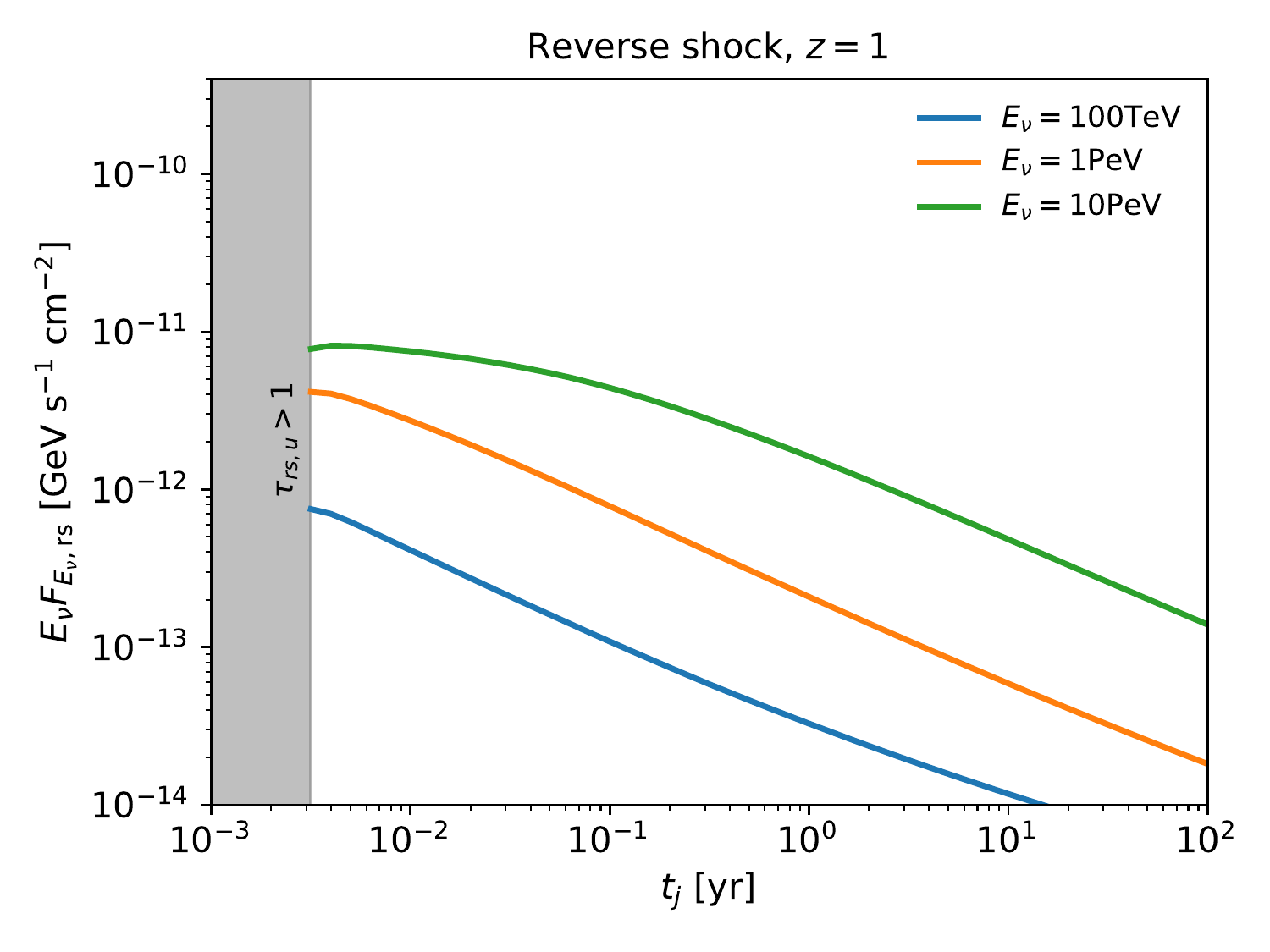}
    \caption{Muon neutrino fluxes versus jet time $t_j$ for the CS (up left), IS (up right), FS (bottom left) and RS (bottom right) scenarios. {The optimistic parameters (e.g., $\dot m=10, \ \epsilon_p=0.5$) are used.} The blue, orange and green curves correspond to the specified neutrino energies in the observer's frame $E_\nu=100\rm\ TeV$, $1\ \rm PeV$ and $10\ \rm PeV$. For the FS and RS cases, the neutrino emissions are isotropic and $L_{k,j}$ is used in Eq.~(\ref{eq:neu_spec}) instead of $L_{k,\rm iso}.$ The relativistic jet is on-axis and located at $z=1$.}
    \label{fig:light_curves}
\end{figure*}
%%%%%%%%%%%%%%%%% sec 4

\section{\label{sec:results} High-energy neutrino Emission from shocks in the jets}
\subsection{Neutrino fluences}
\begin{figure*}[]
    \centering
    \includegraphics[width=0.49\textwidth]{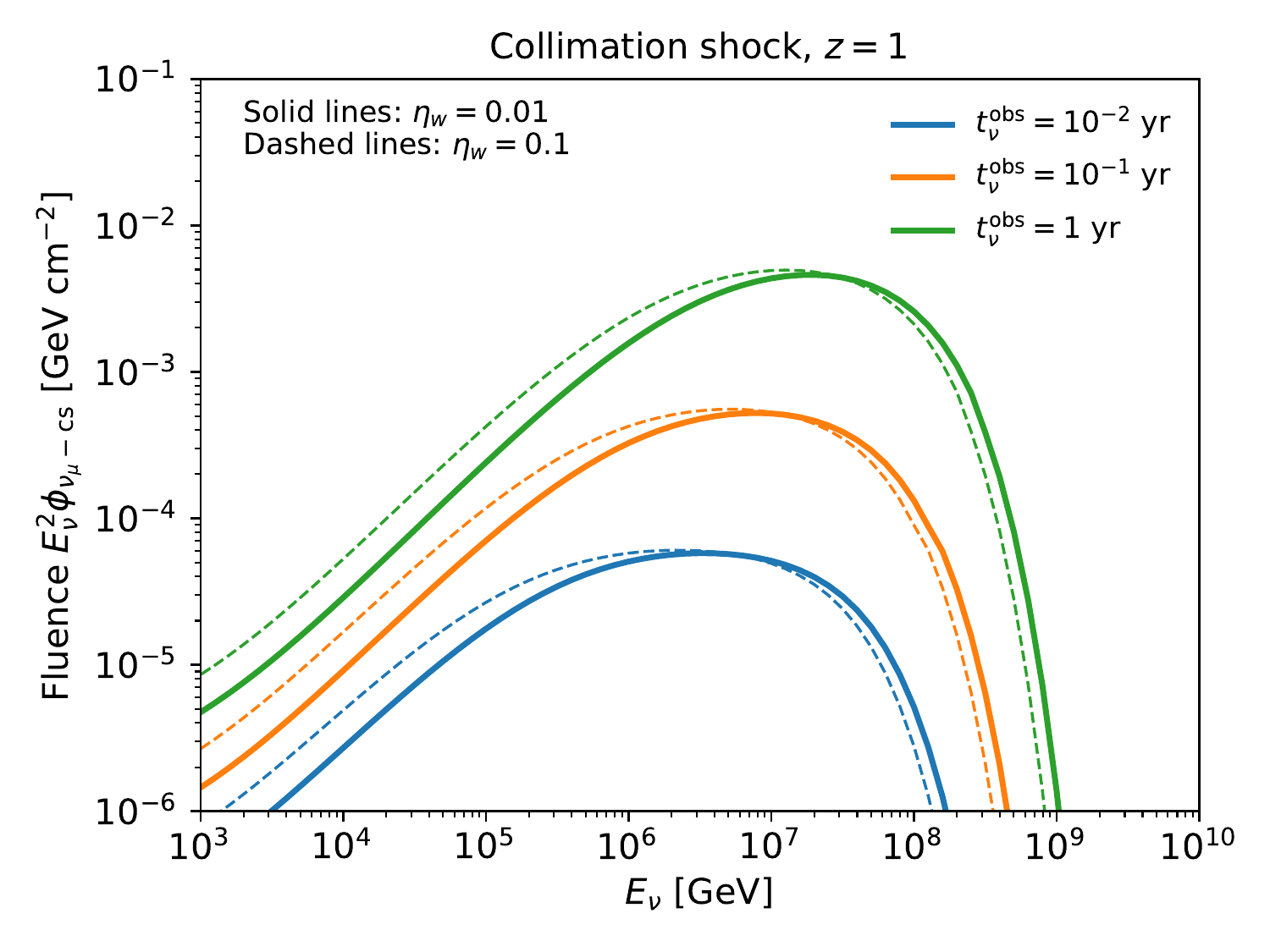}
    \includegraphics[width=0.49\textwidth]{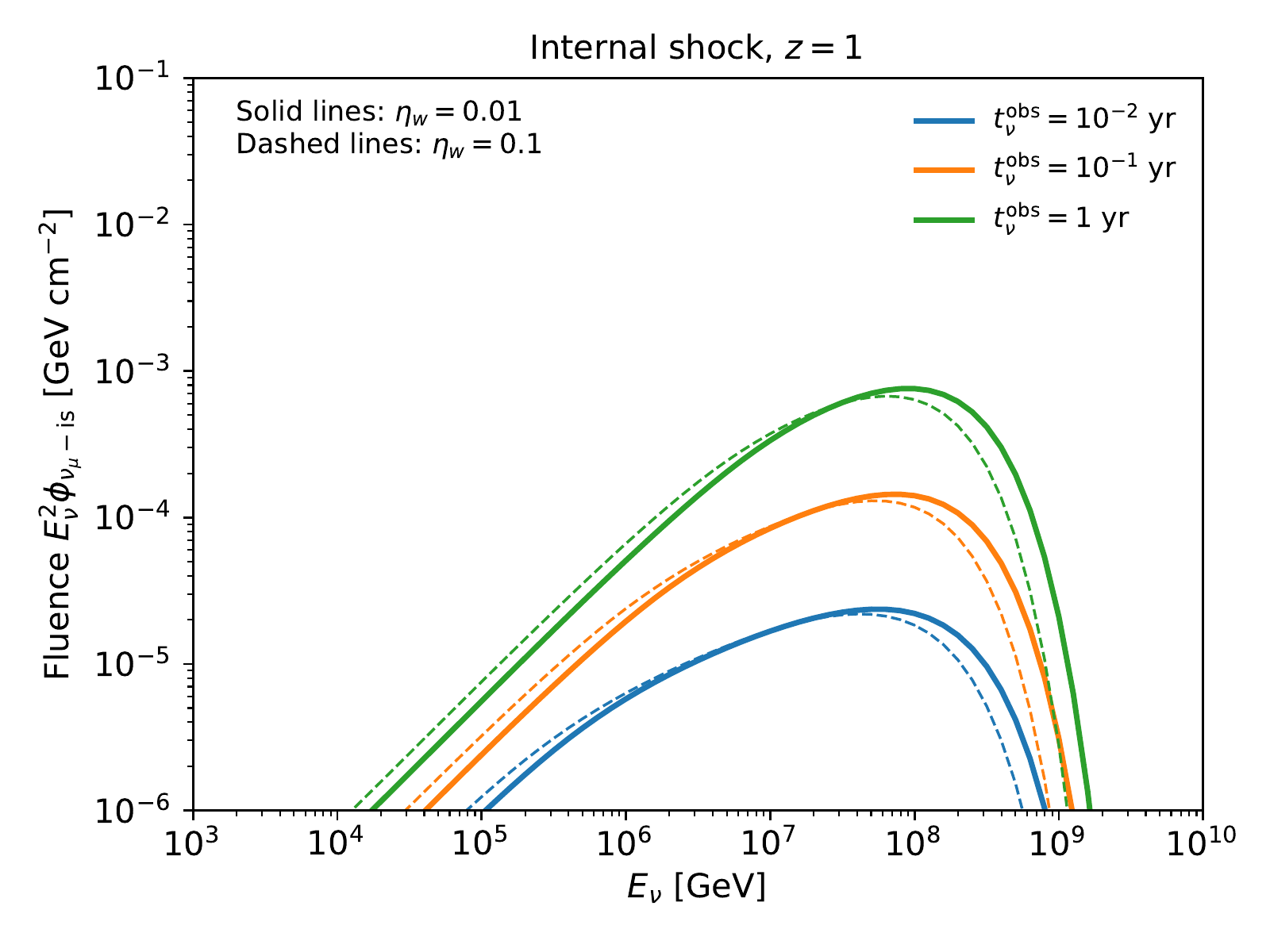}
    \includegraphics[width=0.49\textwidth]{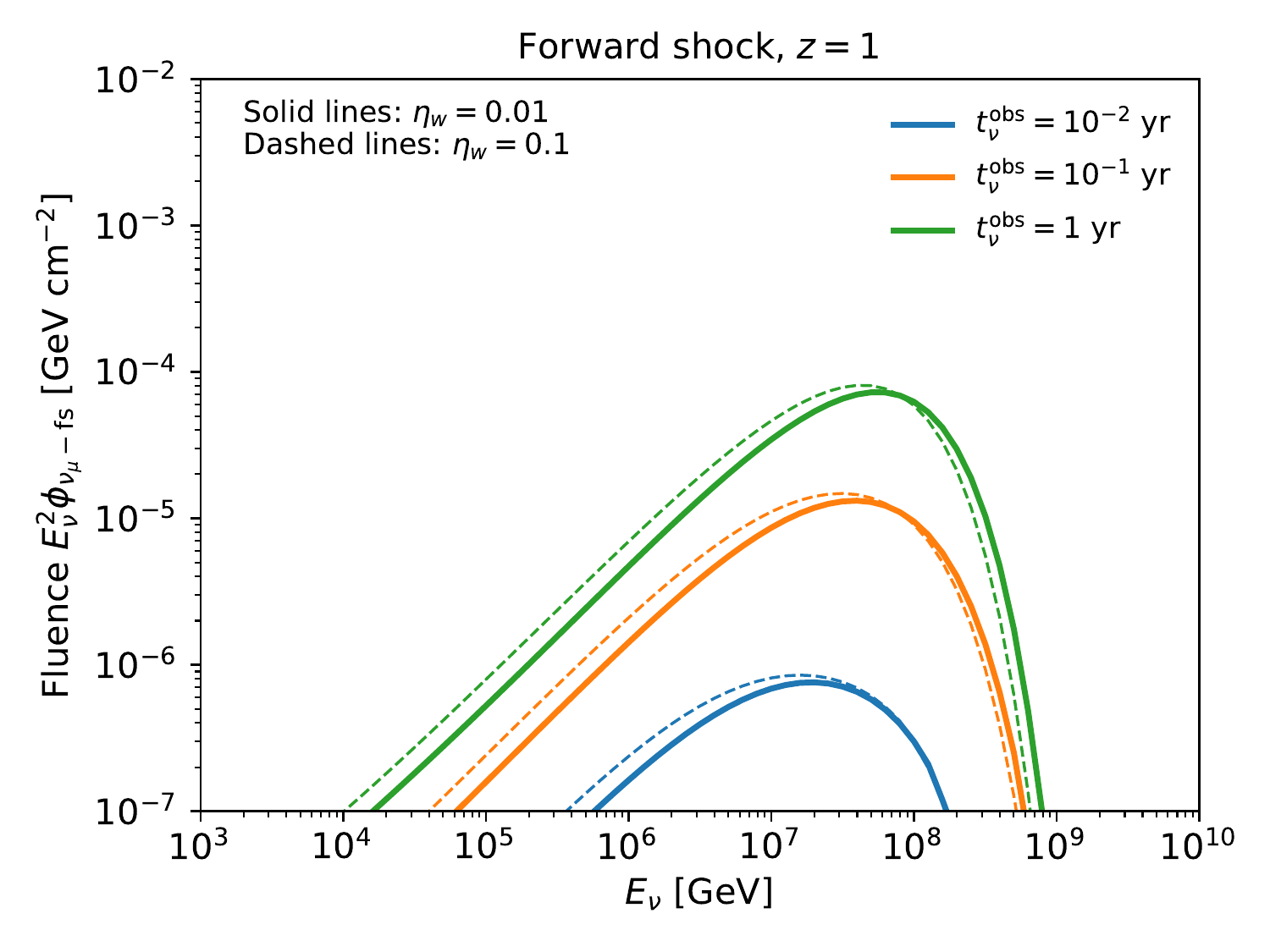}
    \includegraphics[width=0.49\textwidth]{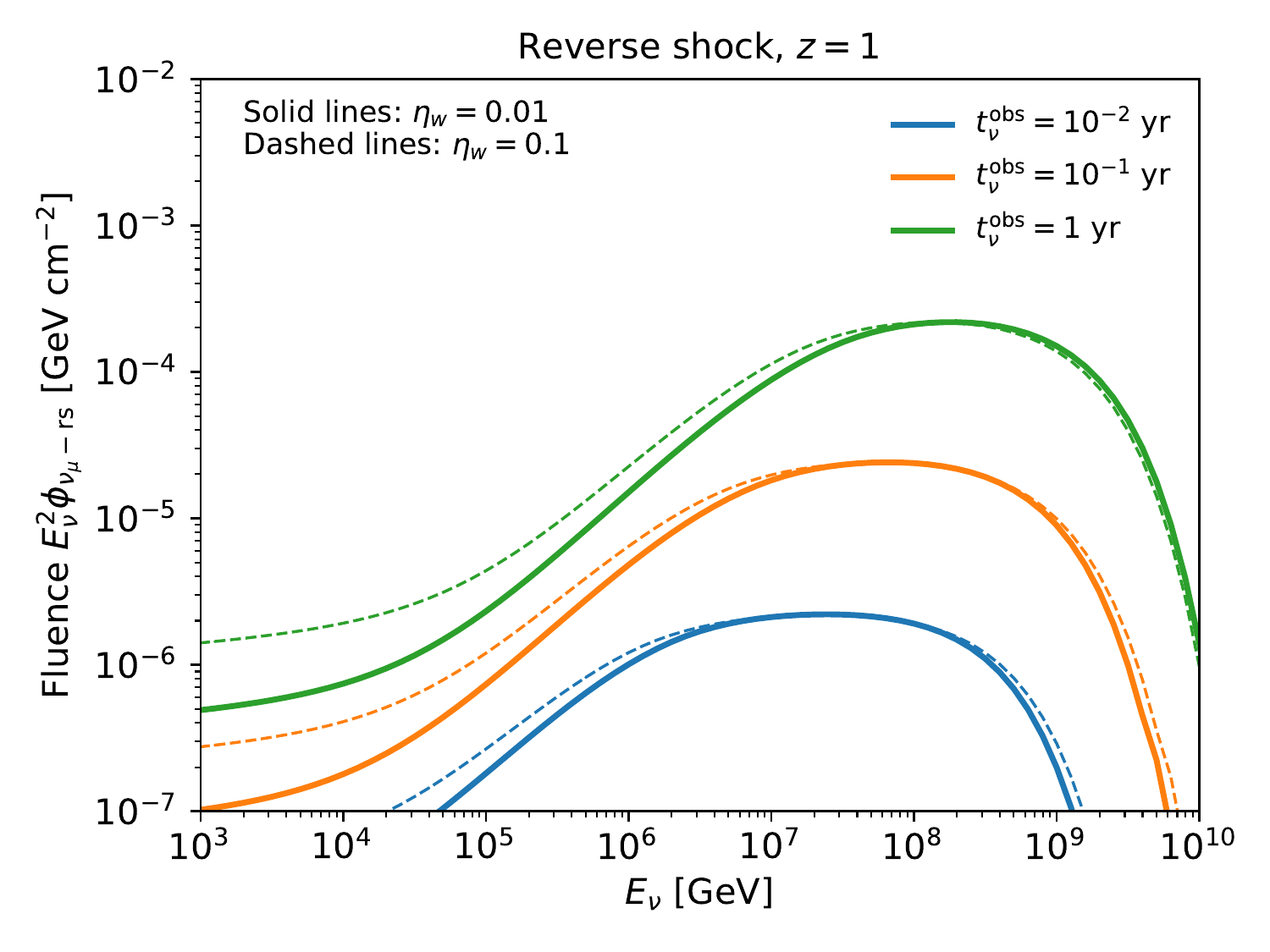}
    \caption{{Observed muon} neutrino fluences for the CS (up left), IS (up right), FS (bottom left) and RS (bottom right) scenarios at various {observation times $t_\nu^{\rm obs}=10^{-2}$ yr (blue lines), $10^{-1}$ yr (orange lines) and $1$ yr (green lines) after the merger}. {The optimistic parameters (e.g., $\dot m=10, \ \epsilon_p=0.5$) are used to obtain these curves.} {The solid lines are obtained from fiducial parameters, e.g., $\eta_w=0.01$, whereas $\eta_w=0.1$ is used for the thin dashed lines as a reference.} For the FS and RS cases, the neutrino emissions are isotropic and $L_{k,j}$ is used in Eq.~(\ref{eq:neu_spec}) instead of $L_{k,\rm iso}.$ The relativistic jet is on-axis and located at $z=1$.}
    \label{fig:neu_fluences}
\end{figure*}

Assuming that the high-energy protons have the canonical shock acceleration spectrum with a spectral index $p=2$ and an exponential cutoff at the maximum proton energy, we obtain the single flavor isotropic neutrino spectrum by pion decay at each site in the observer's frame 
\be\begin{split}
    E_\nu F_{E_\nu, i}\approx &\frac{\epsilon_p L_{k,\rm iso}}{4\pi d_L^2\mathcal C_p}\left(\frac{1}{8}f_{p\gamma-i}+\frac{1}{6}f_{pp-i}\right)f_{\pi,{\rm sup}-i}\\
    &\times H(t_j-t_*)e^{-\frac{\varepsilon_p}{\varepsilon_{p,\rm max}}}|_{E_\nu\approx0.05\varepsilon_p(1+z)^{-1}},
\end{split}
\label{eq:neu_spec}
\ee
where the label $i$=CS, IS, FS or RS represents the site of neutrino production, $\epsilon_p$ is the CR acceleration efficiency, $\mathcal C_p=\ln(\varepsilon_{p,\rm max}/{\varepsilon_{p,\rm min}})$ is the normalization parameter, $\varepsilon_{p,\rm min}\approx \Gamma_{\rm cj}\Gamma_{\rm rel-i}m_pc^2$ is the proton minimum energy in the cosmological comoving frame, $\varepsilon_{p,\rm max}$ is the maximum proton energy, and $d_L$ is the luminosity distance between the source and the observer. {In this paper, we assume efficient baryon loading rate $\epsilon_p=0.5$.} 
Noting that the maximum proton energy is constrained by the cooling energy $\varepsilon_{p,c}$ and the maximum proton energy from acceleration $\varepsilon_{p,\rm acc}$ in the jet comoving frame, we conclude that {$\varepsilon_{p,\rm max}\approx{\Gamma_{cj}}\min[\varepsilon_{p,c},\varepsilon_{p,\rm acc}]$, where {$\varepsilon_{p,c}$ is determined by the equation $t_{p,c}^{-1}+t_{i,\rm dyn}^{-1}=t_{p,\rm acc}^{-1}$}}. For the FS and RS cases, considering that these shocks are initially relativistic and then rapidly decrease to being sub-relativistic as the jet expands, we expect that the corresponding neutrino emissions are not beamed and we replace $L_{k,\rm iso}$ with $L_{k,j}$ in Eq.~(\ref{eq:neu_spec}). In the following text, we show the neutrino light curves and spectra for each site by fixing the luminosity distance to be $d_L=6.7\ \rm Gpc$ ($z=1$); {(see section \ref{sec:detectability} for the reason of this choice).}
Fig. \ref{fig:light_curves} shows the light curves for specified neutrino energies $E_\nu=100\rm\ TeV$ (blue lines), $1\ \rm PeV$ (orange lines) and $10\ \rm PeV$ (green lines). As for the forward shock and the reverse shock, no neutrinos are expected before the onset time $t_*$. {One common feature for all the four light curves is that the neutrino fluxes decreases monotonically in the later time, due to a decreasing $f_{p\gamma}$ resulting from a less denser photon environment.}

{For the convenience of the detectability discussion below, it is useful to calculate the observed cumulative {muon neutrino} fluence at a given time {$t_\nu^{\rm obs}$ after the jet is launched} by integrating the flux over time 
\begin{equation}
    E_\nu^2\phi_{\nu_\mu-i}(t_{\nu}^{\rm obs})=\int_{0}^{ {t_\nu^{\rm obs}/(1+z)}}dt_jE_{\nu}F_{E_{\nu},i}.
    \label{eq:fluence}
\end{equation}}
Cumulative {muon} neutrino fluences for various observation times $t_\nu^{\rm obs}=10^{-2}$ yr, $10^{-1}$ yr and $1$ yr for CS, IS, FS and RS scenarios in the optimistic case are shown in Fig. \ref{fig:neu_fluences}. {From Fig. \ref{fig:neu_fluences}, we find that the neutrino flux from IS is subdominant comparing to that from CS. The main reason is that the comoving photon density at IS is much lower than the CS site, noting that $n_{\gamma,\rm cs}\propto \Gamma_{\rm cj}^{-2}$ whereas $n_{\gamma,\rm is}\propto \Gamma_j^{-2}$.} {The thin dashed lines in Fig. \ref{fig:neu_fluences} depict the corresponding neutrino fluences for a denser circumnuclear material with $\eta_w=0.1$. Comparing with the solid lines, we conclude that the neutrino emission does not sensitively depend on $\eta_w$ and the results obtained from previous assumptions are not sensitive to the uncertainties of the outflow model.} The neutrino fluences of the FS and RS scenarios are clearly lower than for the CS and IS cases since the neutrinos from the FS and RS are not beamed. 

{To calculate the observed flavor ratio, we write down the ratio of neutrino fluences of different flavors at the source $\nu_\mu:\nu_e:\nu_\tau\sim 1:2:0$. According to tri-bimaximal mixing, the observed neutrino fluences after long-distance oscillation is (e.g., Ref.~\cite{harrison2002tri})
\begin{equation}
    \begin{split}
        \phi_{\nu_e}&=\frac{10}{18}\phi_{\nu_e}^0+\frac{4}{18}\left(\phi_{\nu_\mu}^0+\phi_{\nu_\tau}^0\right)\\
        \phi_{\nu_\mu}&=\frac{4}{18}\phi_{\nu_e}^0+\frac{7}{18}\left(\phi_{\nu_\mu}^0+\phi_{\nu_\tau}^0\right)
    \end{split}
\end{equation}
implying that the observed favor ratio is $\nu_\mu:\nu_e:\nu_\tau \sim 1:1:1.$ We need to keep in mind that the flavor ratio may deviate from $1:1:1$ if the muon decay suppression factor becomes less than unity, e.g., $f_{\mu,\rm sup}<1$.
}
\subsection{Detectability}
\label{sec:detectability}
\begin{table*}
    \centering
    \caption{Detectability of jet-induced {muon} neutrino emissions by IceCube (IC) and IceCube-Gen2 (IC-Gen2)}
    Neutrino event number $\mathcal N_i$ ($t_\nu^{\rm obs}=1\rm \ yr$) for an on-axis source at $d_L=$6.7 Gpc $(z=1)$ 
    \begin{ruledtabular}
    \begin{tabular}{c|ccc|ccc}
        &\multicolumn{3}{c|}{$\ $Optimistic parameters$\ $}&\multicolumn{3}{c}{Conservative parameters}\\
        &\multicolumn{3}{c|}{$\dot m=10,\ L_{k,j}\simeq3.4\times10^{46}\ \rm erg\ s^{-1}$,$\ \epsilon_p=0.5,\ h=0.3$}&\multicolumn{3}{c}{$\dot m=0.1,\ L_{k,j}\simeq3.4\times10^{44}\ {\rm erg\ s^{-1}},\ \epsilon_p=0.5$, $h=0.01$}\\
        \hline
        Scenario & IC (up+hor) & IC (down) & IC-Gen2 (up+hor) & IC (up+hor) & IC (down) & IC-Gen2 (up+hor)\\
         \hline
        CS &0.031 &$0.027$ &0.21 &$1.2\times10^{-4}$ &$9.5\times10^{-5}$&7.6$\times10^{-4}$\\
        IS &$1.3\times10^{-3}$&$1.1\times10^{-3}$&$7.0\times10^{-3}$&$2.6\times10^{-6}$&$2.3\times10^{-6}$&$1.5\times10^{-5}$\\
        FS &$2.6\times10^{-4}$&$2.1\times10^{-4}$&$1.4\times10^{-3}$&$8.1\times10^{-6}$&$6.2\times10^{-6}$&$4.1\times10^{-5}$\\
        RS &$8.8\times10^{-4}$&$7.2\times10^{-4}$&$4.7\times10^{-3}$&$3.6\times10^{-5}$&$2.7\times10^{-5}$&$1.8\times10^{-4}$ \\
     \end{tabular}
    \end{ruledtabular}
    
\smallskip
Neutrino event number $\mathcal N_i$ ($t_\nu^{\rm obs}=10\rm \ yr$) for an on-axis source at $d_L=$6.7 Gpc $(z=1)$ 
    \begin{ruledtabular}
    \begin{tabular}{c|ccc|ccc}
        &\multicolumn{3}{c|}{$\ $Optimistic parameters$\ $}&\multicolumn{3}{c}{Conservative parameters}\\
        &\multicolumn{3}{c|}{$\dot m=10,\ L_{k,j}\simeq3.4\times10^{46}\ \rm erg\ s^{-1}$,$\ \epsilon_p=0.5,\ h=0.3$}&\multicolumn{3}{c}{$\dot m=0.1,\ L_{k,j}\simeq3.4\times10^{44}\ {\rm erg\ s^{-1}},\ \epsilon_p=0.5$, $h=0.01$}\\
        \hline
        Scenario & IC (up+hor) & IC (down) & IC-Gen2 (up+hor) & IC (up+hor) & IC (down) & IC-Gen2 (up+hor)\\
         \hline
        CS &0.17 &$0.14$ &1.04 &$6.1\times10^{-4}$ &$4.7\times10^{-4}$&3.2$\times10^{-3}$\\
        IS &$6.9\times10^{-3}$&$4.9\times10^{-3}$&$2.6\times10^{-3}$&$1.01\times10^{-5}$&$8.9\times10^{-6}$&6.1$\times10^{-5}$\\
        FS &$1.1\times10^{-3}$&$8.4\times10^{-4}$&$5.1\times10^{-3}$&$3.1\times10^{-5}$&$2.4\times10^{-5}$&$2.4\times10^{-4}$\\
        RS &$3.3\times10^{-3}$&$2.9\times10^{-3}$&$1.9\times10^{-2}$&$1.4\times10^{-4}$&$9.8\times10^{-5}$&$8.1\times10^{-4}$ \\
     \end{tabular}
    \end{ruledtabular}
\smallskip
    Neutrino detection rate $\dot N_{\nu,i}$ for SMBH mergers within the LISA detection range $z\lesssim6$ $[\rm yr^{-1}]$
    \begin{ruledtabular}
    \begin{tabular}{c|ccc|ccc}
        %&\multicolumn{3}{c|}{$\dot m=5,\ L_{k,j}\simeq2.93\times10^{46}\ \rm erg\ s^{-1}$}&\multicolumn{3}{c}{$\dot m=0.5,\ L_{k,j}\simeq2.93\times10^{45}\ \rm erg\ s^{-1}$}\\
        &\multicolumn{3}{c|}{$\ $Optimistic parameters$\ $}&\multicolumn{3}{c}{Conservative parameters}\\
        &\multicolumn{3}{c|}{$\dot m=10,\ L_{k,j}\simeq3.4\times10^{46}\ \rm erg\ s^{-1}$,$\ \epsilon_p=0.5,\ h=0.3$}&\multicolumn{3}{c}{$\dot m=0.1,\ L_{k,j}\simeq3.4\times10^{44}\ {\rm erg\ s^{-1}},\ \epsilon_p=0.5$, $h=0.01$}\\
        \hline
        Scenario & IC (up+hor) & IC (down) & IC-Gen2 (up+hor) & IC (up+hor) & IC (down) & IC-Gen2 (up+hor)\\
         \hline
        CS &0.019&$0.014$&0.16&$8.2\times10^{-5}$&$4.3\times10^{-5}$&3.7$\times10^{-4}$\\
        IS &$9.1\times10^{-4}$&$7.8\times10^{-4}$&$4.2\times10^{-3}$&$1.7\times10^{-6}$&$1.3\times10^{-6}$&$9.5\times10^{-6}$\\
        FS &$2.6\times10^{-3}$&$1.8\times10^{-3}$&$0.013$&$9.6\times10^{-5}$&$7.2\times10^{-5}$&$4.1\times10^{-4}$\\
        RS &$0.011$&8.4$\times10^{-3}$&0.044&$3.5\times10^{-4}$&$1.9\times10^{-4}$&$2.1\times10^{-3}$ \\
     \end{tabular}
    \end{ruledtabular}
    \label{tab:event_rate}
\end{table*}
Using the {muon} neutrino fluence $\phi_{\nu_\mu-i}$ at $t_\nu^{\rm obs}$ and the detector effective area $A_{\rm eff}(\delta,E_\nu)$, we estimate the observed muon neutrino event number to be 
\begin{equation}
    \mathcal N_i(t_\nu^{\rm obs})=\int\phi_{\nu_\mu-i}A_{\rm eff}(\delta,E_\nu)dE_\nu,
    \label{eq:event_num}
\end{equation}
where $A_{\rm eff}$ typically depends on the declination $\delta$. For IceCube (IC), the effective areas for 79- and 86-string configurations are similar and we use the $A_{\rm eff}$ shown in  \cite{ICeffectivearea} to calculate the 1-year event numbers of downgoing and upgoing+horizontal neutrinos. In the future, foreseeing a substantial expansion of the detector size, IceCube-Gen2 is expected to have a larger effective area \cite{aartsen2014icecube}. Here we assume that the effective area of IceCube-Gen2 (IC-Gen2) is a factor of $10^{2/3}$ larger that that of IceCube. The threshold neutrino energies for IceCube and IceCube-Gen2 are fixed to be 0.1 TeV and 1 TeV respectively. In our case, we focus on the detectability of track events considering that the effective area for shower events is much smaller that that of track events. Note that we only consider the contribution of upgoing+horizontal neutrinos. KM3NeT, a network of deep underwater neutrino detectors that will be constructed in the Mediterranean Sea \cite{adrian2016letter}, will cover the southern sky and will further enhance the discovery potential of the jets produced by SMBH mergers as neutrino sources in the near future.

We calculate the expected one-year, e.g., $t_\nu^{\rm obs}=1 \rm\ yr$, neutrino detection numbers of the CS, IS, FS and RS scenarios for an on-axis merger event located at $z=1$ ($\sim6.7 $ Gpc) with the parameters used before, $\dot m=10,\ \epsilon_p=0.5,\ \eta_w=10^{-2},\ \eta_j=3,\ \epsilon_e=0.1$, $\epsilon_B=0.01$, $\Gamma_j=10$ and $\Gamma_{cj}=\theta_j^{-1}=3$. The results are summarized in the upper part of Table \ref{tab:event_rate}. {Correspondingly, the middle panel of Table \ref{tab:event_rate} shows the expected event numbers for IceCube and IceCube-Gen2 in the 10-year operation (e.g., $t_{\nu}^{\rm obs}=10\ \rm yr$).} One caveat is that the accretion rate as well as the jet luminosity $L_{k,j}$ might be optimistic for SMBH mergers. Hence, we show also the results for a conservative case with a sub-Eddington accretion rate $\dot m=0.1$ and the same baryon loading factor $\epsilon_p=0.5$, for the purposes of comparison. In this case, the other parameters are unchanged except for modifying the disk scale height to $h=0.01$, which is consistent with thin disk models of low mass accretion rates. 
The event numbers in the upper and middle parts of Table \ref{tab:event_rate} demonstrate that IceCube-Gen2 could detect $\gtrsim1$ events from an on-axis source located at $z=1$ in a 10-year operation period, whereas the detection is difficult for IceCube.

It is also useful to discuss the neutrino detection rate for all SMBH mergers within a certain comoving volume $\mathcal V(z_{\rm lim})$ at redshift $z_{\rm lim}$. Given the SMBH merger rate $\mathcal R(z)$, the number of mergers per unit comoving volume per unit time, and assuming that all SMBH mergers are identical, we obtain the average neutrino detection rate per year from the $i$-th component~\cite{Murase:2007ar,murase2016constraining}
\begin{equation}\begin{split}
    \dot N_{\nu,i}(<z_{\rm lim})=& \frac{c}{H_0}f_i(\theta_j)\Delta\Omega_{\rm sur}\\
    &\times\int_0^{z_{\rm lim}}dz\frac{ P_{m\geq1}(\mathcal N_i|_{t_\nu^{\rm obs}=\rm 1\ yr})\mathcal R(z)d_L^2}{(1+z)^3\sqrt{\Omega_m(1+z)^3+\Omega_\Lambda}},
\end{split}
    \label{eq:event_rate}
\end{equation}
where $f_i(\theta_j)$ is the probability of on-axis mergers and the solid angle is $\Delta_{\rm sur}\approx 2\pi$ for the upgoing+horizontal detections, and $P_{m\geq1}(\mathcal N_i|_{t_\nu^{\rm obs}=\rm 1\ yr})=1-\exp(-\mathcal N_i|_{t_\nu^{\rm obs}=\rm 1\ yr})$ is the probability that a sigle source at $z$ produces nonzero neutrino events. 
For $i=$ CS and IS, the neutrino emission is beamed and we conclude that $f_i(\theta_j)=\theta_j^2/2$, whereas $f_i=1$ corresponds to the isotropic FS and RS. 
Note that the critical redshift that satisfies $\frac{1}{2}\theta_j^2\mathcal R(z)\mathcal V(z)\times 1\ \rm yr\sim1$ is $z\sim1$, within which one may expect one on-axis merger in one year. 
Simulations based on the history of dark matter halo mergers \cite{menou2001merger,erickcek2006supermassive} and the history of seed black hole growth \cite{micic2007supermassive} have predicted the redshift evolution of SMBH merger rate, and we use the results of Ref.~\cite{micic2007supermassive} for $\mathcal R(z)$.  
%These works show that there are approximately 10-20 SMBH mergers per year in the range $z<1$ and the exact value depends on the details and preconditions of the simulation. 

It has been expected that LISA can detect SMBHs up to high redshifts (see, e.g., Ref.~\cite{berti2006lisa} and references therein). SMBH binary coalescences at high redshifts ($z\gtrsim2-3$) dominate the total event rate, whereas approximately 10\% of the event rate may come from the mergers at redshifts with $z\lesssim1$ ~\cite{Haehnelt:1994wt,menou2001merger,enoki2004gravitational,berti2006lisa}. The cumulative LISA event rate is expected to be $\sim10\ \rm yr^{-1}$~\cite{berti2006lisa}. But the number is subjected to large uncertainties coming from binary formation models. For example, Ref.~\cite{enoki2004gravitational} gives $\sim1\ \rm yr^{-1}$ for $M_{\rm BH}\sim10^6~M_\odot$. 
We are interested in the neutrino detection rate from SMBH mergers detected by LISA, i.e., GW+neutrino detection rate. 
Combining Eqs.~(\ref{eq:event_num}) and (\ref{eq:event_rate}), we present the neutrino detection rates for SMBH mergers by setting $z_{\rm lim}=6$ (given that LISA can detect such high-redshift SMBH mergers) in the lower part of Table \ref{tab:event_rate}. 
From the neutrino detection rates and event numbers presented in Table \ref{tab:event_rate}, we find that it may be challenging for IceCube-Gen2 to detect neutrinos from LISA-detected SMBH mergers with conservative parameters ($\dot m=0.1$). On the other hand, if the LISA-detected binary SMBH systems are super-Eddington accreters (e.g., $\dot m=10$) before and after the merger, the resulting neutrino emission from the jet-induced shocks may be detected by IceCube-Gen2 within a decade. Note that the atmospheric neutrino background would be negligibly small even for a time window of $t_{\nu}^{\rm obs}\sim1$~yr because the neutrino energy is expected to be very high.  
% Hence, we roughly estimate the GW + neutrino ten-year ($t_\nu^{\rm obs}=10\ \rm yr$) coincident number for LISA and IceCube-Gen2, $\sum_i\int dz \mathcal N_i(t_\nu^{\rm obs})|_zf_i\mathcal R_{\rm LISA}(z)t_\nu^{\rm obs}\sim0.2$, where we use the optimistic parameter set.

\begin{figure}
    \includegraphics[width=0.50\textwidth]{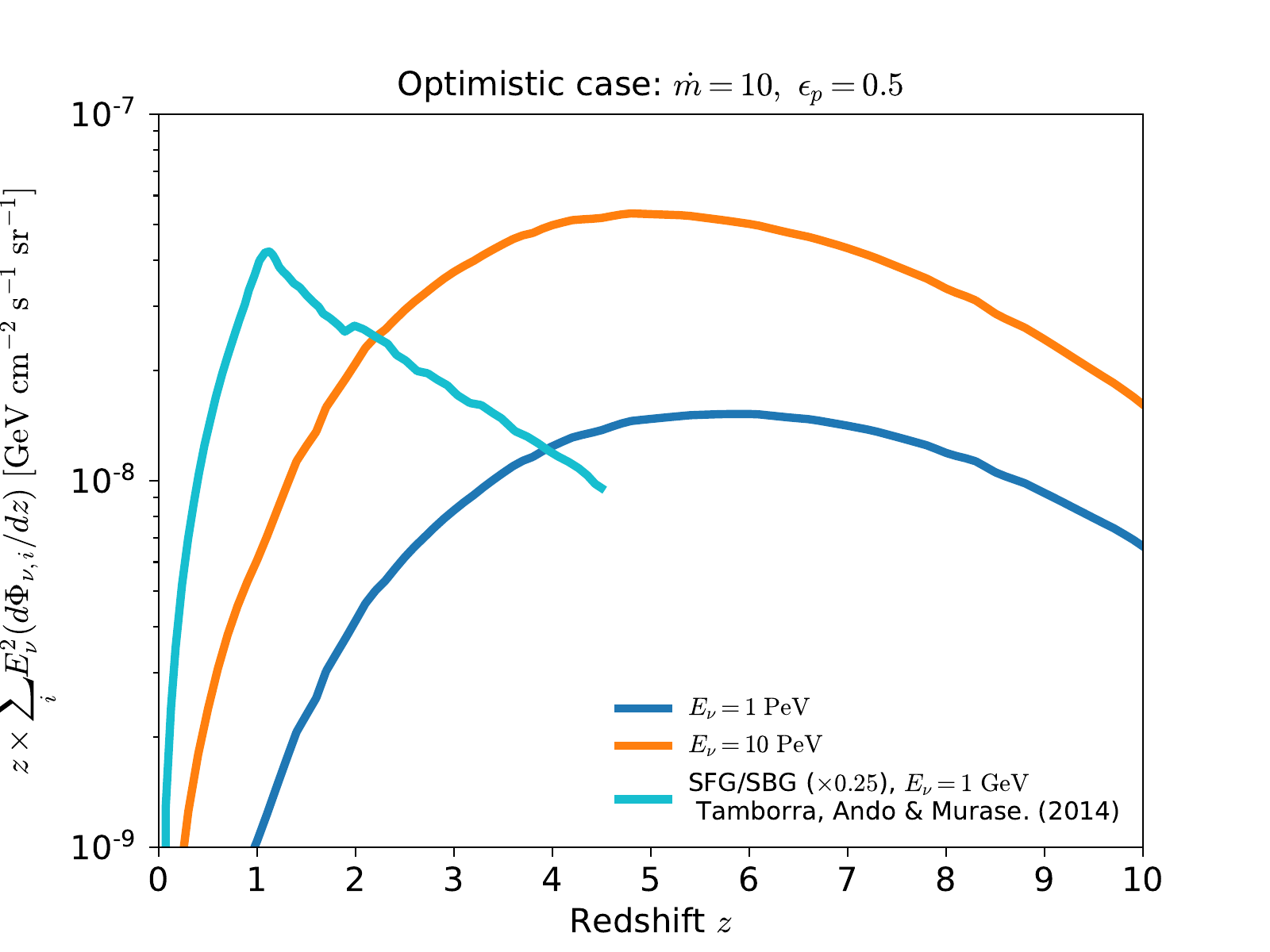}
    \caption{Differential contributions to the diffuse neutrino intensity $z\times \sum_iE_\nu^2 (d\Phi_{{\nu,i}}/dz)$ for the optimistic case at $E_\nu=$1 PeV (blue line) and 10 PeV (orange line). The cyan line depicts the the contributions ($\times0.25$) from starforming/starburst galaxies (SFG/SBG) \cite{tamborra2014star} at $E_\nu=$1 GeV.}
    \label{fig:neu_background}
\end{figure}
\begin{figure}[h]
\includegraphics[width=0.5\textwidth]{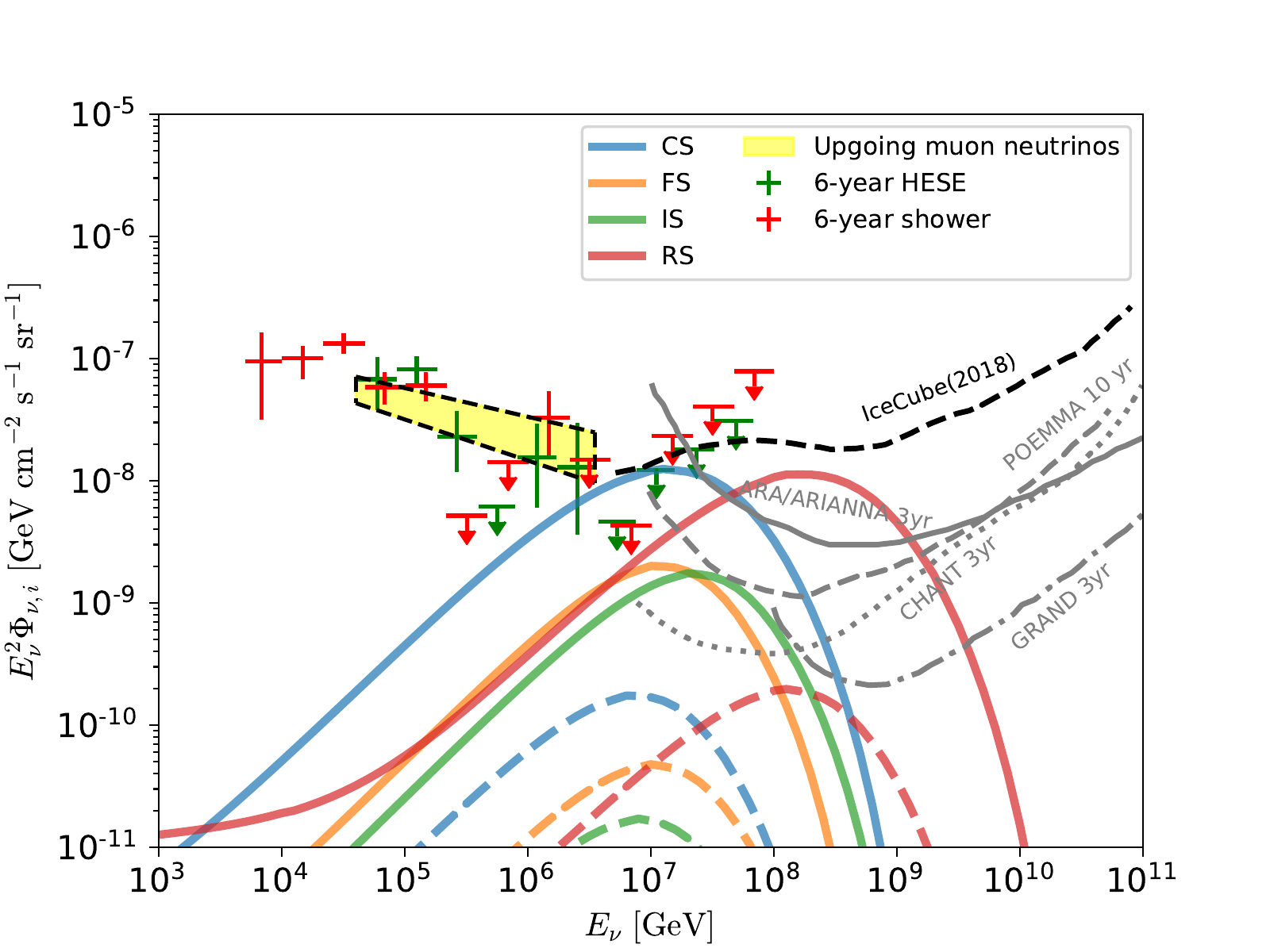}
\caption{Redshift-integrated {all-flavor diffuse neutrino flux expected} from relativistic jets in SMBH mergers. The CS, IS, FS and RS components are illustrated as blue, orange, green and red lines. The solid and dashed lines respectively correspond to the optimistic ($\dot m=10,\ \epsilon_p=0.5$) and conservative ($\dot m=0.1,\ \epsilon_p=0.5$) cases. The fiducial value $\eta_w=0.01$ is adopted for both cases.  {Parameters for these two cases are listed in table \ref{tab:event_rate}. For each case, we use $t_\nu=100\ \rm yr$ as the {rest-frame duration of the neutrino emission in the jets}.} {The 90\% C.L. Sensitivities of current (black-dashed; IceCube \cite{aartsen2018differential}) and some future ultrahigh-energy neutrino detectors (gray lines; ARA/ARIANNA, POEMMA, CHANT, GRAND) are also shown.}}
\label{fig:diffuse_flux}
\end{figure}

\subsection{Cumulative neutrino background}
It is useful to evaluate the contribution of SMBH mergers to the diffuse neutrino background and to check if this model can alleviate the tension between the diffuse neutrino and the gamma-ray backgrounds. In the scenario of jet induced neutrinos, the all-flavor diffuse neutrino flux from each site is calculated via~\cite{murase2014diffuse}
\begin{equation}
        \begin{split}
            E_\nu^2\Phi_{\nu,i}=&\frac{c}{4\pi H_0}\int dz\int^{t_\nu} dt_j\frac{\mathcal R(z)}{(1+z)^2\sqrt{\Omega_m(1+z)^3+\Omega_\Lambda}}\\
            &\times\left(\frac{3}{8}f_{p\gamma-i}+\frac{1}{2}f_{pp-i}\right)f_{\pi,{\rm sup}-i}\frac{\theta_j^2\epsilon_pL_{k,\rm iso}}{2\mathcal C_p}\\
            &\times H(t_j-t_*)e^{-\frac{\varepsilon_p}{\varepsilon_{p,\rm max}}},%|_{E_\nu\approx0.05\varepsilon_p(1+z)^{-1}}
    \end{split}
    \label{eq:neu_background}
\end{equation}
where the summation takes all neutrino production sites into account. From the light curves in Fig.~\ref{fig:light_curves}, we find that the neutrino emissions can last as long as one hundred years. To calculate the contribution to the diffuse neutrino background, we treat these jets as {long-duration} neutrino sources and take the rest-frame jet time to be 100 yr in the integral. {The Fig.~\ref{fig:neu_background} illustrates the differential contributions to the diffuse neutrino intensity, $z\times\sum_i E_\nu^2 (d\Phi_{{\nu,i}}/dz)$, for the optimistic parameters at $E_\nu=$ 1 PeV (blue line) and 10 PeV (orange lines). The fiducial parameter $\eta_w=0.01$ is used to obtain these curves. For the purpose of comparison, we also show in cyan the contribution ($\times0.25$) to $E_\nu=1\rm GeV$ neutrino background from starforming/starburst galaxies (SBG/SFG) \cite{tamborra2014star}.} Using the redshift evolution of SMBH merger rate $\mathcal R(z)$ provided by Ref.~\cite{micic2007supermassive}, we show the diffuse neutrino fluxes from each shock site for optimistic and conservative cases in Fig.~\ref{fig:diffuse_flux}. In this figure, the yellow area, green and red data points corresponds to the diffuse neutrino fluxes deduced
from upgoing muon neutrinos, six-year high-energy start events (HESE) analysis and six-year shower analysis \cite{stettner2019measurement,aartsen2015combined,aartsen2016observation,aartsen2020characteristics}, respectively. The results obtained from Eq.~(\ref{eq:neu_background}) is consistent with the analytical estimation~\cite{murase2014diffuse}
\begin{equation}
\begin{aligned}[t]
    E_\nu^2\Phi_{\nu}&\sim \frac{c}{4\pi H_0}\frac{3}{8}f_{p\gamma}f_{\pi,\rm sup}\xi_z \mathcal R|_{z=0} t_\nu \mathcal C_p^{-1}\epsilon_pL_{k,j}\\
    &\sim 10^{-8}~{\rm GeV\ cm^{-2}\ s^{-1}\ sr^{-1}}\\
    &\times(\epsilon_p\dot m/5)f_{p\gamma}\left(\frac{\xi_z}{12}\right)\left(\frac{t_{\nu,\rm eff} \mathcal R|_{z=0}}{0.11\rm Gpc^{-3}}\right),\\
    \end{aligned}
    \label{eq:background_analytical}
\end{equation}
where $f_{p\gamma}$ is close to unity at $E_\nu\sim10\rm PeV$ in the effective duration $t_{\nu,\rm eff}=10\rm\ yr$, $\mathcal C_p\simeq 15-20$ depends on the jet time and $\xi_z$ is the redshift evolution parameter (see e.g., Ref.~\cite{waxman1998high}). {Here, the analytical estimation is energy-dependent, since at different $E_\nu$, the effective neutrino emission time $t_{\nu,\rm eff}$, during which $f_{p\gamma}$ remains close to unity, strongly depends on the neutrino energy according to the light curves in Fig.~\ref{fig:light_curves}.} {From this figure we find that the CS and RS contribute to the diffuse neutrino flux roughly in the same level. The main reason is that these sites can continuously produce very high-energy neutrinos in a longer duration, e.g., $\sim 10$ yr (see the green curves in Figure \ref{fig:light_curves}). Moreover, since the dynamic time of the reverse shock $t_{\rm rs,dyn}\approx R_h/(\beta_hc)$ is longer than that of the collimation shock, $t_{\rm cs,dyn}\approx R_{\rm cs}/(\Gamma_{\rm cj}c)$, the reverse shock scenario predicts higher-energy neutrinos (in the EeV range).} 

One simplification in Eq.~(\ref{eq:neu_background}) is that all sources have the same physical conditions and share the same set of parameters throughout the universe. However, in reality, the situation is more complicated. 
Nevertheless, one can infer that the jet-induced neutrino emissions from SMBH mergers could significantly contribute to the diffuse neutrino flux {in the very high-energy range, i.e., $E_\nu\gtrsim 1$ PeV}, if the optimistic parameters are applied. 
%Considering the uncertainties, e.g., normalization factor $\mathcal R_0$, the emission time $t_\nu$ and the redshift evolution parameters $\xi_z$, the diffuse neutrino flux predicted by our model may vary by a factor of $\sim2$.

Since SMBH mergers are promising emitters of ultrahigh-energy neutrinos, these sources will become important candidates for future neutrino detectors, such as the giant radio array for neutrino detection (GRAND \cite{martineau2017giant}), Cherenkov from astrophysical neutrinos telescope (CHANT \cite{neronov2017sensitivity}), Probe Of Extreme Multi-Messenger Astrophysics (POEMMA \cite{venters2019poemma}), Askaryan Radio Array (ARA \cite{allison2012design}) and Antarctic
Ross Ice Shelf Antenna Neutrino Array (ARIANNA \cite{barwick2015first}). An absence of detection can in return constrain the jet luminosity/accretion rate and the source distribution. Typically, the source density and jet luminosity are constrained by the nondetection of multiplet sources \cite{murase2016constraining,senno2017high,ackermann2019astrophysics}. However, such multiplet constraint is very stringent in the energy range $E_\nu\sim30-100$ TeV (see e.g., Ref.~\cite{murase2016constraining}) and becomes very weak for $E_\nu\gtrsim10$ PeV. In this work, the neutrino emission concentrates in the ultrahigh-energy band, e.g., 10 PeV - 1 EeV, implying that the our model can avoid the multiplet constraint.

From the previous sections we find that the neutrino fluxes produced through the $pp$ process are negligible compared to that from $p\gamma$ interactions, implying a low contribution to the gamma-ray background in the GeV-TeV range covered by the $Fermi$ large Area Telescope (LAT). Most importantly, $p\gamma$ interactions in our model mainly produce very high-energy neutrinos of energies greater than 100 TeV. The accompanied very high-energy gamma rays can avoid the constraint from $Fermi$ LAT, since the gamma-ray constraint is stringent for neutrinos in the range 10-100 TeV if the source is dominated by $p\gamma$ interactions \cite{murase2016hidden}.  
On the other hand, according to the redshift evolution of the SMBH merger rate {and the differential contributions to the diffuse neutrino intensity shown in Fig.~\ref{fig:neu_background}, the sources located at high redshifts $z\sim4-6$ contribute a significant fraction of the cumulative neutrino background, and the sources are fast evolving objects with a redshift evolution parameter $\xi_z\sim12$. In this case the very high-energy gamma rays produced through $\pi^0$ decay can be sufficiently attenuated through $\gamma\gamma$ interactions with the extragalactic background light (EBL) and the cosmic microwave background (CMB; see, e.g., Ref.~\cite{franceschini2017extragalactic} for the optical depth). 
Hence, this model can significantly contribute to the very high-energy ($\gtrsim1$ PeV) diffuse neutrino background without violating the gamma-ray background observed by $Fermi$ LAT (cf. Figs.~5 and 6 of Ref.~\cite{Xiao:2016rvd}).

\section{\label{sec:summary}Summary and Discussion}
In this work, we studied jet-induced neutrino emission from SMBH mergers under the assumption that the jet is launched after the merger and it subsequently propagates inside the premerger circumnuclear formed by the disk wind which precedes the merger. 
We showed that with optimistic but plausible parameters, the overall neutrino emission from four different shock sites, CS, IS, FS and RS, can be detected by IceCube-Gen2 within ten years of operation. If the accretion rate of the newborn SMBHs are sub-Eddington, e.g., $\dot m=0.1$, it may be challenging to detect neutrinos even with IceCube-Gen2 because of the low SMBH merger rate in the local Universe. On the other hand, the expected rapid redshift evolution rate of SMBH mergers implies that they could be promising sources that contribute to the diffuse neutrino background. In the previous section, we found that even using the conservative parameters the SMBH merger scenario can significantly contribute to the diffuse neutrino background flux in the 1-100~PeV range. 
Importantly our model mainly produces very high-energy neutrinos of $E_\nu\gtrsim 1$~PeV via $p\gamma$ interactions, making it possible to simultaneously avoid the gamma-ray constraints. 

As noted before, one crucial parameter of the model is the mass accretion rate $\dot M_{\rm BH}=\dot m\dot M_{\rm Edd}$ since it determines the jet luminosity $L_{k,j}$. Many simulations have shown that the ratio of the mass loss rate by the wind and the accretion rate, $\eta_w=\dot M_w/\dot M_{\rm BH}$, strongly depends on the accretion rate, which implies that the density of the circumnuclear material is also sensitive to the accretion rate. In reality, the mass accretion rates before and after the merger may range from extreme sub-Eddington cases (e.g., $\dot m\sim10^{-4}$) to extreme super-Eddington cases ($\dot m\sim100$), depending on the model of accretion disks. We adopted the moderately super- and sub-Eddington accretion rates as fiducial values. With such assumptions, the ratio $\eta_w\sim10^{-2}$ used in our calculation is justified by the global three-dimensional radiation MHD simulations {\cite{Jiang2019a,Jiang2019,Ohsuga2009,firstm87}}. Our results show that with the reasonably optimistic parameters, $\dot m=10$ and $\epsilon_p=0.5$, it is possible for IceCube-Gen2 to see neutrinos from SMBH mergers within the operation of approximately ten years if the jet opening angle $\theta_j\sim0.3$ is comparable with that of AGN. 

Noting that a SMBH coalescence will produce strong GWs that will be detected by LISA, we discussed the expected coincident detection rates of both neutrinos and GWs. From the bottom part of Table \ref{tab:event_rate}, we found that it would be possible for LISA and IceCube-Gen2 to make a coincident detection of SMBH mergers within the observation of five to ten years in the optimistic case. One advantage of this model is that we can use the GW detection as the alert of the post-merger neutrino emission. The time lag between the GW burst and the prompt neutrino emission is approximately $\sim10^{-3}-10^{-2}\ \rm yr$ (hours to days, similar as $t_{\rm m}$ and/or $t_{\rm vis}$ in \S\ref{sec:phys-condition}), depending on the properties of the circumbinary disk. Since currently there does not exist an accurate function to describe the redshift and mass dependence of SMBH merger rate, the single-mass approximation adopted here will unavoidably leads to uncertainties in equation \ref{eq:event_rate}. In the future, the GW detections of SMBH mergers will shed more light on our understanding toward such systems and then our model can provide more accurate predictions on the GW+neutrino coincident detection rate.

The relativistic jets of SMBH mergers can also produce detectable electromagnetic emission, analogous to that of GRB afterglows. High-energy electrons that are accelerated in the relativistic shocks caused by the jets will produce high-energy photon emission through synchrotron radiation and inverse-Compton scattering. The recent detection of the IceCube-170922A neutrino coincident with the flaring blazar TXS 0506+056 shows that EM+neutrino multimessenger analyses are coming on stage and will play an increasingly important role in the future astronomy. It has been argued that the outburst signature of TXS 0506+056 could be caused by a ``binary" of two host galaxies and/or their SMBHs \cite{kun2019very,britzen2019cosmic} {(in which periodic neutrino emission can be expected by the jet precession~\cite{de2020recurrent})}, although the radio signatures may also be explained by structured jets~\cite{Ros:2019bgo}. In a continuation of this work, we will explore the electromagnetic signatures of the relativistic jets of SMBH mergers, which together with the results presented paper will provide more complete insights into the multimessenger study of SMBH mergers. 

\begin{acknowledgments}
We would like to thank Julia Becker Tjus, Ali Kheirandish and B. Theodore Zhang for fruitful discussions and comments. The work of K.M. is supported by the Alfred P. Sloan Foundation, NSF Grant No.~AST-1908689, and KAKENHI No.~20H01901. The work of S.S.K is supported by JSPS Research Fellowship and KAKENHI No. 19J00198. C.C.Y. and P.M. acknowledge support from the Eberly Foundation.
\end{acknowledgments}
\bibliography{smbh_merger.bib}

\begin{thebibliography}{122}%
\makeatletter
\providecommand \@ifxundefined [1]{%
 \@ifx{#1\undefined}
}%
\providecommand \@ifnum [1]{%
 \ifnum #1\expandafter \@firstoftwo
 \else \expandafter \@secondoftwo
 \fi
}%
\providecommand \@ifx [1]{%
 \ifx #1\expandafter \@firstoftwo
 \else \expandafter \@secondoftwo
 \fi
}%
\providecommand \natexlab [1]{#1}%
\providecommand \enquote  [1]{``#1''}%
\providecommand \bibnamefont  [1]{#1}%
\providecommand \bibfnamefont [1]{#1}%
\providecommand \citenamefont [1]{#1}%
\providecommand \href@noop [0]{\@secondoftwo}%
\providecommand \href [0]{\begingroup \@sanitize@url \@href}%
\providecommand \@href[1]{\@@startlink{#1}\@@href}%
\providecommand \@@href[1]{\endgroup#1\@@endlink}%
\providecommand \@sanitize@url [0]{\catcode `\\12\catcode `\$12\catcode
  `\&12\catcode `\#12\catcode `\^12\catcode `\_12\catcode `\%12\relax}%
\providecommand \@@startlink[1]{}%
\providecommand \@@endlink[0]{}%
\providecommand \url  [0]{\begingroup\@sanitize@url \@url }%
\providecommand \@url [1]{\endgroup\@href {#1}{\urlprefix }}%
\providecommand \urlprefix  [0]{URL }%
\providecommand \Eprint [0]{\href }%
\providecommand \doibase [0]{http://dx.doi.org/}%
\providecommand \selectlanguage [0]{\@gobble}%
\providecommand \bibinfo  [0]{\@secondoftwo}%
\providecommand \bibfield  [0]{\@secondoftwo}%
\providecommand \translation [1]{[#1]}%
\providecommand \BibitemOpen [0]{}%
\providecommand \bibitemStop [0]{}%
\providecommand \bibitemNoStop [0]{.\EOS\space}%
\providecommand \EOS [0]{\spacefactor3000\relax}%
\providecommand \BibitemShut  [1]{\csname bibitem#1\endcsname}%
\let\auto@bib@innerbib\@empty
%</preamble>
\bibitem [{\citenamefont {Abbott}\ \emph
  {et~al.}(2017{\natexlab{a}})\citenamefont {Abbott}, \citenamefont {Abbott},
  \citenamefont {Abbott}, \citenamefont {Acernese}, \citenamefont {Ackley},
  \citenamefont {Adams}, \citenamefont {Adams}, \citenamefont {Addesso},
  \citenamefont {Adhikari}, \citenamefont {Adya} \emph
  {et~al.}}]{abbott2017gw170817}%
  \BibitemOpen
  \bibfield  {author} {\bibinfo {author} {\bibfnamefont {B.~P.}\ \bibnamefont
  {Abbott}}, \bibinfo {author} {\bibfnamefont {R.}~\bibnamefont {Abbott}},
  \bibinfo {author} {\bibfnamefont {T.}~\bibnamefont {Abbott}}, \bibinfo
  {author} {\bibfnamefont {F.}~\bibnamefont {Acernese}}, \bibinfo {author}
  {\bibfnamefont {K.}~\bibnamefont {Ackley}}, \bibinfo {author} {\bibfnamefont
  {C.}~\bibnamefont {Adams}}, \bibinfo {author} {\bibfnamefont
  {T.}~\bibnamefont {Adams}}, \bibinfo {author} {\bibfnamefont
  {P.}~\bibnamefont {Addesso}}, \bibinfo {author} {\bibfnamefont
  {R.}~\bibnamefont {Adhikari}}, \bibinfo {author} {\bibfnamefont
  {V.}~\bibnamefont {Adya}},  \emph {et~al.},\ }\href@noop {} {\bibfield
  {journal} {\bibinfo  {journal} {Physical Review Letters}\ }\textbf {\bibinfo
  {volume} {119}},\ \bibinfo {pages} {161101} (\bibinfo {year}
  {2017}{\natexlab{a}})}\BibitemShut {NoStop}%
\bibitem [{\citenamefont {Abbott}\ \emph
  {et~al.}(2017{\natexlab{b}})\citenamefont {Abbott}, \citenamefont {Bloemen},
  \citenamefont {Canizares}, \citenamefont {Falcke}, \citenamefont {Fender},
  \citenamefont {Ghosh}, \citenamefont {Groot}, \citenamefont {Hinderer},
  \citenamefont {H{\"o}randel}, \citenamefont {Jonker} \emph
  {et~al.}}]{abbott2017multi}%
  \BibitemOpen
  \bibfield  {author} {\bibinfo {author} {\bibfnamefont {B.~P.}\ \bibnamefont
  {Abbott}}, \bibinfo {author} {\bibfnamefont {S.}~\bibnamefont {Bloemen}},
  \bibinfo {author} {\bibfnamefont {P.}~\bibnamefont {Canizares}}, \bibinfo
  {author} {\bibfnamefont {H.}~\bibnamefont {Falcke}}, \bibinfo {author}
  {\bibfnamefont {R.}~\bibnamefont {Fender}}, \bibinfo {author} {\bibfnamefont
  {S.}~\bibnamefont {Ghosh}}, \bibinfo {author} {\bibfnamefont
  {P.}~\bibnamefont {Groot}}, \bibinfo {author} {\bibfnamefont
  {T.}~\bibnamefont {Hinderer}}, \bibinfo {author} {\bibfnamefont
  {J.}~\bibnamefont {H{\"o}randel}}, \bibinfo {author} {\bibfnamefont
  {P.}~\bibnamefont {Jonker}},  \emph {et~al.},\ }\href {\doibase
  10.3847/2041-8213/aa91c9} {\bibfield  {journal} {\bibinfo  {journal} {The
  Astrophysical Journal Letters}\ }\textbf {\bibinfo {volume} {848}},\ \bibinfo
  {eid} {L12} (\bibinfo {year} {2017}{\natexlab{b}})}\BibitemShut {NoStop}%
\bibitem [{\citenamefont {Abbott}\ \emph {et~al.}(2016)\citenamefont {Abbott},
  \citenamefont {Abbott}, \citenamefont {Abbott}, \citenamefont {Abernathy},
  \citenamefont {Acernese}, \citenamefont {Ackley}, \citenamefont {Adams},
  \citenamefont {Adams}, \citenamefont {Addesso}, \citenamefont {Adhikari}
  \emph {et~al.}}]{abbott2016observation}%
  \BibitemOpen
  \bibfield  {author} {\bibinfo {author} {\bibfnamefont {B.~P.}\ \bibnamefont
  {Abbott}}, \bibinfo {author} {\bibfnamefont {R.}~\bibnamefont {Abbott}},
  \bibinfo {author} {\bibfnamefont {T.}~\bibnamefont {Abbott}}, \bibinfo
  {author} {\bibfnamefont {M.}~\bibnamefont {Abernathy}}, \bibinfo {author}
  {\bibfnamefont {F.}~\bibnamefont {Acernese}}, \bibinfo {author}
  {\bibfnamefont {K.}~\bibnamefont {Ackley}}, \bibinfo {author} {\bibfnamefont
  {C.}~\bibnamefont {Adams}}, \bibinfo {author} {\bibfnamefont
  {T.}~\bibnamefont {Adams}}, \bibinfo {author} {\bibfnamefont
  {P.}~\bibnamefont {Addesso}}, \bibinfo {author} {\bibfnamefont
  {R.}~\bibnamefont {Adhikari}},  \emph {et~al.},\ }\href@noop {} {\bibfield
  {journal} {\bibinfo  {journal} {Physical review letters}\ }\textbf {\bibinfo
  {volume} {116}},\ \bibinfo {pages} {061102} (\bibinfo {year}
  {2016})}\BibitemShut {NoStop}%
\bibitem [{\citenamefont {Abbott}\ \emph
  {et~al.}(2017{\natexlab{c}})\citenamefont {Abbott}, \citenamefont {Abbott},
  \citenamefont {Abbott}, \citenamefont {Acernese}, \citenamefont {Ackley},
  \citenamefont {Adams}, \citenamefont {Adams}, \citenamefont {Addesso},
  \citenamefont {Adhikari}, \citenamefont {Adya} \emph
  {et~al.}}]{abbott2017gw170814}%
  \BibitemOpen
  \bibfield  {author} {\bibinfo {author} {\bibfnamefont {B.~P.}\ \bibnamefont
  {Abbott}}, \bibinfo {author} {\bibfnamefont {R.}~\bibnamefont {Abbott}},
  \bibinfo {author} {\bibfnamefont {T.}~\bibnamefont {Abbott}}, \bibinfo
  {author} {\bibfnamefont {F.}~\bibnamefont {Acernese}}, \bibinfo {author}
  {\bibfnamefont {K.}~\bibnamefont {Ackley}}, \bibinfo {author} {\bibfnamefont
  {C.}~\bibnamefont {Adams}}, \bibinfo {author} {\bibfnamefont
  {T.}~\bibnamefont {Adams}}, \bibinfo {author} {\bibfnamefont
  {P.}~\bibnamefont {Addesso}}, \bibinfo {author} {\bibfnamefont
  {R.}~\bibnamefont {Adhikari}}, \bibinfo {author} {\bibfnamefont
  {V.}~\bibnamefont {Adya}},  \emph {et~al.},\ }\href@noop {} {\bibfield
  {journal} {\bibinfo  {journal} {Physical review letters}\ }\textbf {\bibinfo
  {volume} {119}},\ \bibinfo {pages} {141101} (\bibinfo {year}
  {2017}{\natexlab{c}})}\BibitemShut {NoStop}%
\bibitem [{\citenamefont {Murase}\ and\ \citenamefont
  {Bartos}(2019)}]{Murase:2019tjj}%
  \BibitemOpen
  \bibfield  {author} {\bibinfo {author} {\bibfnamefont {K.}~\bibnamefont
  {Murase}}\ and\ \bibinfo {author} {\bibfnamefont {I.}~\bibnamefont
  {Bartos}},\ }\href {\doibase 10.1146/annurev-nucl-101918-023510} {\bibfield
  {journal} {\bibinfo  {journal} {Ann. Rev. Nucl. Part. Sci.}\ }\textbf
  {\bibinfo {volume} {69}},\ \bibinfo {pages} {477} (\bibinfo {year} {2019})},\
  \Eprint {http://arxiv.org/abs/1907.12506} {arXiv:1907.12506 [astro-ph.HE]}
  \BibitemShut {NoStop}%
\bibitem [{\citenamefont {Albert}\ \emph {et~al.}(2017)\citenamefont {Albert},
  \citenamefont {Andr{\'e}}, \citenamefont {Anghinolfi}, \citenamefont {Ardid},
  \citenamefont {Aubert}, \citenamefont {Aublin}, \citenamefont {Avgitas},
  \citenamefont {Baret}, \citenamefont {Barrios-Mart{\'\i}}, \citenamefont
  {Basa} \emph {et~al.}}]{albert2017search}%
  \BibitemOpen
  \bibfield  {author} {\bibinfo {author} {\bibfnamefont {A.}~\bibnamefont
  {Albert}}, \bibinfo {author} {\bibfnamefont {M.}~\bibnamefont {Andr{\'e}}},
  \bibinfo {author} {\bibfnamefont {M.}~\bibnamefont {Anghinolfi}}, \bibinfo
  {author} {\bibfnamefont {M.}~\bibnamefont {Ardid}}, \bibinfo {author}
  {\bibfnamefont {J.-J.}\ \bibnamefont {Aubert}}, \bibinfo {author}
  {\bibfnamefont {J.}~\bibnamefont {Aublin}}, \bibinfo {author} {\bibfnamefont
  {T.}~\bibnamefont {Avgitas}}, \bibinfo {author} {\bibfnamefont
  {B.}~\bibnamefont {Baret}}, \bibinfo {author} {\bibfnamefont
  {J.}~\bibnamefont {Barrios-Mart{\'\i}}}, \bibinfo {author} {\bibfnamefont
  {S.}~\bibnamefont {Basa}},  \emph {et~al.},\ }\href@noop {} {\bibfield
  {journal} {\bibinfo  {journal} {arXiv preprint arXiv:1710.05839}\ } (\bibinfo
  {year} {2017})}\BibitemShut {NoStop}%
\bibitem [{\citenamefont {{Albert}}\ \emph {et~al.}(2019)\citenamefont
  {{Albert}}, \citenamefont {{Andr{\'e}}}, \citenamefont {{Anghinolfi}},
  \citenamefont {{Ardid}}, \citenamefont {{Aubert}} \emph
  {et~al.}}]{2019ApJ...870..134A}%
  \BibitemOpen
  \bibfield  {author} {\bibinfo {author} {\bibfnamefont {A.}~\bibnamefont
  {{Albert}}}, \bibinfo {author} {\bibfnamefont {M.}~\bibnamefont
  {{Andr{\'e}}}}, \bibinfo {author} {\bibfnamefont {M.}~\bibnamefont
  {{Anghinolfi}}}, \bibinfo {author} {\bibfnamefont {M.}~\bibnamefont
  {{Ardid}}}, \bibinfo {author} {\bibnamefont {{Aubert}}},  \emph {et~al.},\
  }\href {\doibase 10.3847/1538-4357/aaf21d} {\bibfield  {journal} {\bibinfo
  {journal} {Astrophysical Journal}\ }\textbf {\bibinfo {volume} {870}},\
  \bibinfo {eid} {134} (\bibinfo {year} {2019})}\BibitemShut {NoStop}%
\bibitem [{\citenamefont {Kimura}\ \emph {et~al.}(2017)\citenamefont {Kimura},
  \citenamefont {Murase}, \citenamefont {M{\'e}sz{\'a}ros},\ and\ \citenamefont
  {Kiuchi}}]{kimura2017high}%
  \BibitemOpen
  \bibfield  {author} {\bibinfo {author} {\bibfnamefont {S.~S.}\ \bibnamefont
  {Kimura}}, \bibinfo {author} {\bibfnamefont {K.}~\bibnamefont {Murase}},
  \bibinfo {author} {\bibfnamefont {P.}~\bibnamefont {M{\'e}sz{\'a}ros}}, \
  and\ \bibinfo {author} {\bibfnamefont {K.}~\bibnamefont {Kiuchi}},\
  }\href@noop {} {\bibfield  {journal} {\bibinfo  {journal} {The Astrophysical
  Journal Letters}\ }\textbf {\bibinfo {volume} {848}},\ \bibinfo {pages} {L4}
  (\bibinfo {year} {2017})}\BibitemShut {NoStop}%
\bibitem [{\citenamefont {Kimura}\ \emph {et~al.}(2018)\citenamefont {Kimura},
  \citenamefont {Murase}, \citenamefont {Bartos}, \citenamefont {Ioka},
  \citenamefont {Heng},\ and\ \citenamefont
  {M{\'e}sz{\'a}ros}}]{kimura2018transejecta}%
  \BibitemOpen
  \bibfield  {author} {\bibinfo {author} {\bibfnamefont {S.~S.}\ \bibnamefont
  {Kimura}}, \bibinfo {author} {\bibfnamefont {K.}~\bibnamefont {Murase}},
  \bibinfo {author} {\bibfnamefont {I.}~\bibnamefont {Bartos}}, \bibinfo
  {author} {\bibfnamefont {K.}~\bibnamefont {Ioka}}, \bibinfo {author}
  {\bibfnamefont {I.~S.}\ \bibnamefont {Heng}}, \ and\ \bibinfo {author}
  {\bibfnamefont {P.}~\bibnamefont {M{\'e}sz{\'a}ros}},\ }\href@noop {}
  {\bibfield  {journal} {\bibinfo  {journal} {Physical Review D}\ }\textbf
  {\bibinfo {volume} {98}},\ \bibinfo {pages} {043020} (\bibinfo {year}
  {2018})}\BibitemShut {NoStop}%
\bibitem [{\citenamefont {Fang}\ and\ \citenamefont
  {Metzger}(2017)}]{fang2017high}%
  \BibitemOpen
  \bibfield  {author} {\bibinfo {author} {\bibfnamefont {K.}~\bibnamefont
  {Fang}}\ and\ \bibinfo {author} {\bibfnamefont {B.~D.}\ \bibnamefont
  {Metzger}},\ }\href@noop {} {\bibfield  {journal} {\bibinfo  {journal} {The
  Astrophysical Journal}\ }\textbf {\bibinfo {volume} {849}},\ \bibinfo {pages}
  {153} (\bibinfo {year} {2017})}\BibitemShut {NoStop}%
\bibitem [{\citenamefont {Decoene}\ \emph {et~al.}(2020)\citenamefont
  {Decoene}, \citenamefont {Gu{\'e}pin}, \citenamefont {Fang}, \citenamefont
  {Kotera},\ and\ \citenamefont {Metzger}}]{decoene2020high}%
  \BibitemOpen
  \bibfield  {author} {\bibinfo {author} {\bibfnamefont {V.}~\bibnamefont
  {Decoene}}, \bibinfo {author} {\bibfnamefont {C.}~\bibnamefont {Gu{\'e}pin}},
  \bibinfo {author} {\bibfnamefont {K.}~\bibnamefont {Fang}}, \bibinfo {author}
  {\bibfnamefont {K.}~\bibnamefont {Kotera}}, \ and\ \bibinfo {author}
  {\bibfnamefont {B.~D.}\ \bibnamefont {Metzger}},\ }\href@noop {} {\bibfield
  {journal} {\bibinfo  {journal} {Journal of Cosmology and Astroparticle
  Physics}\ }\textbf {\bibinfo {volume} {2020}},\ \bibinfo {pages} {045}
  (\bibinfo {year} {2020})}\BibitemShut {NoStop}%
\bibitem [{\citenamefont {Aartsen}\ \emph
  {et~al.}(2018{\natexlab{a}})\citenamefont {Aartsen} \emph
  {et~al.}}]{telescope2018multimessenger}%
  \BibitemOpen
  \bibfield  {author} {\bibinfo {author} {\bibfnamefont {M.~G.}\ \bibnamefont
  {Aartsen}} \emph {et~al.},\ }\href@noop {} {\bibfield  {journal} {\bibinfo
  {journal} {Science}\ }\textbf {\bibinfo {volume} {361}},\ \bibinfo {pages}
  {eaat1378} (\bibinfo {year} {2018}{\natexlab{a}})}\BibitemShut {NoStop}%
\bibitem [{\citenamefont {Keivani}\ \emph {et~al.}(2018)\citenamefont
  {Keivani}, \citenamefont {Murase}, \citenamefont {Petropoulou}, \citenamefont
  {Fox}, \citenamefont {Cenko}, \citenamefont {Chaty}, \citenamefont {Coleiro},
  \citenamefont {DeLaunay}, \citenamefont {Dimitrakoudis}, \citenamefont
  {Evans} \emph {et~al.}}]{keivani2018multimessenger}%
  \BibitemOpen
  \bibfield  {author} {\bibinfo {author} {\bibfnamefont {A.}~\bibnamefont
  {Keivani}}, \bibinfo {author} {\bibfnamefont {K.}~\bibnamefont {Murase}},
  \bibinfo {author} {\bibfnamefont {M.}~\bibnamefont {Petropoulou}}, \bibinfo
  {author} {\bibfnamefont {D.~B.}\ \bibnamefont {Fox}}, \bibinfo {author}
  {\bibfnamefont {S.}~\bibnamefont {Cenko}}, \bibinfo {author} {\bibfnamefont
  {S.}~\bibnamefont {Chaty}}, \bibinfo {author} {\bibfnamefont
  {A.}~\bibnamefont {Coleiro}}, \bibinfo {author} {\bibfnamefont
  {J.}~\bibnamefont {DeLaunay}}, \bibinfo {author} {\bibfnamefont
  {S.}~\bibnamefont {Dimitrakoudis}}, \bibinfo {author} {\bibfnamefont
  {P.}~\bibnamefont {Evans}},  \emph {et~al.},\ }\href@noop {} {\bibfield
  {journal} {\bibinfo  {journal} {The Astrophysical Journal}\ }\textbf
  {\bibinfo {volume} {864}},\ \bibinfo {pages} {84} (\bibinfo {year}
  {2018})}\BibitemShut {NoStop}%
\bibitem [{\citenamefont {Murase}\ \emph {et~al.}(2018)\citenamefont {Murase},
  \citenamefont {Oikonomou},\ and\ \citenamefont
  {Petropoulou}}]{Murase:2018iyl}%
  \BibitemOpen
  \bibfield  {author} {\bibinfo {author} {\bibfnamefont {K.}~\bibnamefont
  {Murase}}, \bibinfo {author} {\bibfnamefont {F.}~\bibnamefont {Oikonomou}}, \
  and\ \bibinfo {author} {\bibfnamefont {M.}~\bibnamefont {Petropoulou}},\
  }\href {\doibase 10.3847/1538-4357/aada00} {\bibfield  {journal} {\bibinfo
  {journal} {Astrophys. J.}\ }\textbf {\bibinfo {volume} {865}},\ \bibinfo
  {pages} {124} (\bibinfo {year} {2018})},\ \Eprint
  {http://arxiv.org/abs/1807.04748} {arXiv:1807.04748 [astro-ph.HE]}
  \BibitemShut {NoStop}%
\bibitem [{\citenamefont {Ansoldi}\ \emph {et~al.}(2018)\citenamefont
  {Ansoldi}, \citenamefont {Antonelli}, \citenamefont {Arcaro}, \citenamefont
  {Baack}, \citenamefont {Babi{\'c}}, \citenamefont {Banerjee}, \citenamefont
  {Bangale}, \citenamefont {de~Almeida}, \citenamefont {Barrio}, \citenamefont
  {Gonz{\'a}lez} \emph {et~al.}}]{ansoldi2018blazar}%
  \BibitemOpen
  \bibfield  {author} {\bibinfo {author} {\bibfnamefont {S.}~\bibnamefont
  {Ansoldi}}, \bibinfo {author} {\bibfnamefont {L.~A.}\ \bibnamefont
  {Antonelli}}, \bibinfo {author} {\bibfnamefont {C.}~\bibnamefont {Arcaro}},
  \bibinfo {author} {\bibfnamefont {D.}~\bibnamefont {Baack}}, \bibinfo
  {author} {\bibfnamefont {A.}~\bibnamefont {Babi{\'c}}}, \bibinfo {author}
  {\bibfnamefont {B.}~\bibnamefont {Banerjee}}, \bibinfo {author}
  {\bibfnamefont {P.}~\bibnamefont {Bangale}}, \bibinfo {author} {\bibfnamefont
  {U.~B.}\ \bibnamefont {de~Almeida}}, \bibinfo {author} {\bibfnamefont
  {J.~A.}\ \bibnamefont {Barrio}}, \bibinfo {author} {\bibfnamefont {J.~B.}\
  \bibnamefont {Gonz{\'a}lez}},  \emph {et~al.},\ }\href@noop {} {\bibfield
  {journal} {\bibinfo  {journal} {The Astrophysical Journal Letters}\ }\textbf
  {\bibinfo {volume} {863}},\ \bibinfo {pages} {L10} (\bibinfo {year}
  {2018})}\BibitemShut {NoStop}%
\bibitem [{\citenamefont {Padovani}\ \emph {et~al.}(2018)\citenamefont
  {Padovani}, \citenamefont {Giommi}, \citenamefont {Resconi}, \citenamefont
  {Glauch}, \citenamefont {Arsioli}, \citenamefont {Sahakyan},\ and\
  \citenamefont {Huber}}]{padovani2018dissecting}%
  \BibitemOpen
  \bibfield  {author} {\bibinfo {author} {\bibfnamefont {P.}~\bibnamefont
  {Padovani}}, \bibinfo {author} {\bibfnamefont {P.}~\bibnamefont {Giommi}},
  \bibinfo {author} {\bibfnamefont {E.}~\bibnamefont {Resconi}}, \bibinfo
  {author} {\bibfnamefont {T.}~\bibnamefont {Glauch}}, \bibinfo {author}
  {\bibfnamefont {B.}~\bibnamefont {Arsioli}}, \bibinfo {author} {\bibfnamefont
  {N.}~\bibnamefont {Sahakyan}}, \ and\ \bibinfo {author} {\bibfnamefont
  {M.}~\bibnamefont {Huber}},\ }\href@noop {} {\bibfield  {journal} {\bibinfo
  {journal} {Monthly Notices of the Royal Astronomical Society}\ }\textbf
  {\bibinfo {volume} {480}},\ \bibinfo {pages} {192} (\bibinfo {year}
  {2018})}\BibitemShut {NoStop}%
\bibitem [{\citenamefont {Cerruti}\ \emph {et~al.}(2019)\citenamefont
  {Cerruti}, \citenamefont {Zech}, \citenamefont {Boisson}, \citenamefont
  {Emery}, \citenamefont {Inoue},\ and\ \citenamefont
  {Lenain}}]{cerruti2019leptohadronic}%
  \BibitemOpen
  \bibfield  {author} {\bibinfo {author} {\bibfnamefont {M.}~\bibnamefont
  {Cerruti}}, \bibinfo {author} {\bibfnamefont {A.}~\bibnamefont {Zech}},
  \bibinfo {author} {\bibfnamefont {C.}~\bibnamefont {Boisson}}, \bibinfo
  {author} {\bibfnamefont {G.}~\bibnamefont {Emery}}, \bibinfo {author}
  {\bibfnamefont {S.}~\bibnamefont {Inoue}}, \ and\ \bibinfo {author}
  {\bibfnamefont {J.}~\bibnamefont {Lenain}},\ }\href@noop {} {\bibfield
  {journal} {\bibinfo  {journal} {Monthly Notices of the Royal Astronomical
  Society: Letters}\ }\textbf {\bibinfo {volume} {483}},\ \bibinfo {pages}
  {L12} (\bibinfo {year} {2019})}\BibitemShut {NoStop}%
\bibitem [{\citenamefont {Gao}\ \emph {et~al.}(2019)\citenamefont {Gao},
  \citenamefont {Fedynitch}, \citenamefont {Winter},\ and\ \citenamefont
  {Pohl}}]{gao2019modelling}%
  \BibitemOpen
  \bibfield  {author} {\bibinfo {author} {\bibfnamefont {S.}~\bibnamefont
  {Gao}}, \bibinfo {author} {\bibfnamefont {A.}~\bibnamefont {Fedynitch}},
  \bibinfo {author} {\bibfnamefont {W.}~\bibnamefont {Winter}}, \ and\ \bibinfo
  {author} {\bibfnamefont {M.}~\bibnamefont {Pohl}},\ }\href@noop {} {\bibfield
   {journal} {\bibinfo  {journal} {Nature Astronomy}\ }\textbf {\bibinfo
  {volume} {3}},\ \bibinfo {pages} {88} (\bibinfo {year} {2019})}\BibitemShut
  {NoStop}%
\bibitem [{\citenamefont {Reimer}\ \emph {et~al.}(2019)\citenamefont {Reimer},
  \citenamefont {B{\"o}ttcher},\ and\ \citenamefont
  {Buson}}]{reimer2019cascading}%
  \BibitemOpen
  \bibfield  {author} {\bibinfo {author} {\bibfnamefont {A.}~\bibnamefont
  {Reimer}}, \bibinfo {author} {\bibfnamefont {M.}~\bibnamefont
  {B{\"o}ttcher}}, \ and\ \bibinfo {author} {\bibfnamefont {S.}~\bibnamefont
  {Buson}},\ }\href@noop {} {\bibfield  {journal} {\bibinfo  {journal} {The
  Astrophysical Journal}\ }\textbf {\bibinfo {volume} {881}},\ \bibinfo {pages}
  {46} (\bibinfo {year} {2019})}\BibitemShut {NoStop}%
\bibitem [{\citenamefont {Rodrigues}\ \emph {et~al.}(2019)\citenamefont
  {Rodrigues}, \citenamefont {Gao}, \citenamefont {Fedynitch}, \citenamefont
  {Palladino},\ and\ \citenamefont {Winter}}]{rodrigues2019leptohadronic}%
  \BibitemOpen
  \bibfield  {author} {\bibinfo {author} {\bibfnamefont {X.}~\bibnamefont
  {Rodrigues}}, \bibinfo {author} {\bibfnamefont {S.}~\bibnamefont {Gao}},
  \bibinfo {author} {\bibfnamefont {A.}~\bibnamefont {Fedynitch}}, \bibinfo
  {author} {\bibfnamefont {A.}~\bibnamefont {Palladino}}, \ and\ \bibinfo
  {author} {\bibfnamefont {W.}~\bibnamefont {Winter}},\ }\href@noop {}
  {\bibfield  {journal} {\bibinfo  {journal} {The Astrophysical Journal
  Letters}\ }\textbf {\bibinfo {volume} {874}},\ \bibinfo {pages} {L29}
  (\bibinfo {year} {2019})}\BibitemShut {NoStop}%
\bibitem [{\citenamefont {Petropoulou}\ \emph {et~al.}(2020)\citenamefont
  {Petropoulou} \emph {et~al.}}]{Petropoulou:2019zqp}%
  \BibitemOpen
  \bibfield  {author} {\bibinfo {author} {\bibfnamefont {M.}~\bibnamefont
  {Petropoulou}} \emph {et~al.},\ }\href {\doibase 10.3847/1538-4357/ab76d0}
  {\bibfield  {journal} {\bibinfo  {journal} {Astrophys. J.}\ }\textbf
  {\bibinfo {volume} {891}},\ \bibinfo {pages} {115} (\bibinfo {year}
  {2020})},\ \Eprint {http://arxiv.org/abs/1911.04010} {arXiv:1911.04010
  [astro-ph.HE]} \BibitemShut {NoStop}%
\bibitem [{\citenamefont {Aartsen}\ \emph
  {et~al.}(2013{\natexlab{a}})\citenamefont {Aartsen}, \citenamefont {Abbasi},
  \citenamefont {Abdou}, \citenamefont {Ackermann}, \citenamefont {Adams},
  \citenamefont {Aguilar}, \citenamefont {Ahlers}, \citenamefont {Altmann},
  \citenamefont {Auffenberg}, \citenamefont {Bai} \emph
  {et~al.}}]{aartsen2013first}%
  \BibitemOpen
  \bibfield  {author} {\bibinfo {author} {\bibfnamefont {M.~G.}\ \bibnamefont
  {Aartsen}}, \bibinfo {author} {\bibfnamefont {R.}~\bibnamefont {Abbasi}},
  \bibinfo {author} {\bibfnamefont {Y.}~\bibnamefont {Abdou}}, \bibinfo
  {author} {\bibfnamefont {M.}~\bibnamefont {Ackermann}}, \bibinfo {author}
  {\bibfnamefont {J.}~\bibnamefont {Adams}}, \bibinfo {author} {\bibfnamefont
  {J.}~\bibnamefont {Aguilar}}, \bibinfo {author} {\bibfnamefont
  {M.}~\bibnamefont {Ahlers}}, \bibinfo {author} {\bibfnamefont
  {D.}~\bibnamefont {Altmann}}, \bibinfo {author} {\bibfnamefont
  {J.}~\bibnamefont {Auffenberg}}, \bibinfo {author} {\bibfnamefont
  {X.}~\bibnamefont {Bai}},  \emph {et~al.},\ }\href@noop {} {\bibfield
  {journal} {\bibinfo  {journal} {Physical review letters}\ }\textbf {\bibinfo
  {volume} {111}},\ \bibinfo {pages} {021103} (\bibinfo {year}
  {2013}{\natexlab{a}})}\BibitemShut {NoStop}%
\bibitem [{\citenamefont {Aartsen}\ \emph
  {et~al.}(2013{\natexlab{b}})\citenamefont {Aartsen}, \citenamefont {Abbasi},
  \citenamefont {Abdou}, \citenamefont {Ackermann}, \citenamefont {Adams},
  \citenamefont {Aguilar}, \citenamefont {Ahlers}, \citenamefont {Altmann},
  \citenamefont {Auffenberg}, \citenamefont {Bai} \emph
  {et~al.}}]{icecube2013evidence}%
  \BibitemOpen
  \bibfield  {author} {\bibinfo {author} {\bibfnamefont {M.~G.}\ \bibnamefont
  {Aartsen}}, \bibinfo {author} {\bibfnamefont {R.}~\bibnamefont {Abbasi}},
  \bibinfo {author} {\bibfnamefont {Y.}~\bibnamefont {Abdou}}, \bibinfo
  {author} {\bibfnamefont {M.}~\bibnamefont {Ackermann}}, \bibinfo {author}
  {\bibfnamefont {J.}~\bibnamefont {Adams}}, \bibinfo {author} {\bibfnamefont
  {J.}~\bibnamefont {Aguilar}}, \bibinfo {author} {\bibfnamefont
  {M.}~\bibnamefont {Ahlers}}, \bibinfo {author} {\bibfnamefont
  {D.}~\bibnamefont {Altmann}}, \bibinfo {author} {\bibfnamefont
  {J.}~\bibnamefont {Auffenberg}}, \bibinfo {author} {\bibfnamefont
  {X.}~\bibnamefont {Bai}},  \emph {et~al.},\ }\href@noop {} {\bibfield
  {journal} {\bibinfo  {journal} {Science}\ }\textbf {\bibinfo {volume}
  {342}},\ \bibinfo {pages} {1242856} (\bibinfo {year}
  {2013}{\natexlab{b}})}\BibitemShut {NoStop}%
\bibitem [{\citenamefont {Aartsen}\ \emph
  {et~al.}(2014{\natexlab{a}})\citenamefont {Aartsen}, \citenamefont
  {Ackermann}, \citenamefont {Adams}, \citenamefont {Aguilar}, \citenamefont
  {Ahlers}, \citenamefont {Ahrens}, \citenamefont {Altmann}, \citenamefont
  {Anderson}, \citenamefont {Arguelles}, \citenamefont {Arlen} \emph
  {et~al.}}]{aartsen2014observation}%
  \BibitemOpen
  \bibfield  {author} {\bibinfo {author} {\bibfnamefont {M.}~\bibnamefont
  {Aartsen}}, \bibinfo {author} {\bibfnamefont {M.}~\bibnamefont {Ackermann}},
  \bibinfo {author} {\bibfnamefont {J.}~\bibnamefont {Adams}}, \bibinfo
  {author} {\bibfnamefont {J.}~\bibnamefont {Aguilar}}, \bibinfo {author}
  {\bibfnamefont {M.}~\bibnamefont {Ahlers}}, \bibinfo {author} {\bibfnamefont
  {M.}~\bibnamefont {Ahrens}}, \bibinfo {author} {\bibfnamefont
  {D.}~\bibnamefont {Altmann}}, \bibinfo {author} {\bibfnamefont
  {T.}~\bibnamefont {Anderson}}, \bibinfo {author} {\bibfnamefont
  {C.}~\bibnamefont {Arguelles}}, \bibinfo {author} {\bibfnamefont
  {T.}~\bibnamefont {Arlen}},  \emph {et~al.},\ }\href@noop {} {\bibfield
  {journal} {\bibinfo  {journal} {Physical review letters}\ }\textbf {\bibinfo
  {volume} {113}},\ \bibinfo {pages} {101101} (\bibinfo {year}
  {2014}{\natexlab{a}})}\BibitemShut {NoStop}%
\bibitem [{\citenamefont {Aartsen}\ \emph {et~al.}(2015)\citenamefont
  {Aartsen}, \citenamefont {Abraham}, \citenamefont {Ackermann}, \citenamefont
  {Adams}, \citenamefont {Aguilar}, \citenamefont {Ahlers}, \citenamefont
  {Ahrens}, \citenamefont {Altmann}, \citenamefont {Anderson}, \citenamefont
  {Archinger} \emph {et~al.}}]{aartsen2015combined}%
  \BibitemOpen
  \bibfield  {author} {\bibinfo {author} {\bibfnamefont {M.}~\bibnamefont
  {Aartsen}}, \bibinfo {author} {\bibfnamefont {K.}~\bibnamefont {Abraham}},
  \bibinfo {author} {\bibfnamefont {M.}~\bibnamefont {Ackermann}}, \bibinfo
  {author} {\bibfnamefont {J.}~\bibnamefont {Adams}}, \bibinfo {author}
  {\bibfnamefont {J.~A.}\ \bibnamefont {Aguilar}}, \bibinfo {author}
  {\bibfnamefont {M.}~\bibnamefont {Ahlers}}, \bibinfo {author} {\bibfnamefont
  {M.}~\bibnamefont {Ahrens}}, \bibinfo {author} {\bibfnamefont
  {D.}~\bibnamefont {Altmann}}, \bibinfo {author} {\bibfnamefont
  {T.}~\bibnamefont {Anderson}}, \bibinfo {author} {\bibfnamefont
  {M.}~\bibnamefont {Archinger}},  \emph {et~al.},\ }\href@noop {} {\bibfield
  {journal} {\bibinfo  {journal} {The Astrophysical Journal}\ }\textbf
  {\bibinfo {volume} {809}},\ \bibinfo {pages} {98} (\bibinfo {year}
  {2015})}\BibitemShut {NoStop}%
\bibitem [{\citenamefont {Aartsen}\ \emph {et~al.}(2016)\citenamefont
  {Aartsen}, \citenamefont {Abraham}, \citenamefont {Ackermann}, \citenamefont
  {Adams}, \citenamefont {Aguilar}, \citenamefont {Ahlers}, \citenamefont
  {Ahrens}, \citenamefont {Altmann}, \citenamefont {Andeen}, \citenamefont
  {Anderson} \emph {et~al.}}]{aartsen2016observation}%
  \BibitemOpen
  \bibfield  {author} {\bibinfo {author} {\bibfnamefont {M.}~\bibnamefont
  {Aartsen}}, \bibinfo {author} {\bibfnamefont {K.}~\bibnamefont {Abraham}},
  \bibinfo {author} {\bibfnamefont {M.}~\bibnamefont {Ackermann}}, \bibinfo
  {author} {\bibfnamefont {J.}~\bibnamefont {Adams}}, \bibinfo {author}
  {\bibfnamefont {J.}~\bibnamefont {Aguilar}}, \bibinfo {author} {\bibfnamefont
  {M.}~\bibnamefont {Ahlers}}, \bibinfo {author} {\bibfnamefont
  {M.}~\bibnamefont {Ahrens}}, \bibinfo {author} {\bibfnamefont
  {D.}~\bibnamefont {Altmann}}, \bibinfo {author} {\bibfnamefont
  {K.}~\bibnamefont {Andeen}}, \bibinfo {author} {\bibfnamefont
  {T.}~\bibnamefont {Anderson}},  \emph {et~al.},\ }\href@noop {} {\bibfield
  {journal} {\bibinfo  {journal} {The Astrophysical Journal}\ }\textbf
  {\bibinfo {volume} {833}},\ \bibinfo {pages} {3} (\bibinfo {year}
  {2016})}\BibitemShut {NoStop}%
\bibitem [{\citenamefont {Aartsen}\ \emph {et~al.}(2020)\citenamefont
  {Aartsen}, \citenamefont {Ackermann}, \citenamefont {Adams}, \citenamefont
  {Aguilar}, \citenamefont {Ahlers}, \citenamefont {Ahrens}, \citenamefont
  {Alispach}, \citenamefont {Andeen}, \citenamefont {Anderson}, \citenamefont
  {Ansseau} \emph {et~al.}}]{aartsen2020characteristics}%
  \BibitemOpen
  \bibfield  {author} {\bibinfo {author} {\bibfnamefont {M.}~\bibnamefont
  {Aartsen}}, \bibinfo {author} {\bibfnamefont {M.}~\bibnamefont {Ackermann}},
  \bibinfo {author} {\bibfnamefont {J.}~\bibnamefont {Adams}}, \bibinfo
  {author} {\bibfnamefont {J.}~\bibnamefont {Aguilar}}, \bibinfo {author}
  {\bibfnamefont {M.}~\bibnamefont {Ahlers}}, \bibinfo {author} {\bibfnamefont
  {M.}~\bibnamefont {Ahrens}}, \bibinfo {author} {\bibfnamefont
  {C.}~\bibnamefont {Alispach}}, \bibinfo {author} {\bibfnamefont
  {K.}~\bibnamefont {Andeen}}, \bibinfo {author} {\bibfnamefont
  {T.}~\bibnamefont {Anderson}}, \bibinfo {author} {\bibfnamefont
  {I.}~\bibnamefont {Ansseau}},  \emph {et~al.},\ }\href@noop {} {\bibfield
  {journal} {\bibinfo  {journal} {arXiv preprint arXiv:2001.09520}\ } (\bibinfo
  {year} {2020})}\BibitemShut {NoStop}%
\bibitem [{\citenamefont {Ahlers}\ and\ \citenamefont
  {Halzen}(2018)}]{ahlers2018opening}%
  \BibitemOpen
  \bibfield  {author} {\bibinfo {author} {\bibfnamefont {M.}~\bibnamefont
  {Ahlers}}\ and\ \bibinfo {author} {\bibfnamefont {F.}~\bibnamefont
  {Halzen}},\ }\href@noop {} {\bibfield  {journal} {\bibinfo  {journal}
  {Progress in Particle and Nuclear Physics}\ }\textbf {\bibinfo {volume}
  {102}},\ \bibinfo {pages} {73} (\bibinfo {year} {2018})}\BibitemShut
  {NoStop}%
\bibitem [{\citenamefont {M\'esz\'aros}\ \emph {et~al.}(2019)\citenamefont
  {M\'esz\'aros}, \citenamefont {Fox}, \citenamefont {Hanna},\ and\
  \citenamefont {Murase}}]{Meszaros:2019xej}%
  \BibitemOpen
  \bibfield  {author} {\bibinfo {author} {\bibfnamefont {P.}~\bibnamefont
  {M\'esz\'aros}}, \bibinfo {author} {\bibfnamefont {D.~B.}\ \bibnamefont
  {Fox}}, \bibinfo {author} {\bibfnamefont {C.}~\bibnamefont {Hanna}}, \ and\
  \bibinfo {author} {\bibfnamefont {K.}~\bibnamefont {Murase}},\ }\href
  {\doibase 10.1038/s42254-019-0101-z} {\bibfield  {journal} {\bibinfo
  {journal} {Nature Rev. Phys.}\ }\textbf {\bibinfo {volume} {1}},\ \bibinfo
  {pages} {585} (\bibinfo {year} {2019})},\ \Eprint
  {http://arxiv.org/abs/1906.10212} {arXiv:1906.10212 [astro-ph.HE]}
  \BibitemShut {NoStop}%
\bibitem [{\citenamefont {Murase}\ \emph {et~al.}(2014)\citenamefont {Murase},
  \citenamefont {Inoue},\ and\ \citenamefont {Dermer}}]{murase2014diffuse}%
  \BibitemOpen
  \bibfield  {author} {\bibinfo {author} {\bibfnamefont {K.}~\bibnamefont
  {Murase}}, \bibinfo {author} {\bibfnamefont {Y.}~\bibnamefont {Inoue}}, \
  and\ \bibinfo {author} {\bibfnamefont {C.~D.}\ \bibnamefont {Dermer}},\
  }\href@noop {} {\bibfield  {journal} {\bibinfo  {journal} {Physical Review
  D}\ }\textbf {\bibinfo {volume} {90}},\ \bibinfo {pages} {023007} (\bibinfo
  {year} {2014})}\BibitemShut {NoStop}%
\bibitem [{\citenamefont {Dermer}\ \emph {et~al.}(2014)\citenamefont {Dermer},
  \citenamefont {Murase},\ and\ \citenamefont {Inoue}}]{dermer2014photopion}%
  \BibitemOpen
  \bibfield  {author} {\bibinfo {author} {\bibfnamefont {C.~D.}\ \bibnamefont
  {Dermer}}, \bibinfo {author} {\bibfnamefont {K.}~\bibnamefont {Murase}}, \
  and\ \bibinfo {author} {\bibfnamefont {Y.}~\bibnamefont {Inoue}},\
  }\href@noop {} {\bibfield  {journal} {\bibinfo  {journal} {Journal of High
  Energy Astrophysics}\ }\textbf {\bibinfo {volume} {3}},\ \bibinfo {pages}
  {29} (\bibinfo {year} {2014})}\BibitemShut {NoStop}%
\bibitem [{\citenamefont {Padovani}\ \emph {et~al.}(2015)\citenamefont
  {Padovani}, \citenamefont {Petropoulou}, \citenamefont {Giommi},\ and\
  \citenamefont {Resconi}}]{padovani2015simplified}%
  \BibitemOpen
  \bibfield  {author} {\bibinfo {author} {\bibfnamefont {P.}~\bibnamefont
  {Padovani}}, \bibinfo {author} {\bibfnamefont {M.}~\bibnamefont
  {Petropoulou}}, \bibinfo {author} {\bibfnamefont {P.}~\bibnamefont {Giommi}},
  \ and\ \bibinfo {author} {\bibfnamefont {E.}~\bibnamefont {Resconi}},\
  }\href@noop {} {\bibfield  {journal} {\bibinfo  {journal} {Monthly Notices of
  the Royal Astronomical Society}\ }\textbf {\bibinfo {volume} {452}},\
  \bibinfo {pages} {1877} (\bibinfo {year} {2015})}\BibitemShut {NoStop}%
\bibitem [{\citenamefont {Petropoulou}\ \emph {et~al.}(2015)\citenamefont
  {Petropoulou}, \citenamefont {Dimitrakoudis}, \citenamefont {Padovani},
  \citenamefont {Mastichiadis},\ and\ \citenamefont
  {Resconi}}]{petropoulou2015photohadronic}%
  \BibitemOpen
  \bibfield  {author} {\bibinfo {author} {\bibfnamefont {M.}~\bibnamefont
  {Petropoulou}}, \bibinfo {author} {\bibfnamefont {S.}~\bibnamefont
  {Dimitrakoudis}}, \bibinfo {author} {\bibfnamefont {P.}~\bibnamefont
  {Padovani}}, \bibinfo {author} {\bibfnamefont {A.}~\bibnamefont
  {Mastichiadis}}, \ and\ \bibinfo {author} {\bibfnamefont {E.}~\bibnamefont
  {Resconi}},\ }\href@noop {} {\bibfield  {journal} {\bibinfo  {journal}
  {Monthly Notices of the Royal Astronomical Society}\ }\textbf {\bibinfo
  {volume} {448}},\ \bibinfo {pages} {2412} (\bibinfo {year}
  {2015})}\BibitemShut {NoStop}%
\bibitem [{\citenamefont {Yuan}\ \emph {et~al.}(2020)\citenamefont {Yuan},
  \citenamefont {Murase},\ and\ \citenamefont
  {M{\'e}sz{\'a}ros}}]{yuan2020complementarity}%
  \BibitemOpen
  \bibfield  {author} {\bibinfo {author} {\bibfnamefont {C.}~\bibnamefont
  {Yuan}}, \bibinfo {author} {\bibfnamefont {K.}~\bibnamefont {Murase}}, \ and\
  \bibinfo {author} {\bibfnamefont {P.}~\bibnamefont {M{\'e}sz{\'a}ros}},\
  }\href@noop {} {\bibfield  {journal} {\bibinfo  {journal} {The Astrophysical
  Journal}\ }\textbf {\bibinfo {volume} {890}},\ \bibinfo {pages} {25}
  (\bibinfo {year} {2020})}\BibitemShut {NoStop}%
\bibitem [{\citenamefont {Stecker}\ \emph {et~al.}(1991)\citenamefont
  {Stecker}, \citenamefont {Done}, \citenamefont {Salamon},\ and\ \citenamefont
  {Sommers}}]{stecker1991high}%
  \BibitemOpen
  \bibfield  {author} {\bibinfo {author} {\bibfnamefont {F.~W.}\ \bibnamefont
  {Stecker}}, \bibinfo {author} {\bibfnamefont {C.}~\bibnamefont {Done}},
  \bibinfo {author} {\bibfnamefont {M.~H.}\ \bibnamefont {Salamon}}, \ and\
  \bibinfo {author} {\bibfnamefont {P.}~\bibnamefont {Sommers}},\ }\href@noop
  {} {\bibfield  {journal} {\bibinfo  {journal} {Physical Review Letters}\
  }\textbf {\bibinfo {volume} {66}},\ \bibinfo {pages} {2697} (\bibinfo {year}
  {1991})}\BibitemShut {NoStop}%
\bibitem [{\citenamefont {Kimura}\ \emph {et~al.}(2015)\citenamefont {Kimura},
  \citenamefont {Murase},\ and\ \citenamefont {Toma}}]{kimura2015neutrino}%
  \BibitemOpen
  \bibfield  {author} {\bibinfo {author} {\bibfnamefont {S.~S.}\ \bibnamefont
  {Kimura}}, \bibinfo {author} {\bibfnamefont {K.}~\bibnamefont {Murase}}, \
  and\ \bibinfo {author} {\bibfnamefont {K.}~\bibnamefont {Toma}},\ }\href@noop
  {} {\bibfield  {journal} {\bibinfo  {journal} {The Astrophysical Journal}\
  }\textbf {\bibinfo {volume} {806}},\ \bibinfo {pages} {159} (\bibinfo {year}
  {2015})}\BibitemShut {NoStop}%
\bibitem [{\citenamefont {Murase}\ \emph {et~al.}(2016)\citenamefont {Murase},
  \citenamefont {Guetta},\ and\ \citenamefont {Ahlers}}]{murase2016hidden}%
  \BibitemOpen
  \bibfield  {author} {\bibinfo {author} {\bibfnamefont {K.}~\bibnamefont
  {Murase}}, \bibinfo {author} {\bibfnamefont {D.}~\bibnamefont {Guetta}}, \
  and\ \bibinfo {author} {\bibfnamefont {M.}~\bibnamefont {Ahlers}},\
  }\href@noop {} {\bibfield  {journal} {\bibinfo  {journal} {Physical Review
  Letters}\ }\textbf {\bibinfo {volume} {116}},\ \bibinfo {pages} {071101}
  (\bibinfo {year} {2016})}\BibitemShut {NoStop}%
\bibitem [{\citenamefont {Murase}\ \emph {et~al.}(2020)\citenamefont {Murase},
  \citenamefont {Kimura},\ and\ \citenamefont {Meszaros}}]{murase2019hidden}%
  \BibitemOpen
  \bibfield  {author} {\bibinfo {author} {\bibfnamefont {K.}~\bibnamefont
  {Murase}}, \bibinfo {author} {\bibfnamefont {S.~S.}\ \bibnamefont {Kimura}},
  \ and\ \bibinfo {author} {\bibfnamefont {P.}~\bibnamefont {Meszaros}},\
  }\href {\doibase 10.1103/PhysRevLett.125.011101} {\bibfield  {journal}
  {\bibinfo  {journal} {Phys. Rev. Lett.}\ }\textbf {\bibinfo {volume} {125}},\
  \bibinfo {pages} {011101} (\bibinfo {year} {2020})},\ \Eprint
  {http://arxiv.org/abs/1904.04226} {arXiv:1904.04226 [astro-ph.HE]}
  \BibitemShut {NoStop}%
\bibitem [{\citenamefont {Murase}\ \emph {et~al.}(2008)\citenamefont {Murase},
  \citenamefont {Inoue},\ and\ \citenamefont {Nagataki}}]{murase2008cosmic}%
  \BibitemOpen
  \bibfield  {author} {\bibinfo {author} {\bibfnamefont {K.}~\bibnamefont
  {Murase}}, \bibinfo {author} {\bibfnamefont {S.}~\bibnamefont {Inoue}}, \
  and\ \bibinfo {author} {\bibfnamefont {S.}~\bibnamefont {Nagataki}},\
  }\href@noop {} {\bibfield  {journal} {\bibinfo  {journal} {The Astrophysical
  Journal Letters}\ }\textbf {\bibinfo {volume} {689}},\ \bibinfo {pages}
  {L105} (\bibinfo {year} {2008})}\BibitemShut {NoStop}%
\bibitem [{\citenamefont {Murase}\ \emph {et~al.}(2013)\citenamefont {Murase},
  \citenamefont {Ahlers},\ and\ \citenamefont {Lacki}}]{murase2013testing}%
  \BibitemOpen
  \bibfield  {author} {\bibinfo {author} {\bibfnamefont {K.}~\bibnamefont
  {Murase}}, \bibinfo {author} {\bibfnamefont {M.}~\bibnamefont {Ahlers}}, \
  and\ \bibinfo {author} {\bibfnamefont {B.~C.}\ \bibnamefont {Lacki}},\
  }\href@noop {} {\bibfield  {journal} {\bibinfo  {journal} {Physical Review
  D}\ }\textbf {\bibinfo {volume} {88}},\ \bibinfo {pages} {121301} (\bibinfo
  {year} {2013})}\BibitemShut {NoStop}%
\bibitem [{\citenamefont {Fang}\ and\ \citenamefont
  {Murase}(2018)}]{fang2018linking}%
  \BibitemOpen
  \bibfield  {author} {\bibinfo {author} {\bibfnamefont {K.}~\bibnamefont
  {Fang}}\ and\ \bibinfo {author} {\bibfnamefont {K.}~\bibnamefont {Murase}},\
  }\href@noop {} {\bibfield  {journal} {\bibinfo  {journal} {Nature Physics}\
  }\textbf {\bibinfo {volume} {14}},\ \bibinfo {pages} {396} (\bibinfo {year}
  {2018})}\BibitemShut {NoStop}%
\bibitem [{\citenamefont {Loeb}\ and\ \citenamefont
  {Waxman}(2006)}]{loeb2006cumulative}%
  \BibitemOpen
  \bibfield  {author} {\bibinfo {author} {\bibfnamefont {A.}~\bibnamefont
  {Loeb}}\ and\ \bibinfo {author} {\bibfnamefont {E.}~\bibnamefont {Waxman}},\
  }\href@noop {} {\bibfield  {journal} {\bibinfo  {journal} {Journal of
  Cosmology and Astroparticle Physics}\ }\textbf {\bibinfo {volume} {2006}},\
  \bibinfo {pages} {003} (\bibinfo {year} {2006})}\BibitemShut {NoStop}%
\bibitem [{\citenamefont {Senno}\ \emph {et~al.}(2015)\citenamefont {Senno},
  \citenamefont {M{\'e}sz{\'a}ros}, \citenamefont {Murase}, \citenamefont
  {Baerwald},\ and\ \citenamefont {Rees}}]{senno2015extragalactic}%
  \BibitemOpen
  \bibfield  {author} {\bibinfo {author} {\bibfnamefont {N.}~\bibnamefont
  {Senno}}, \bibinfo {author} {\bibfnamefont {P.}~\bibnamefont
  {M{\'e}sz{\'a}ros}}, \bibinfo {author} {\bibfnamefont {K.}~\bibnamefont
  {Murase}}, \bibinfo {author} {\bibfnamefont {P.}~\bibnamefont {Baerwald}}, \
  and\ \bibinfo {author} {\bibfnamefont {M.~J.}\ \bibnamefont {Rees}},\
  }\href@noop {} {\bibfield  {journal} {\bibinfo  {journal} {The Astrophysical
  Journal}\ }\textbf {\bibinfo {volume} {806}},\ \bibinfo {pages} {24}
  (\bibinfo {year} {2015})}\BibitemShut {NoStop}%
\bibitem [{\citenamefont {Liu}\ \emph {et~al.}(2018)\citenamefont {Liu},
  \citenamefont {Murase}, \citenamefont {Inoue}, \citenamefont {Ge},\ and\
  \citenamefont {Wang}}]{Liu:2017bjr}%
  \BibitemOpen
  \bibfield  {author} {\bibinfo {author} {\bibfnamefont {R.-Y.}\ \bibnamefont
  {Liu}}, \bibinfo {author} {\bibfnamefont {K.}~\bibnamefont {Murase}},
  \bibinfo {author} {\bibfnamefont {S.}~\bibnamefont {Inoue}}, \bibinfo
  {author} {\bibfnamefont {C.}~\bibnamefont {Ge}}, \ and\ \bibinfo {author}
  {\bibfnamefont {X.-Y.}\ \bibnamefont {Wang}},\ }\href {\doibase
  10.3847/1538-4357/aaba74} {\bibfield  {journal} {\bibinfo  {journal}
  {Astrophys. J.}\ }\textbf {\bibinfo {volume} {858}},\ \bibinfo {pages} {9}
  (\bibinfo {year} {2018})},\ \Eprint {http://arxiv.org/abs/1712.10168}
  {arXiv:1712.10168 [astro-ph.HE]} \BibitemShut {NoStop}%
\bibitem [{\citenamefont {Kashiyama}\ and\ \citenamefont
  {M{\'e}sz{\'a}ros}(2014)}]{kashiyama2014galaxy}%
  \BibitemOpen
  \bibfield  {author} {\bibinfo {author} {\bibfnamefont {K.}~\bibnamefont
  {Kashiyama}}\ and\ \bibinfo {author} {\bibfnamefont {P.}~\bibnamefont
  {M{\'e}sz{\'a}ros}},\ }\href@noop {} {\bibfield  {journal} {\bibinfo
  {journal} {The Astrophysical Journal Letters}\ }\textbf {\bibinfo {volume}
  {790}},\ \bibinfo {pages} {L14} (\bibinfo {year} {2014})}\BibitemShut
  {NoStop}%
\bibitem [{\citenamefont {Yuan}\ \emph {et~al.}(2018)\citenamefont {Yuan},
  \citenamefont {M{\'e}sz{\'a}ros}, \citenamefont {Murase},\ and\ \citenamefont
  {Jeong}}]{yuan2018cumulative}%
  \BibitemOpen
  \bibfield  {author} {\bibinfo {author} {\bibfnamefont {C.}~\bibnamefont
  {Yuan}}, \bibinfo {author} {\bibfnamefont {P.}~\bibnamefont
  {M{\'e}sz{\'a}ros}}, \bibinfo {author} {\bibfnamefont {K.}~\bibnamefont
  {Murase}}, \ and\ \bibinfo {author} {\bibfnamefont {D.}~\bibnamefont
  {Jeong}},\ }\href@noop {} {\bibfield  {journal} {\bibinfo  {journal} {The
  Astrophysical Journal}\ }\textbf {\bibinfo {volume} {857}},\ \bibinfo {pages}
  {50} (\bibinfo {year} {2018})}\BibitemShut {NoStop}%
\bibitem [{\citenamefont {Ackermann}\ \emph {et~al.}(2015)\citenamefont
  {Ackermann}, \citenamefont {Ajello}, \citenamefont {Albert}, \citenamefont
  {Atwood}, \citenamefont {Baldini}, \citenamefont {Ballet}, \citenamefont
  {Barbiellini}, \citenamefont {Bastieri}, \citenamefont {Bechtol},
  \citenamefont {Bellazzini} \emph {et~al.}}]{ackermann2015spectrum}%
  \BibitemOpen
  \bibfield  {author} {\bibinfo {author} {\bibfnamefont {M.}~\bibnamefont
  {Ackermann}}, \bibinfo {author} {\bibfnamefont {M.}~\bibnamefont {Ajello}},
  \bibinfo {author} {\bibfnamefont {A.}~\bibnamefont {Albert}}, \bibinfo
  {author} {\bibfnamefont {W.}~\bibnamefont {Atwood}}, \bibinfo {author}
  {\bibfnamefont {L.}~\bibnamefont {Baldini}}, \bibinfo {author} {\bibfnamefont
  {J.}~\bibnamefont {Ballet}}, \bibinfo {author} {\bibfnamefont
  {G.}~\bibnamefont {Barbiellini}}, \bibinfo {author} {\bibfnamefont
  {D.}~\bibnamefont {Bastieri}}, \bibinfo {author} {\bibfnamefont
  {K.}~\bibnamefont {Bechtol}}, \bibinfo {author} {\bibfnamefont
  {R.}~\bibnamefont {Bellazzini}},  \emph {et~al.},\ }\href@noop {} {\bibfield
  {journal} {\bibinfo  {journal} {The Astrophysical Journal}\ }\textbf
  {\bibinfo {volume} {799}},\ \bibinfo {pages} {86} (\bibinfo {year}
  {2015})}\BibitemShut {NoStop}%
\bibitem [{\citenamefont {Ackermann}\ \emph {et~al.}(2016)\citenamefont
  {Ackermann}, \citenamefont {Ajello}, \citenamefont {Albert}, \citenamefont
  {Atwood}, \citenamefont {Baldini}, \citenamefont {Ballet}, \citenamefont
  {Barbiellini}, \citenamefont {Bastieri}, \citenamefont {Bechtol},
  \citenamefont {Bellazzini} \emph {et~al.}}]{ackermann2016resolving}%
  \BibitemOpen
  \bibfield  {author} {\bibinfo {author} {\bibfnamefont {M.}~\bibnamefont
  {Ackermann}}, \bibinfo {author} {\bibfnamefont {M.}~\bibnamefont {Ajello}},
  \bibinfo {author} {\bibfnamefont {A.}~\bibnamefont {Albert}}, \bibinfo
  {author} {\bibfnamefont {W.}~\bibnamefont {Atwood}}, \bibinfo {author}
  {\bibfnamefont {L.}~\bibnamefont {Baldini}}, \bibinfo {author} {\bibfnamefont
  {J.}~\bibnamefont {Ballet}}, \bibinfo {author} {\bibfnamefont
  {G.}~\bibnamefont {Barbiellini}}, \bibinfo {author} {\bibfnamefont
  {D.}~\bibnamefont {Bastieri}}, \bibinfo {author} {\bibfnamefont
  {K.}~\bibnamefont {Bechtol}}, \bibinfo {author} {\bibfnamefont
  {R.}~\bibnamefont {Bellazzini}},  \emph {et~al.},\ }\href@noop {} {\bibfield
  {journal} {\bibinfo  {journal} {Physical Review Letters}\ }\textbf {\bibinfo
  {volume} {116}},\ \bibinfo {pages} {151105} (\bibinfo {year}
  {2016})}\BibitemShut {NoStop}%
\bibitem [{\citenamefont {Richstone}\ \emph {et~al.}(1998)\citenamefont
  {Richstone}, \citenamefont {Ajhar}, \citenamefont {Bender}, \citenamefont
  {Bower}, \citenamefont {Dressler}, \citenamefont {Faber}, \citenamefont
  {Filippenko}, \citenamefont {Gebhardt}, \citenamefont {Green}, \citenamefont
  {Ho} \emph {et~al.}}]{richstone1998supermassive}%
  \BibitemOpen
  \bibfield  {author} {\bibinfo {author} {\bibfnamefont {D.}~\bibnamefont
  {Richstone}}, \bibinfo {author} {\bibfnamefont {E.}~\bibnamefont {Ajhar}},
  \bibinfo {author} {\bibfnamefont {R.}~\bibnamefont {Bender}}, \bibinfo
  {author} {\bibfnamefont {G.}~\bibnamefont {Bower}}, \bibinfo {author}
  {\bibfnamefont {A.}~\bibnamefont {Dressler}}, \bibinfo {author}
  {\bibfnamefont {S.}~\bibnamefont {Faber}}, \bibinfo {author} {\bibfnamefont
  {A.}~\bibnamefont {Filippenko}}, \bibinfo {author} {\bibfnamefont
  {K.}~\bibnamefont {Gebhardt}}, \bibinfo {author} {\bibfnamefont
  {R.}~\bibnamefont {Green}}, \bibinfo {author} {\bibfnamefont
  {L.}~\bibnamefont {Ho}},  \emph {et~al.},\ }\href@noop {} {\bibfield
  {journal} {\bibinfo  {journal} {nature}\ }\textbf {\bibinfo {volume} {395}},\
  \bibinfo {pages} {A14} (\bibinfo {year} {1998})}\BibitemShut {NoStop}%
\bibitem [{\citenamefont {Begelman}\ \emph {et~al.}(1980)\citenamefont
  {Begelman}, \citenamefont {Blandford},\ and\ \citenamefont
  {Rees}}]{begelman1980massive}%
  \BibitemOpen
  \bibfield  {author} {\bibinfo {author} {\bibfnamefont {M.~C.}\ \bibnamefont
  {Begelman}}, \bibinfo {author} {\bibfnamefont {R.~D.}\ \bibnamefont
  {Blandford}}, \ and\ \bibinfo {author} {\bibfnamefont {M.~J.}\ \bibnamefont
  {Rees}},\ }\href@noop {} {\bibfield  {journal} {\bibinfo  {journal} {Nature}\
  }\textbf {\bibinfo {volume} {287}},\ \bibinfo {pages} {307} (\bibinfo {year}
  {1980})}\BibitemShut {NoStop}%
\bibitem [{\citenamefont {Kormendy}\ and\ \citenamefont
  {Ho}(2013)}]{kormendy2013coevolution}%
  \BibitemOpen
  \bibfield  {author} {\bibinfo {author} {\bibfnamefont {J.}~\bibnamefont
  {Kormendy}}\ and\ \bibinfo {author} {\bibfnamefont {L.~C.}\ \bibnamefont
  {Ho}},\ }\href@noop {} {\bibfield  {journal} {\bibinfo  {journal} {Annual
  Review of Astronomy and Astrophysics}\ }\textbf {\bibinfo {volume} {51}},\
  \bibinfo {pages} {511} (\bibinfo {year} {2013})}\BibitemShut {NoStop}%
\bibitem [{\citenamefont {Amaro-Seoane}\ \emph {et~al.}(2017)\citenamefont
  {Amaro-Seoane}, \citenamefont {Audley}, \citenamefont {Babak}, \citenamefont
  {Baker}, \citenamefont {Barausse}, \citenamefont {Bender}, \citenamefont
  {Berti}, \citenamefont {Binetruy}, \citenamefont {Born}, \citenamefont
  {Bortoluzzi} \emph {et~al.}}]{amaro2017laser}%
  \BibitemOpen
  \bibfield  {author} {\bibinfo {author} {\bibfnamefont {P.}~\bibnamefont
  {Amaro-Seoane}}, \bibinfo {author} {\bibfnamefont {H.}~\bibnamefont
  {Audley}}, \bibinfo {author} {\bibfnamefont {S.}~\bibnamefont {Babak}},
  \bibinfo {author} {\bibfnamefont {J.}~\bibnamefont {Baker}}, \bibinfo
  {author} {\bibfnamefont {E.}~\bibnamefont {Barausse}}, \bibinfo {author}
  {\bibfnamefont {P.}~\bibnamefont {Bender}}, \bibinfo {author} {\bibfnamefont
  {E.}~\bibnamefont {Berti}}, \bibinfo {author} {\bibfnamefont
  {P.}~\bibnamefont {Binetruy}}, \bibinfo {author} {\bibfnamefont
  {M.}~\bibnamefont {Born}}, \bibinfo {author} {\bibfnamefont {D.}~\bibnamefont
  {Bortoluzzi}},  \emph {et~al.},\ }\href@noop {} {\bibfield  {journal}
  {\bibinfo  {journal} {arXiv preprint arXiv:1702.00786}\ } (\bibinfo {year}
  {2017})}\BibitemShut {NoStop}%
\bibitem [{\citenamefont {Barnes}\ and\ \citenamefont
  {Hernquist}(1996)}]{barnes1996transformations}%
  \BibitemOpen
  \bibfield  {author} {\bibinfo {author} {\bibfnamefont {J.~E.}\ \bibnamefont
  {Barnes}}\ and\ \bibinfo {author} {\bibfnamefont {L.}~\bibnamefont
  {Hernquist}},\ }\href@noop {} {\bibfield  {journal} {\bibinfo  {journal} {The
  Astrophysical Journal}\ }\textbf {\bibinfo {volume} {471}},\ \bibinfo {pages}
  {115} (\bibinfo {year} {1996})}\BibitemShut {NoStop}%
\bibitem [{\citenamefont {Milosavljevi{\'c}}\ and\ \citenamefont
  {Phinney}(2005)}]{milosavljevic2005afterglow}%
  \BibitemOpen
  \bibfield  {author} {\bibinfo {author} {\bibfnamefont {M.}~\bibnamefont
  {Milosavljevi{\'c}}}\ and\ \bibinfo {author} {\bibfnamefont {E.~S.}\
  \bibnamefont {Phinney}},\ }\href@noop {} {\bibfield  {journal} {\bibinfo
  {journal} {The Astrophysical Journal Letters}\ }\textbf {\bibinfo {volume}
  {622}},\ \bibinfo {pages} {L93} (\bibinfo {year} {2005})}\BibitemShut
  {NoStop}%
\bibitem [{\citenamefont {Narayan}\ \emph {et~al.}(2012)\citenamefont
  {Narayan}, \citenamefont {Sadowski}, \citenamefont {Penna},\ and\
  \citenamefont {Kulkarni}}]{narayan2012grmhd}%
  \BibitemOpen
  \bibfield  {author} {\bibinfo {author} {\bibfnamefont {R.}~\bibnamefont
  {Narayan}}, \bibinfo {author} {\bibfnamefont {A.}~\bibnamefont {Sadowski}},
  \bibinfo {author} {\bibfnamefont {R.~F.}\ \bibnamefont {Penna}}, \ and\
  \bibinfo {author} {\bibfnamefont {A.~K.}\ \bibnamefont {Kulkarni}},\
  }\href@noop {} {\bibfield  {journal} {\bibinfo  {journal} {Monthly Notices of
  the Royal Astronomical Society}\ }\textbf {\bibinfo {volume} {426}},\
  \bibinfo {pages} {3241} (\bibinfo {year} {2012})}\BibitemShut {NoStop}%
\bibitem [{\citenamefont {Yuan}\ \emph {et~al.}(2012)\citenamefont {Yuan},
  \citenamefont {Bu},\ and\ \citenamefont {Wu}}]{yuan2012numerical}%
  \BibitemOpen
  \bibfield  {author} {\bibinfo {author} {\bibfnamefont {F.}~\bibnamefont
  {Yuan}}, \bibinfo {author} {\bibfnamefont {D.}~\bibnamefont {Bu}}, \ and\
  \bibinfo {author} {\bibfnamefont {M.}~\bibnamefont {Wu}},\ }\href@noop {}
  {\bibfield  {journal} {\bibinfo  {journal} {The Astrophysical Journal}\
  }\textbf {\bibinfo {volume} {761}},\ \bibinfo {pages} {130} (\bibinfo {year}
  {2012})}\BibitemShut {NoStop}%
\bibitem [{\citenamefont {Sadowski}\ \emph {et~al.}(2013)\citenamefont
  {Sadowski}, \citenamefont {Narayan}, \citenamefont {Penna},\ and\
  \citenamefont {Zhu}}]{skadowski2013energy}%
  \BibitemOpen
  \bibfield  {author} {\bibinfo {author} {\bibfnamefont {A.}~\bibnamefont
  {Sadowski}}, \bibinfo {author} {\bibfnamefont {R.}~\bibnamefont {Narayan}},
  \bibinfo {author} {\bibfnamefont {R.}~\bibnamefont {Penna}}, \ and\ \bibinfo
  {author} {\bibfnamefont {Y.}~\bibnamefont {Zhu}},\ }\href@noop {} {\bibfield
  {journal} {\bibinfo  {journal} {Monthly Notices of the Royal Astronomical
  Society}\ }\textbf {\bibinfo {volume} {436}},\ \bibinfo {pages} {3856}
  (\bibinfo {year} {2013})}\BibitemShut {NoStop}%
\bibitem [{\citenamefont {Blandford}\ and\ \citenamefont
  {Znajek}(1977)}]{blandford1977electromagnetic}%
  \BibitemOpen
  \bibfield  {author} {\bibinfo {author} {\bibfnamefont {R.~D.}\ \bibnamefont
  {Blandford}}\ and\ \bibinfo {author} {\bibfnamefont {R.~L.}\ \bibnamefont
  {Znajek}},\ }\href@noop {} {\bibfield  {journal} {\bibinfo  {journal}
  {Monthly Notices of the Royal Astronomical Society}\ }\textbf {\bibinfo
  {volume} {179}},\ \bibinfo {pages} {433} (\bibinfo {year}
  {1977})}\BibitemShut {NoStop}%
\bibitem [{\citenamefont {Ramirez-Ruiz}\ \emph {et~al.}(2002)\citenamefont
  {Ramirez-Ruiz}, \citenamefont {Celotti},\ and\ \citenamefont
  {Rees}}]{ramirez2002events}%
  \BibitemOpen
  \bibfield  {author} {\bibinfo {author} {\bibfnamefont {E.}~\bibnamefont
  {Ramirez-Ruiz}}, \bibinfo {author} {\bibfnamefont {A.}~\bibnamefont
  {Celotti}}, \ and\ \bibinfo {author} {\bibfnamefont {M.~J.}\ \bibnamefont
  {Rees}},\ }\href@noop {} {\bibfield  {journal} {\bibinfo  {journal} {Monthly
  Notices of the Royal Astronomical Society}\ }\textbf {\bibinfo {volume}
  {337}},\ \bibinfo {pages} {1349} (\bibinfo {year} {2002})}\BibitemShut
  {NoStop}%
\bibitem [{\citenamefont {Zhang}\ \emph {et~al.}(2003)\citenamefont {Zhang},
  \citenamefont {Woosley},\ and\ \citenamefont
  {MacFadyen}}]{zhang2003relativistic}%
  \BibitemOpen
  \bibfield  {author} {\bibinfo {author} {\bibfnamefont {W.}~\bibnamefont
  {Zhang}}, \bibinfo {author} {\bibfnamefont {S.}~\bibnamefont {Woosley}}, \
  and\ \bibinfo {author} {\bibfnamefont {A.}~\bibnamefont {MacFadyen}},\
  }\href@noop {} {\bibfield  {journal} {\bibinfo  {journal} {The Astrophysical
  Journal}\ }\textbf {\bibinfo {volume} {586}},\ \bibinfo {pages} {356}
  (\bibinfo {year} {2003})}\BibitemShut {NoStop}%
\bibitem [{\citenamefont {Matzner}(2003)}]{matzner2003supernova}%
  \BibitemOpen
  \bibfield  {author} {\bibinfo {author} {\bibfnamefont {C.~D.}\ \bibnamefont
  {Matzner}},\ }\href@noop {} {\bibfield  {journal} {\bibinfo  {journal}
  {Monthly Notices of the Royal Astronomical Society}\ }\textbf {\bibinfo
  {volume} {345}},\ \bibinfo {pages} {575} (\bibinfo {year}
  {2003})}\BibitemShut {NoStop}%
\bibitem [{\citenamefont {Bromberg}\ \emph {et~al.}(2011)\citenamefont
  {Bromberg}, \citenamefont {Nakar}, \citenamefont {Piran} \emph
  {et~al.}}]{bromberg2011propagation}%
  \BibitemOpen
  \bibfield  {author} {\bibinfo {author} {\bibfnamefont {O.}~\bibnamefont
  {Bromberg}}, \bibinfo {author} {\bibfnamefont {E.}~\bibnamefont {Nakar}},
  \bibinfo {author} {\bibfnamefont {T.}~\bibnamefont {Piran}},  \emph
  {et~al.},\ }\href@noop {} {\bibfield  {journal} {\bibinfo  {journal} {The
  Astrophysical Journal}\ }\textbf {\bibinfo {volume} {740}},\ \bibinfo {pages}
  {100} (\bibinfo {year} {2011})}\BibitemShut {NoStop}%
\bibitem [{\citenamefont {Mizuta}\ and\ \citenamefont
  {Ioka}(2013)}]{mizuta2013opening}%
  \BibitemOpen
  \bibfield  {author} {\bibinfo {author} {\bibfnamefont {A.}~\bibnamefont
  {Mizuta}}\ and\ \bibinfo {author} {\bibfnamefont {K.}~\bibnamefont {Ioka}},\
  }\href@noop {} {\bibfield  {journal} {\bibinfo  {journal} {The Astrophysical
  Journal}\ }\textbf {\bibinfo {volume} {777}},\ \bibinfo {pages} {162}
  (\bibinfo {year} {2013})}\BibitemShut {NoStop}%
\bibitem [{\citenamefont {Nakar}\ and\ \citenamefont
  {Piran}(2016)}]{nakar2016observable}%
  \BibitemOpen
  \bibfield  {author} {\bibinfo {author} {\bibfnamefont {E.}~\bibnamefont
  {Nakar}}\ and\ \bibinfo {author} {\bibfnamefont {T.}~\bibnamefont {Piran}},\
  }\href@noop {} {\bibfield  {journal} {\bibinfo  {journal} {The Astrophysical
  Journal}\ }\textbf {\bibinfo {volume} {834}},\ \bibinfo {pages} {28}
  (\bibinfo {year} {2016})}\BibitemShut {NoStop}%
\bibitem [{\citenamefont {Lazzati}\ \emph {et~al.}(2017)\citenamefont
  {Lazzati}, \citenamefont {Deich}, \citenamefont {Morsony},\ and\
  \citenamefont {Workman}}]{lazzati2017off}%
  \BibitemOpen
  \bibfield  {author} {\bibinfo {author} {\bibfnamefont {D.}~\bibnamefont
  {Lazzati}}, \bibinfo {author} {\bibfnamefont {A.}~\bibnamefont {Deich}},
  \bibinfo {author} {\bibfnamefont {B.~J.}\ \bibnamefont {Morsony}}, \ and\
  \bibinfo {author} {\bibfnamefont {J.~C.}\ \bibnamefont {Workman}},\
  }\href@noop {} {\bibfield  {journal} {\bibinfo  {journal} {Monthly Notices of
  the Royal Astronomical Society}\ }\textbf {\bibinfo {volume} {471}},\
  \bibinfo {pages} {1652} (\bibinfo {year} {2017})}\BibitemShut {NoStop}%
\bibitem [{\citenamefont {Matsumoto}\ and\ \citenamefont
  {Kimura}(2018)}]{matsumoto2018delayed}%
  \BibitemOpen
  \bibfield  {author} {\bibinfo {author} {\bibfnamefont {T.}~\bibnamefont
  {Matsumoto}}\ and\ \bibinfo {author} {\bibfnamefont {S.~S.}\ \bibnamefont
  {Kimura}},\ }\href@noop {} {\bibfield  {journal} {\bibinfo  {journal} {The
  Astrophysical Journal Letters}\ }\textbf {\bibinfo {volume} {866}},\ \bibinfo
  {pages} {L16} (\bibinfo {year} {2018})}\BibitemShut {NoStop}%
\bibitem [{\citenamefont {Lyutikov}(2019)}]{10.1093/mnras/stz3044}%
  \BibitemOpen
  \bibfield  {author} {\bibinfo {author} {\bibfnamefont {M.}~\bibnamefont
  {Lyutikov}},\ }\href@noop {} {\bibfield  {journal} {\bibinfo  {journal}
  {Monthly Notices of the Royal Astronomical Society}\ }\textbf {\bibinfo
  {volume} {491}},\ \bibinfo {pages} {483} (\bibinfo {year}
  {2019})}\BibitemShut {NoStop}%
\bibitem [{\citenamefont {Hamidani}\ \emph {et~al.}(2020)\citenamefont
  {Hamidani}, \citenamefont {Kiuchi},\ and\ \citenamefont
  {Ioka}}]{hamidani2020jet}%
  \BibitemOpen
  \bibfield  {author} {\bibinfo {author} {\bibfnamefont {H.}~\bibnamefont
  {Hamidani}}, \bibinfo {author} {\bibfnamefont {K.}~\bibnamefont {Kiuchi}}, \
  and\ \bibinfo {author} {\bibfnamefont {K.}~\bibnamefont {Ioka}},\ }\href@noop
  {} {\bibfield  {journal} {\bibinfo  {journal} {Monthly Notices of the Royal
  Astronomical Society}\ }\textbf {\bibinfo {volume} {491}},\ \bibinfo {pages}
  {3192} (\bibinfo {year} {2020})}\BibitemShut {NoStop}%
\bibitem [{\citenamefont {Rees}\ and\ \citenamefont
  {M{\'e}sz{\'a}ros}(1994)}]{rees1994unsteady}%
  \BibitemOpen
  \bibfield  {author} {\bibinfo {author} {\bibfnamefont {M.~J.}\ \bibnamefont
  {Rees}}\ and\ \bibinfo {author} {\bibfnamefont {P.}~\bibnamefont
  {M{\'e}sz{\'a}ros}},\ }\href@noop {} {\bibfield  {journal} {\bibinfo
  {journal} {arXiv preprint astro-ph/9404038}\ } (\bibinfo {year}
  {1994})}\BibitemShut {NoStop}%
\bibitem [{\citenamefont {Armitage}\ and\ \citenamefont
  {Natarajan}(2002)}]{armitage2002accretion}%
  \BibitemOpen
  \bibfield  {author} {\bibinfo {author} {\bibfnamefont {P.~J.}\ \bibnamefont
  {Armitage}}\ and\ \bibinfo {author} {\bibfnamefont {P.}~\bibnamefont
  {Natarajan}},\ }\href@noop {} {\bibfield  {journal} {\bibinfo  {journal} {The
  Astrophysical Journal Letters}\ }\textbf {\bibinfo {volume} {567}},\ \bibinfo
  {pages} {L9} (\bibinfo {year} {2002})}\BibitemShut {NoStop}%
\bibitem [{\citenamefont {Escala}\ \emph {et~al.}(2005)\citenamefont {Escala},
  \citenamefont {Larson}, \citenamefont {Coppi},\ and\ \citenamefont
  {Mardones}}]{escala2005role}%
  \BibitemOpen
  \bibfield  {author} {\bibinfo {author} {\bibfnamefont {A.}~\bibnamefont
  {Escala}}, \bibinfo {author} {\bibfnamefont {R.~B.}\ \bibnamefont {Larson}},
  \bibinfo {author} {\bibfnamefont {P.~S.}\ \bibnamefont {Coppi}}, \ and\
  \bibinfo {author} {\bibfnamefont {D.}~\bibnamefont {Mardones}},\ }\href@noop
  {} {\bibfield  {journal} {\bibinfo  {journal} {The Astrophysical Journal}\
  }\textbf {\bibinfo {volume} {630}},\ \bibinfo {pages} {152} (\bibinfo {year}
  {2005})}\BibitemShut {NoStop}%
\bibitem [{\citenamefont {Dotti}\ \emph {et~al.}(2007)\citenamefont {Dotti},
  \citenamefont {Colpi}, \citenamefont {Haardt},\ and\ \citenamefont
  {Mayer}}]{dotti2007supermassive}%
  \BibitemOpen
  \bibfield  {author} {\bibinfo {author} {\bibfnamefont {M.}~\bibnamefont
  {Dotti}}, \bibinfo {author} {\bibfnamefont {M.}~\bibnamefont {Colpi}},
  \bibinfo {author} {\bibfnamefont {F.}~\bibnamefont {Haardt}}, \ and\ \bibinfo
  {author} {\bibfnamefont {L.}~\bibnamefont {Mayer}},\ }\href@noop {}
  {\bibfield  {journal} {\bibinfo  {journal} {Monthly Notices of the Royal
  Astronomical Society}\ }\textbf {\bibinfo {volume} {379}},\ \bibinfo {pages}
  {956} (\bibinfo {year} {2007})}\BibitemShut {NoStop}%
\bibitem [{\citenamefont {Pringle}(1981)}]{pringle1981accretion}%
  \BibitemOpen
  \bibfield  {author} {\bibinfo {author} {\bibfnamefont {J.}~\bibnamefont
  {Pringle}},\ }\href@noop {} {\bibfield  {journal} {\bibinfo  {journal}
  {Annual review of astronomy and astrophysics}\ }\textbf {\bibinfo {volume}
  {19}},\ \bibinfo {pages} {137} (\bibinfo {year} {1981})}\BibitemShut
  {NoStop}%
\bibitem [{\citenamefont {D'Ascoli}\ \emph {et~al.}(2018)\citenamefont
  {D'Ascoli}, \citenamefont {Noble}, \citenamefont {Bowen}, \citenamefont
  {Campanelli}, \citenamefont {Krolik},\ and\ \citenamefont
  {Mewes}}]{DAscoli2018}%
  \BibitemOpen
  \bibfield  {author} {\bibinfo {author} {\bibfnamefont {S.}~\bibnamefont
  {D'Ascoli}}, \bibinfo {author} {\bibfnamefont {S.~C.}\ \bibnamefont {Noble}},
  \bibinfo {author} {\bibfnamefont {D.~B.}\ \bibnamefont {Bowen}}, \bibinfo
  {author} {\bibfnamefont {M.}~\bibnamefont {Campanelli}}, \bibinfo {author}
  {\bibfnamefont {J.~H.}\ \bibnamefont {Krolik}}, \ and\ \bibinfo {author}
  {\bibfnamefont {V.}~\bibnamefont {Mewes}},\ }\href {\doibase
  10.3847/1538-4357/aad8b4} {\bibfield  {journal} {\bibinfo  {journal} {The
  Astrophysical Journal}\ }\textbf {\bibinfo {volume} {865}},\ \bibinfo {pages}
  {140} (\bibinfo {year} {2018})},\ \Eprint {http://arxiv.org/abs/1806.05697}
  {arXiv:1806.05697} \BibitemShut {NoStop}%
\bibitem [{\citenamefont {Shapiro}\ and\ \citenamefont
  {Teukolsky}(2008)}]{shapiro2008black}%
  \BibitemOpen
  \bibfield  {author} {\bibinfo {author} {\bibfnamefont {S.~L.}\ \bibnamefont
  {Shapiro}}\ and\ \bibinfo {author} {\bibfnamefont {S.~A.}\ \bibnamefont
  {Teukolsky}},\ }\href@noop {} {\emph {\bibinfo {title} {Black holes, white
  dwarfs, and neutron stars: The physics of compact objects}}}\ (\bibinfo
  {publisher} {John Wiley \& Sons},\ \bibinfo {year} {2008})\BibitemShut
  {NoStop}%
\bibitem [{\citenamefont {Farris}\ \emph {et~al.}(2015)\citenamefont {Farris},
  \citenamefont {Duffell}, \citenamefont {MacFadyen},\ and\ \citenamefont
  {Haiman}}]{farris2015binary}%
  \BibitemOpen
  \bibfield  {author} {\bibinfo {author} {\bibfnamefont {B.~D.}\ \bibnamefont
  {Farris}}, \bibinfo {author} {\bibfnamefont {P.}~\bibnamefont {Duffell}},
  \bibinfo {author} {\bibfnamefont {A.~I.}\ \bibnamefont {MacFadyen}}, \ and\
  \bibinfo {author} {\bibfnamefont {Z.}~\bibnamefont {Haiman}},\ }\href@noop {}
  {\bibfield  {journal} {\bibinfo  {journal} {Monthly Notices of the Royal
  Astronomical Society: Letters}\ }\textbf {\bibinfo {volume} {447}},\ \bibinfo
  {pages} {L80} (\bibinfo {year} {2015})}\BibitemShut {NoStop}%
\bibitem [{\citenamefont {Jiang}\ \emph
  {et~al.}(2019{\natexlab{a}})\citenamefont {Jiang}, \citenamefont {Stone},\
  and\ \citenamefont {Davis}}]{Jiang2019a}%
  \BibitemOpen
  \bibfield  {author} {\bibinfo {author} {\bibfnamefont {Y.-F.}\ \bibnamefont
  {Jiang}}, \bibinfo {author} {\bibfnamefont {J.~M.}\ \bibnamefont {Stone}}, \
  and\ \bibinfo {author} {\bibfnamefont {S.~W.}\ \bibnamefont {Davis}},\ }\href
  {\doibase 10.3847/1538-4357/ab29ff} {\bibfield  {journal} {\bibinfo
  {journal} {The Astrophysical Journal}\ }\textbf {\bibinfo {volume} {880}},\
  \bibinfo {pages} {67} (\bibinfo {year} {2019}{\natexlab{a}})},\ \Eprint
  {http://arxiv.org/abs/1709.02845} {arXiv:1709.02845} \BibitemShut {NoStop}%
\bibitem [{\citenamefont {Jiang}\ \emph
  {et~al.}(2019{\natexlab{b}})\citenamefont {Jiang}, \citenamefont {Blaes},
  \citenamefont {Stone},\ and\ \citenamefont {Davis}}]{Jiang2019}%
  \BibitemOpen
  \bibfield  {author} {\bibinfo {author} {\bibfnamefont {Y.-F.}\ \bibnamefont
  {Jiang}}, \bibinfo {author} {\bibfnamefont {O.}~\bibnamefont {Blaes}},
  \bibinfo {author} {\bibfnamefont {J.~M.}\ \bibnamefont {Stone}}, \ and\
  \bibinfo {author} {\bibfnamefont {S.~W.}\ \bibnamefont {Davis}},\ }\href
  {\doibase 10.3847/1538-4357/ab4a00} {\bibfield  {journal} {\bibinfo
  {journal} {The Astrophysical Journal}\ }\textbf {\bibinfo {volume} {885}},\
  \bibinfo {pages} {144} (\bibinfo {year} {2019}{\natexlab{b}})},\ \Eprint
  {http://arxiv.org/abs/1904.01674} {arXiv:1904.01674} \BibitemShut {NoStop}%
\bibitem [{\citenamefont {Ohsuga}\ \emph {et~al.}(2009)\citenamefont {Ohsuga},
  \citenamefont {Mlneshige}, \citenamefont {Mori},\ and\ \citenamefont
  {Kato}}]{Ohsuga2009}%
  \BibitemOpen
  \bibfield  {author} {\bibinfo {author} {\bibfnamefont {K.}~\bibnamefont
  {Ohsuga}}, \bibinfo {author} {\bibfnamefont {S.}~\bibnamefont {Mlneshige}},
  \bibinfo {author} {\bibfnamefont {M.}~\bibnamefont {Mori}}, \ and\ \bibinfo
  {author} {\bibfnamefont {Y.}~\bibnamefont {Kato}},\ }\href {\doibase
  10.1093/pasj/61.3.L7} {\bibfield  {journal} {\bibinfo  {journal}
  {Publications of the Astronomical Society of Japan}\ }\textbf {\bibinfo
  {volume} {61}},\ \bibinfo {pages} {7} (\bibinfo {year} {2009})},\ \Eprint
  {http://arxiv.org/abs/0903.5364} {arXiv:0903.5364} \BibitemShut {NoStop}%
\bibitem [{\citenamefont {Akiyama}\ \emph {et~al.}(2019)\citenamefont
  {Akiyama}, \citenamefont {Alberdi}, \citenamefont {Alef} \emph
  {et~al.}}]{firstm87}%
  \BibitemOpen
  \bibfield  {author} {\bibinfo {author} {\bibfnamefont {K.}~\bibnamefont
  {Akiyama}}, \bibinfo {author} {\bibfnamefont {A.}~\bibnamefont {Alberdi}},
  \bibinfo {author} {\bibfnamefont {W.}~\bibnamefont {Alef}},  \emph {et~al.},\
  }\href {\doibase 10.3847/2041-8213/ab0f43} {\bibfield  {journal} {\bibinfo
  {journal} {The Astrophysical Journal}\ }\textbf {\bibinfo {volume} {875}},\
  \bibinfo {pages} {L5} (\bibinfo {year} {2019})}\BibitemShut {NoStop}%
\bibitem [{\citenamefont {Tchekhovskoy}\ \emph {et~al.}(2011)\citenamefont
  {Tchekhovskoy}, \citenamefont {Narayan},\ and\ \citenamefont
  {McKinney}}]{tchekhovskoy2011efficient}%
  \BibitemOpen
  \bibfield  {author} {\bibinfo {author} {\bibfnamefont {A.}~\bibnamefont
  {Tchekhovskoy}}, \bibinfo {author} {\bibfnamefont {R.}~\bibnamefont
  {Narayan}}, \ and\ \bibinfo {author} {\bibfnamefont {J.~C.}\ \bibnamefont
  {McKinney}},\ }\href@noop {} {\bibfield  {journal} {\bibinfo  {journal}
  {Monthly Notices of the Royal Astronomical Society: Letters}\ }\textbf
  {\bibinfo {volume} {418}},\ \bibinfo {pages} {L79} (\bibinfo {year}
  {2011})}\BibitemShut {NoStop}%
\bibitem [{\citenamefont {Harrison}\ \emph {et~al.}(2018)\citenamefont
  {Harrison}, \citenamefont {Gottlieb},\ and\ \citenamefont
  {Nakar}}]{harrison2018numerically}%
  \BibitemOpen
  \bibfield  {author} {\bibinfo {author} {\bibfnamefont {R.}~\bibnamefont
  {Harrison}}, \bibinfo {author} {\bibfnamefont {O.}~\bibnamefont {Gottlieb}},
  \ and\ \bibinfo {author} {\bibfnamefont {E.}~\bibnamefont {Nakar}},\
  }\href@noop {} {\bibfield  {journal} {\bibinfo  {journal} {Monthly Notices of
  the Royal Astronomical Society}\ }\textbf {\bibinfo {volume} {477}},\
  \bibinfo {pages} {2128} (\bibinfo {year} {2018})}\BibitemShut {NoStop}%
\bibitem [{\citenamefont {Gottlieb}\ \emph {et~al.}(2019)\citenamefont
  {Gottlieb}, \citenamefont {Levinson},\ and\ \citenamefont
  {Nakar}}]{gottlieb2019high}%
  \BibitemOpen
  \bibfield  {author} {\bibinfo {author} {\bibfnamefont {O.}~\bibnamefont
  {Gottlieb}}, \bibinfo {author} {\bibfnamefont {A.}~\bibnamefont {Levinson}},
  \ and\ \bibinfo {author} {\bibfnamefont {E.}~\bibnamefont {Nakar}},\
  }\href@noop {} {\bibfield  {journal} {\bibinfo  {journal} {Monthly Notices of
  the Royal Astronomical Society}\ }\textbf {\bibinfo {volume} {488}},\
  \bibinfo {pages} {1416} (\bibinfo {year} {2019})}\BibitemShut {NoStop}%
\bibitem [{\citenamefont {Budnik}\ \emph {et~al.}(2010)\citenamefont {Budnik},
  \citenamefont {Katz}, \citenamefont {Sagiv},\ and\ \citenamefont
  {Waxman}}]{budnik2010relativistic}%
  \BibitemOpen
  \bibfield  {author} {\bibinfo {author} {\bibfnamefont {R.}~\bibnamefont
  {Budnik}}, \bibinfo {author} {\bibfnamefont {B.}~\bibnamefont {Katz}},
  \bibinfo {author} {\bibfnamefont {A.}~\bibnamefont {Sagiv}}, \ and\ \bibinfo
  {author} {\bibfnamefont {E.}~\bibnamefont {Waxman}},\ }\href@noop {}
  {\bibfield  {journal} {\bibinfo  {journal} {The Astrophysical Journal}\
  }\textbf {\bibinfo {volume} {725}},\ \bibinfo {pages} {63} (\bibinfo {year}
  {2010})}\BibitemShut {NoStop}%
\bibitem [{\citenamefont {Nakar}\ \emph {et~al.}(2012)\citenamefont {Nakar}
  \emph {et~al.}}]{nakar2012relativistic}%
  \BibitemOpen
  \bibfield  {author} {\bibinfo {author} {\bibfnamefont {E.}~\bibnamefont
  {Nakar}} \emph {et~al.},\ }\href@noop {} {\bibfield  {journal} {\bibinfo
  {journal} {The Astrophysical Journal}\ }\textbf {\bibinfo {volume} {747}},\
  \bibinfo {pages} {88} (\bibinfo {year} {2012})}\BibitemShut {NoStop}%
\bibitem [{\citenamefont {Murase}\ and\ \citenamefont
  {Ioka}(2013)}]{murase2013tev}%
  \BibitemOpen
  \bibfield  {author} {\bibinfo {author} {\bibfnamefont {K.}~\bibnamefont
  {Murase}}\ and\ \bibinfo {author} {\bibfnamefont {K.}~\bibnamefont {Ioka}},\
  }\href@noop {} {\bibfield  {journal} {\bibinfo  {journal} {Physical Review
  Letters}\ }\textbf {\bibinfo {volume} {111}},\ \bibinfo {pages} {121102}
  (\bibinfo {year} {2013})}\BibitemShut {NoStop}%
\bibitem [{\citenamefont {Sari}\ and\ \citenamefont
  {Esin}(2001)}]{sari2001synchrotron}%
  \BibitemOpen
  \bibfield  {author} {\bibinfo {author} {\bibfnamefont {R.}~\bibnamefont
  {Sari}}\ and\ \bibinfo {author} {\bibfnamefont {A.~A.}\ \bibnamefont
  {Esin}},\ }\href@noop {} {\bibfield  {journal} {\bibinfo  {journal} {The
  Astrophysical Journal}\ }\textbf {\bibinfo {volume} {548}},\ \bibinfo {pages}
  {787} (\bibinfo {year} {2001})}\BibitemShut {NoStop}%
\bibitem [{\citenamefont {Murase}\ \emph {et~al.}(2011)\citenamefont {Murase},
  \citenamefont {Toma}, \citenamefont {Yamazaki},\ and\ \citenamefont
  {M{\'e}sz{\'a}ros}}]{murase2011implications}%
  \BibitemOpen
  \bibfield  {author} {\bibinfo {author} {\bibfnamefont {K.}~\bibnamefont
  {Murase}}, \bibinfo {author} {\bibfnamefont {K.}~\bibnamefont {Toma}},
  \bibinfo {author} {\bibfnamefont {R.}~\bibnamefont {Yamazaki}}, \ and\
  \bibinfo {author} {\bibfnamefont {P.}~\bibnamefont {M{\'e}sz{\'a}ros}},\
  }\href@noop {} {\bibfield  {journal} {\bibinfo  {journal} {The Astrophysical
  Journal}\ }\textbf {\bibinfo {volume} {732}},\ \bibinfo {pages} {77}
  (\bibinfo {year} {2011})}\BibitemShut {NoStop}%
\bibitem [{\citenamefont {Panaitescu}\ and\ \citenamefont
  {Kumar}(2000)}]{panaitescu2000analytic}%
  \BibitemOpen
  \bibfield  {author} {\bibinfo {author} {\bibfnamefont {A.}~\bibnamefont
  {Panaitescu}}\ and\ \bibinfo {author} {\bibfnamefont {P.}~\bibnamefont
  {Kumar}},\ }\href@noop {} {\bibfield  {journal} {\bibinfo  {journal} {The
  Astrophysical Journal}\ }\textbf {\bibinfo {volume} {543}},\ \bibinfo {pages}
  {66} (\bibinfo {year} {2000})}\BibitemShut {NoStop}%
\bibitem [{\citenamefont {Zhang}\ and\ \citenamefont
  {Meszaros}(2001)}]{zhang2001high}%
  \BibitemOpen
  \bibfield  {author} {\bibinfo {author} {\bibfnamefont {B.}~\bibnamefont
  {Zhang}}\ and\ \bibinfo {author} {\bibfnamefont {P.}~\bibnamefont
  {Meszaros}},\ }\href@noop {} {\bibfield  {journal} {\bibinfo  {journal} {The
  Astrophysical Journal}\ }\textbf {\bibinfo {volume} {559}},\ \bibinfo {pages}
  {110} (\bibinfo {year} {2001})}\BibitemShut {NoStop}%
\bibitem [{\citenamefont {Murase}\ and\ \citenamefont
  {Nagataki}(2006)}]{murase2006high}%
  \BibitemOpen
  \bibfield  {author} {\bibinfo {author} {\bibfnamefont {K.}~\bibnamefont
  {Murase}}\ and\ \bibinfo {author} {\bibfnamefont {S.}~\bibnamefont
  {Nagataki}},\ }\href@noop {} {\bibfield  {journal} {\bibinfo  {journal}
  {Physical Review D}\ }\textbf {\bibinfo {volume} {73}},\ \bibinfo {pages}
  {063002} (\bibinfo {year} {2006})}\BibitemShut {NoStop}%
\bibitem [{\citenamefont {Stepney}\ and\ \citenamefont
  {Guilbert}(1983)}]{stepney1983numerical}%
  \BibitemOpen
  \bibfield  {author} {\bibinfo {author} {\bibfnamefont {S.}~\bibnamefont
  {Stepney}}\ and\ \bibinfo {author} {\bibfnamefont {P.~W.}\ \bibnamefont
  {Guilbert}},\ }\href@noop {} {\bibfield  {journal} {\bibinfo  {journal}
  {Monthly Notices of the Royal Astronomical Society}\ }\textbf {\bibinfo
  {volume} {204}},\ \bibinfo {pages} {1269} (\bibinfo {year}
  {1983})}\BibitemShut {NoStop}%
\bibitem [{\citenamefont {Chodorowski}\ \emph {et~al.}(1992)\citenamefont
  {Chodorowski}, \citenamefont {Zdziarski},\ and\ \citenamefont
  {Sikora}}]{chodorowski1992reaction}%
  \BibitemOpen
  \bibfield  {author} {\bibinfo {author} {\bibfnamefont {M.~J.}\ \bibnamefont
  {Chodorowski}}, \bibinfo {author} {\bibfnamefont {A.~A.}\ \bibnamefont
  {Zdziarski}}, \ and\ \bibinfo {author} {\bibfnamefont {M.}~\bibnamefont
  {Sikora}},\ }\href@noop {} {\bibfield  {journal} {\bibinfo  {journal} {The
  Astrophysical Journal}\ }\textbf {\bibinfo {volume} {400}},\ \bibinfo {pages}
  {181} (\bibinfo {year} {1992})}\BibitemShut {NoStop}%
\bibitem [{\citenamefont {Harrison}\ \emph {et~al.}(2002)\citenamefont
  {Harrison}, \citenamefont {Perkins},\ and\ \citenamefont
  {Scott}}]{harrison2002tri}%
  \BibitemOpen
  \bibfield  {author} {\bibinfo {author} {\bibfnamefont {P.~F.}\ \bibnamefont
  {Harrison}}, \bibinfo {author} {\bibfnamefont {D.~H.}\ \bibnamefont
  {Perkins}}, \ and\ \bibinfo {author} {\bibfnamefont {W.}~\bibnamefont
  {Scott}},\ }\href@noop {} {\bibfield  {journal} {\bibinfo  {journal} {Physics
  Letters B}\ }\textbf {\bibinfo {volume} {530}},\ \bibinfo {pages} {167}
  (\bibinfo {year} {2002})}\BibitemShut {NoStop}%
\bibitem [{\citenamefont {Aartsen}\ \emph {et~al.}(2017)\citenamefont
  {Aartsen}, \citenamefont {Ackermann}, \citenamefont {Adams} \emph
  {et~al.}}]{ICeffectivearea}%
  \BibitemOpen
  \bibfield  {author} {\bibinfo {author} {\bibfnamefont {M.~G.}\ \bibnamefont
  {Aartsen}}, \bibinfo {author} {\bibfnamefont {M.}~\bibnamefont {Ackermann}},
  \bibinfo {author} {\bibfnamefont {J.}~\bibnamefont {Adams}},  \emph
  {et~al.},\ }\href {\doibase 10.3847/1538-4357/aa7569} {\bibfield  {journal}
  {\bibinfo  {journal} {The Astrophysical Journal}\ }\textbf {\bibinfo {volume}
  {843}},\ \bibinfo {pages} {112} (\bibinfo {year} {2017})},\ \Eprint
  {http://arxiv.org/abs/1702.06868} {arXiv:1702.06868} \BibitemShut {NoStop}%
\bibitem [{\citenamefont {Aartsen}\ \emph
  {et~al.}(2014{\natexlab{b}})\citenamefont {Aartsen}, \citenamefont
  {Ackermann}, \citenamefont {Adams}, \citenamefont {Aguilar}, \citenamefont
  {Ahlers}, \citenamefont {Ahrens}, \citenamefont {Altmann}, \citenamefont
  {Anderson}, \citenamefont {Anton}, \citenamefont {Arguelles} \emph
  {et~al.}}]{aartsen2014icecube}%
  \BibitemOpen
  \bibfield  {author} {\bibinfo {author} {\bibfnamefont {M.}~\bibnamefont
  {Aartsen}}, \bibinfo {author} {\bibfnamefont {M.}~\bibnamefont {Ackermann}},
  \bibinfo {author} {\bibfnamefont {J.}~\bibnamefont {Adams}}, \bibinfo
  {author} {\bibfnamefont {J.}~\bibnamefont {Aguilar}}, \bibinfo {author}
  {\bibfnamefont {M.}~\bibnamefont {Ahlers}}, \bibinfo {author} {\bibfnamefont
  {M.}~\bibnamefont {Ahrens}}, \bibinfo {author} {\bibfnamefont
  {D.}~\bibnamefont {Altmann}}, \bibinfo {author} {\bibfnamefont
  {T.}~\bibnamefont {Anderson}}, \bibinfo {author} {\bibfnamefont
  {G.}~\bibnamefont {Anton}}, \bibinfo {author} {\bibfnamefont
  {C.}~\bibnamefont {Arguelles}},  \emph {et~al.},\ }\href@noop {} {\bibfield
  {journal} {\bibinfo  {journal} {arXiv preprint arXiv:1412.5106}\ } (\bibinfo
  {year} {2014}{\natexlab{b}})}\BibitemShut {NoStop}%
\bibitem [{\citenamefont {Adrian-Martinez}\ \emph {et~al.}(2016)\citenamefont
  {Adrian-Martinez}, \citenamefont {Ageron}, \citenamefont {Aharonian},
  \citenamefont {Aiello}, \citenamefont {Albert}, \citenamefont {Ameli},
  \citenamefont {Anassontzis}, \citenamefont {Andre}, \citenamefont
  {Androulakis}, \citenamefont {Anghinolfi} \emph {et~al.}}]{adrian2016letter}%
  \BibitemOpen
  \bibfield  {author} {\bibinfo {author} {\bibfnamefont {S.}~\bibnamefont
  {Adrian-Martinez}}, \bibinfo {author} {\bibfnamefont {M.}~\bibnamefont
  {Ageron}}, \bibinfo {author} {\bibfnamefont {F.}~\bibnamefont {Aharonian}},
  \bibinfo {author} {\bibfnamefont {S.}~\bibnamefont {Aiello}}, \bibinfo
  {author} {\bibfnamefont {A.}~\bibnamefont {Albert}}, \bibinfo {author}
  {\bibfnamefont {F.}~\bibnamefont {Ameli}}, \bibinfo {author} {\bibfnamefont
  {E.}~\bibnamefont {Anassontzis}}, \bibinfo {author} {\bibfnamefont
  {M.}~\bibnamefont {Andre}}, \bibinfo {author} {\bibfnamefont
  {G.}~\bibnamefont {Androulakis}}, \bibinfo {author} {\bibfnamefont
  {M.}~\bibnamefont {Anghinolfi}},  \emph {et~al.},\ }\href@noop {} {\bibfield
  {journal} {\bibinfo  {journal} {Journal of Physics G: Nuclear and Particle
  Physics}\ }\textbf {\bibinfo {volume} {43}},\ \bibinfo {pages} {084001}
  (\bibinfo {year} {2016})}\BibitemShut {NoStop}%
\bibitem [{\citenamefont {Murase}\ \emph {et~al.}(2007)\citenamefont {Murase},
  \citenamefont {Asano},\ and\ \citenamefont {Nagataki}}]{Murase:2007ar}%
  \BibitemOpen
  \bibfield  {author} {\bibinfo {author} {\bibfnamefont {K.}~\bibnamefont
  {Murase}}, \bibinfo {author} {\bibfnamefont {K.}~\bibnamefont {Asano}}, \
  and\ \bibinfo {author} {\bibfnamefont {S.}~\bibnamefont {Nagataki}},\ }\href
  {\doibase 10.1086/523031} {\bibfield  {journal} {\bibinfo  {journal}
  {Astrophys. J.}\ }\textbf {\bibinfo {volume} {671}},\ \bibinfo {pages} {1886}
  (\bibinfo {year} {2007})},\ \Eprint {http://arxiv.org/abs/astro-ph/0703759}
  {arXiv:astro-ph/0703759} \BibitemShut {NoStop}%
\bibitem [{\citenamefont {Murase}\ and\ \citenamefont
  {Waxman}(2016)}]{murase2016constraining}%
  \BibitemOpen
  \bibfield  {author} {\bibinfo {author} {\bibfnamefont {K.}~\bibnamefont
  {Murase}}\ and\ \bibinfo {author} {\bibfnamefont {E.}~\bibnamefont
  {Waxman}},\ }\href@noop {} {\bibfield  {journal} {\bibinfo  {journal}
  {Physical Review D}\ }\textbf {\bibinfo {volume} {94}},\ \bibinfo {pages}
  {103006} (\bibinfo {year} {2016})}\BibitemShut {NoStop}%
\bibitem [{\citenamefont {Menou}\ \emph {et~al.}(2001)\citenamefont {Menou},
  \citenamefont {Haiman},\ and\ \citenamefont {Narayanan}}]{menou2001merger}%
  \BibitemOpen
  \bibfield  {author} {\bibinfo {author} {\bibfnamefont {K.}~\bibnamefont
  {Menou}}, \bibinfo {author} {\bibfnamefont {Z.}~\bibnamefont {Haiman}}, \
  and\ \bibinfo {author} {\bibfnamefont {V.~K.}\ \bibnamefont {Narayanan}},\
  }\href@noop {} {\bibfield  {journal} {\bibinfo  {journal} {The Astrophysical
  Journal}\ }\textbf {\bibinfo {volume} {558}},\ \bibinfo {pages} {535}
  (\bibinfo {year} {2001})}\BibitemShut {NoStop}%
\bibitem [{\citenamefont {Erickcek}\ \emph {et~al.}(2006)\citenamefont
  {Erickcek}, \citenamefont {Kamionkowski},\ and\ \citenamefont
  {Benson}}]{erickcek2006supermassive}%
  \BibitemOpen
  \bibfield  {author} {\bibinfo {author} {\bibfnamefont {A.~L.}\ \bibnamefont
  {Erickcek}}, \bibinfo {author} {\bibfnamefont {M.}~\bibnamefont
  {Kamionkowski}}, \ and\ \bibinfo {author} {\bibfnamefont {A.~J.}\
  \bibnamefont {Benson}},\ }\href@noop {} {\bibfield  {journal} {\bibinfo
  {journal} {Monthly Notices of the Royal Astronomical Society}\ }\textbf
  {\bibinfo {volume} {371}},\ \bibinfo {pages} {1992} (\bibinfo {year}
  {2006})}\BibitemShut {NoStop}%
\bibitem [{\citenamefont {Micic}\ \emph {et~al.}(2007)\citenamefont {Micic},
  \citenamefont {Holley-Bockelmann}, \citenamefont {Sigurdsson},\ and\
  \citenamefont {Abel}}]{micic2007supermassive}%
  \BibitemOpen
  \bibfield  {author} {\bibinfo {author} {\bibfnamefont {M.}~\bibnamefont
  {Micic}}, \bibinfo {author} {\bibfnamefont {K.}~\bibnamefont
  {Holley-Bockelmann}}, \bibinfo {author} {\bibfnamefont {S.}~\bibnamefont
  {Sigurdsson}}, \ and\ \bibinfo {author} {\bibfnamefont {T.}~\bibnamefont
  {Abel}},\ }\href@noop {} {\bibfield  {journal} {\bibinfo  {journal} {Monthly
  Notices of the Royal Astronomical Society}\ }\textbf {\bibinfo {volume}
  {380}},\ \bibinfo {pages} {1533} (\bibinfo {year} {2007})}\BibitemShut
  {NoStop}%
\bibitem [{\citenamefont {Berti}(2006)}]{berti2006lisa}%
  \BibitemOpen
  \bibfield  {author} {\bibinfo {author} {\bibfnamefont {E.}~\bibnamefont
  {Berti}},\ }\href@noop {} {\bibfield  {journal} {\bibinfo  {journal}
  {Classical and Quantum Gravity}\ }\textbf {\bibinfo {volume} {23}},\ \bibinfo
  {pages} {S785} (\bibinfo {year} {2006})}\BibitemShut {NoStop}%
\bibitem [{\citenamefont {Haehnelt}(1994)}]{Haehnelt:1994wt}%
  \BibitemOpen
  \bibfield  {author} {\bibinfo {author} {\bibfnamefont {M.~G.}\ \bibnamefont
  {Haehnelt}},\ }\href {\doibase 10.1093/mnras/269.1.199} {\bibfield  {journal}
  {\bibinfo  {journal} {Mon. Not. Roy. Astron. Soc.}\ }\textbf {\bibinfo
  {volume} {269}},\ \bibinfo {pages} {199} (\bibinfo {year} {1994})},\ \Eprint
  {http://arxiv.org/abs/astro-ph/9405032} {arXiv:astro-ph/9405032} \BibitemShut
  {NoStop}%
\bibitem [{\citenamefont {Enoki}\ \emph {et~al.}(2004)\citenamefont {Enoki},
  \citenamefont {Inoue}, \citenamefont {Nagashima},\ and\ \citenamefont
  {Sugiyama}}]{enoki2004gravitational}%
  \BibitemOpen
  \bibfield  {author} {\bibinfo {author} {\bibfnamefont {M.}~\bibnamefont
  {Enoki}}, \bibinfo {author} {\bibfnamefont {K.~T.}\ \bibnamefont {Inoue}},
  \bibinfo {author} {\bibfnamefont {M.}~\bibnamefont {Nagashima}}, \ and\
  \bibinfo {author} {\bibfnamefont {N.}~\bibnamefont {Sugiyama}},\ }\href@noop
  {} {\bibfield  {journal} {\bibinfo  {journal} {The Astrophysical Journal}\
  }\textbf {\bibinfo {volume} {615}},\ \bibinfo {pages} {19} (\bibinfo {year}
  {2004})}\BibitemShut {NoStop}%
\bibitem [{\citenamefont {Tamborra}\ \emph {et~al.}(2014)\citenamefont
  {Tamborra}, \citenamefont {Ando},\ and\ \citenamefont
  {Murase}}]{tamborra2014star}%
  \BibitemOpen
  \bibfield  {author} {\bibinfo {author} {\bibfnamefont {I.}~\bibnamefont
  {Tamborra}}, \bibinfo {author} {\bibfnamefont {S.}~\bibnamefont {Ando}}, \
  and\ \bibinfo {author} {\bibfnamefont {K.}~\bibnamefont {Murase}},\
  }\href@noop {} {\bibfield  {journal} {\bibinfo  {journal} {Journal of
  Cosmology and Astroparticle Physics}\ }\textbf {\bibinfo {volume} {2014}},\
  \bibinfo {pages} {043} (\bibinfo {year} {2014})}\BibitemShut {NoStop}%
\bibitem [{\citenamefont {Aartsen}\ \emph
  {et~al.}(2018{\natexlab{b}})\citenamefont {Aartsen}, \citenamefont
  {Ackermann}, \citenamefont {Adams}, \citenamefont {Aguilar}, \citenamefont
  {Ahlers}, \citenamefont {Ahrens}, \citenamefont {Al~Samarai}, \citenamefont
  {Altmann}, \citenamefont {Andeen}, \citenamefont {Anderson} \emph
  {et~al.}}]{aartsen2018differential}%
  \BibitemOpen
  \bibfield  {author} {\bibinfo {author} {\bibfnamefont {M.}~\bibnamefont
  {Aartsen}}, \bibinfo {author} {\bibfnamefont {M.}~\bibnamefont {Ackermann}},
  \bibinfo {author} {\bibfnamefont {J.}~\bibnamefont {Adams}}, \bibinfo
  {author} {\bibfnamefont {J.}~\bibnamefont {Aguilar}}, \bibinfo {author}
  {\bibfnamefont {M.}~\bibnamefont {Ahlers}}, \bibinfo {author} {\bibfnamefont
  {M.}~\bibnamefont {Ahrens}}, \bibinfo {author} {\bibfnamefont
  {I.}~\bibnamefont {Al~Samarai}}, \bibinfo {author} {\bibfnamefont
  {D.}~\bibnamefont {Altmann}}, \bibinfo {author} {\bibfnamefont
  {K.}~\bibnamefont {Andeen}}, \bibinfo {author} {\bibfnamefont
  {T.}~\bibnamefont {Anderson}},  \emph {et~al.},\ }\href@noop {} {\bibfield
  {journal} {\bibinfo  {journal} {Physical Review D}\ }\textbf {\bibinfo
  {volume} {98}},\ \bibinfo {pages} {062003} (\bibinfo {year}
  {2018}{\natexlab{b}})}\BibitemShut {NoStop}%
\bibitem [{\citenamefont {Stettner}(2019)}]{stettner2019measurement}%
  \BibitemOpen
  \bibfield  {author} {\bibinfo {author} {\bibfnamefont {J.}~\bibnamefont
  {Stettner}},\ }\href@noop {} {\bibfield  {journal} {\bibinfo  {journal}
  {arXiv preprint arXiv:1908.09551}\ } (\bibinfo {year} {2019})}\BibitemShut
  {NoStop}%
\bibitem [{\citenamefont {Waxman}\ and\ \citenamefont
  {Bahcall}(1998)}]{waxman1998high}%
  \BibitemOpen
  \bibfield  {author} {\bibinfo {author} {\bibfnamefont {E.}~\bibnamefont
  {Waxman}}\ and\ \bibinfo {author} {\bibfnamefont {J.}~\bibnamefont
  {Bahcall}},\ }\href@noop {} {\bibfield  {journal} {\bibinfo  {journal}
  {Physical Review D}\ }\textbf {\bibinfo {volume} {59}},\ \bibinfo {pages}
  {023002} (\bibinfo {year} {1998})}\BibitemShut {NoStop}%
\bibitem [{\citenamefont {Martineau-Huynh}\ \emph {et~al.}(2017)\citenamefont
  {Martineau-Huynh}, \citenamefont {Bustamante}, \citenamefont {Carvalho},
  \citenamefont {Charrier}, \citenamefont {De~Jong}, \citenamefont {De~Vries},
  \citenamefont {Fang}, \citenamefont {Feng}, \citenamefont {Finley},
  \citenamefont {Gou} \emph {et~al.}}]{martineau2017giant}%
  \BibitemOpen
  \bibfield  {author} {\bibinfo {author} {\bibfnamefont {O.}~\bibnamefont
  {Martineau-Huynh}}, \bibinfo {author} {\bibfnamefont {M.}~\bibnamefont
  {Bustamante}}, \bibinfo {author} {\bibfnamefont {W.}~\bibnamefont
  {Carvalho}}, \bibinfo {author} {\bibfnamefont {D.}~\bibnamefont {Charrier}},
  \bibinfo {author} {\bibfnamefont {S.}~\bibnamefont {De~Jong}}, \bibinfo
  {author} {\bibfnamefont {K.~D.}\ \bibnamefont {De~Vries}}, \bibinfo {author}
  {\bibfnamefont {K.}~\bibnamefont {Fang}}, \bibinfo {author} {\bibfnamefont
  {Z.}~\bibnamefont {Feng}}, \bibinfo {author} {\bibfnamefont {C.}~\bibnamefont
  {Finley}}, \bibinfo {author} {\bibfnamefont {Q.}~\bibnamefont {Gou}},  \emph
  {et~al.},\ }in\ \href@noop {} {\emph {\bibinfo {booktitle} {EPJ Web of
  Conferences}}},\ Vol.\ \bibinfo {volume} {135}\ (\bibinfo {organization} {EDP
  Sciences},\ \bibinfo {year} {2017})\ p.\ \bibinfo {pages} {02001}\BibitemShut
  {NoStop}%
\bibitem [{\citenamefont {Neronov}\ \emph {et~al.}(2017)\citenamefont
  {Neronov}, \citenamefont {Semikoz}, \citenamefont {Anchordoqui},
  \citenamefont {Adams},\ and\ \citenamefont
  {Olinto}}]{neronov2017sensitivity}%
  \BibitemOpen
  \bibfield  {author} {\bibinfo {author} {\bibfnamefont {A.}~\bibnamefont
  {Neronov}}, \bibinfo {author} {\bibfnamefont {D.~V.}\ \bibnamefont
  {Semikoz}}, \bibinfo {author} {\bibfnamefont {L.~A.}\ \bibnamefont
  {Anchordoqui}}, \bibinfo {author} {\bibfnamefont {J.~H.}\ \bibnamefont
  {Adams}}, \ and\ \bibinfo {author} {\bibfnamefont {A.~V.}\ \bibnamefont
  {Olinto}},\ }\href@noop {} {\bibfield  {journal} {\bibinfo  {journal}
  {Physical Review D}\ }\textbf {\bibinfo {volume} {95}},\ \bibinfo {pages}
  {023004} (\bibinfo {year} {2017})}\BibitemShut {NoStop}%
\bibitem [{\citenamefont {Venters}\ \emph {et~al.}(2019)\citenamefont
  {Venters}, \citenamefont {Reno}, \citenamefont {Krizmanic}, \citenamefont
  {Anchordoqui}, \citenamefont {Gu{\'e}pin},\ and\ \citenamefont
  {Olinto}}]{venters2019poemma}%
  \BibitemOpen
  \bibfield  {author} {\bibinfo {author} {\bibfnamefont {T.~M.}\ \bibnamefont
  {Venters}}, \bibinfo {author} {\bibfnamefont {M.~H.}\ \bibnamefont {Reno}},
  \bibinfo {author} {\bibfnamefont {J.~F.}\ \bibnamefont {Krizmanic}}, \bibinfo
  {author} {\bibfnamefont {L.~A.}\ \bibnamefont {Anchordoqui}}, \bibinfo
  {author} {\bibfnamefont {C.}~\bibnamefont {Gu{\'e}pin}}, \ and\ \bibinfo
  {author} {\bibfnamefont {A.~V.}\ \bibnamefont {Olinto}},\ }\href@noop {}
  {\bibfield  {journal} {\bibinfo  {journal} {arXiv preprint arXiv:1906.07209}\
  } (\bibinfo {year} {2019})}\BibitemShut {NoStop}%
\bibitem [{\citenamefont {Allison}\ \emph {et~al.}(2012)\citenamefont
  {Allison}, \citenamefont {Auffenberg}, \citenamefont {Bard}, \citenamefont
  {Beatty}, \citenamefont {Besson}, \citenamefont {B{\"o}ser}, \citenamefont
  {Chen}, \citenamefont {Chen}, \citenamefont {Connolly}, \citenamefont
  {Davies} \emph {et~al.}}]{allison2012design}%
  \BibitemOpen
  \bibfield  {author} {\bibinfo {author} {\bibfnamefont {P.}~\bibnamefont
  {Allison}}, \bibinfo {author} {\bibfnamefont {J.}~\bibnamefont {Auffenberg}},
  \bibinfo {author} {\bibfnamefont {R.}~\bibnamefont {Bard}}, \bibinfo {author}
  {\bibfnamefont {J.}~\bibnamefont {Beatty}}, \bibinfo {author} {\bibfnamefont
  {D.}~\bibnamefont {Besson}}, \bibinfo {author} {\bibfnamefont
  {S.}~\bibnamefont {B{\"o}ser}}, \bibinfo {author} {\bibfnamefont
  {C.}~\bibnamefont {Chen}}, \bibinfo {author} {\bibfnamefont {P.}~\bibnamefont
  {Chen}}, \bibinfo {author} {\bibfnamefont {A.}~\bibnamefont {Connolly}},
  \bibinfo {author} {\bibfnamefont {J.}~\bibnamefont {Davies}},  \emph
  {et~al.},\ }\href@noop {} {\bibfield  {journal} {\bibinfo  {journal}
  {Astroparticle Physics}\ }\textbf {\bibinfo {volume} {35}},\ \bibinfo {pages}
  {457} (\bibinfo {year} {2012})}\BibitemShut {NoStop}%
\bibitem [{\citenamefont {Barwick}\ \emph {et~al.}(2015)\citenamefont
  {Barwick}, \citenamefont {Berg}, \citenamefont {Besson}, \citenamefont
  {Binder}, \citenamefont {Binns}, \citenamefont {Boersma}, \citenamefont
  {Bose}, \citenamefont {Braun}, \citenamefont {Buckley}, \citenamefont
  {Bugaev} \emph {et~al.}}]{barwick2015first}%
  \BibitemOpen
  \bibfield  {author} {\bibinfo {author} {\bibfnamefont {S.}~\bibnamefont
  {Barwick}}, \bibinfo {author} {\bibfnamefont {E.}~\bibnamefont {Berg}},
  \bibinfo {author} {\bibfnamefont {D.}~\bibnamefont {Besson}}, \bibinfo
  {author} {\bibfnamefont {G.}~\bibnamefont {Binder}}, \bibinfo {author}
  {\bibfnamefont {W.}~\bibnamefont {Binns}}, \bibinfo {author} {\bibfnamefont
  {D.~J.}\ \bibnamefont {Boersma}}, \bibinfo {author} {\bibfnamefont
  {R.}~\bibnamefont {Bose}}, \bibinfo {author} {\bibfnamefont {D.}~\bibnamefont
  {Braun}}, \bibinfo {author} {\bibfnamefont {J.}~\bibnamefont {Buckley}},
  \bibinfo {author} {\bibfnamefont {V.}~\bibnamefont {Bugaev}},  \emph
  {et~al.},\ }\href@noop {} {\bibfield  {journal} {\bibinfo  {journal}
  {Astroparticle Physics}\ }\textbf {\bibinfo {volume} {70}},\ \bibinfo {pages}
  {12} (\bibinfo {year} {2015})}\BibitemShut {NoStop}%
\bibitem [{\citenamefont {Senno}\ \emph {et~al.}(2017)\citenamefont {Senno},
  \citenamefont {Murase},\ and\ \citenamefont
  {M{\'e}sz{\'a}ros}}]{senno2017high}%
  \BibitemOpen
  \bibfield  {author} {\bibinfo {author} {\bibfnamefont {N.}~\bibnamefont
  {Senno}}, \bibinfo {author} {\bibfnamefont {K.}~\bibnamefont {Murase}}, \
  and\ \bibinfo {author} {\bibfnamefont {P.}~\bibnamefont {M{\'e}sz{\'a}ros}},\
  }\href@noop {} {\bibfield  {journal} {\bibinfo  {journal} {The Astrophysical
  Journal}\ }\textbf {\bibinfo {volume} {838}},\ \bibinfo {pages} {3} (\bibinfo
  {year} {2017})}\BibitemShut {NoStop}%
\bibitem [{\citenamefont {Ackermann}\ \emph {et~al.}(2019)\citenamefont
  {Ackermann}, \citenamefont {Ahlers}, \citenamefont {Anchordoqui},
  \citenamefont {Bustamante}, \citenamefont {Connolly}, \citenamefont
  {Deaconu}, \citenamefont {Grant}, \citenamefont {Gorham}, \citenamefont
  {Halzen}, \citenamefont {Karle} \emph {et~al.}}]{ackermann2019astrophysics}%
  \BibitemOpen
  \bibfield  {author} {\bibinfo {author} {\bibfnamefont {M.}~\bibnamefont
  {Ackermann}}, \bibinfo {author} {\bibfnamefont {M.}~\bibnamefont {Ahlers}},
  \bibinfo {author} {\bibfnamefont {L.}~\bibnamefont {Anchordoqui}}, \bibinfo
  {author} {\bibfnamefont {M.}~\bibnamefont {Bustamante}}, \bibinfo {author}
  {\bibfnamefont {A.}~\bibnamefont {Connolly}}, \bibinfo {author}
  {\bibfnamefont {C.}~\bibnamefont {Deaconu}}, \bibinfo {author} {\bibfnamefont
  {D.}~\bibnamefont {Grant}}, \bibinfo {author} {\bibfnamefont
  {P.}~\bibnamefont {Gorham}}, \bibinfo {author} {\bibfnamefont
  {F.}~\bibnamefont {Halzen}}, \bibinfo {author} {\bibfnamefont
  {A.}~\bibnamefont {Karle}},  \emph {et~al.},\ }\href@noop {} {\bibfield
  {journal} {\bibinfo  {journal} {arXiv preprint arXiv:1903.04334}\ } (\bibinfo
  {year} {2019})}\BibitemShut {NoStop}%
\bibitem [{\citenamefont {Franceschini}\ and\ \citenamefont
  {Rodighiero}(2017)}]{franceschini2017extragalactic}%
  \BibitemOpen
  \bibfield  {author} {\bibinfo {author} {\bibfnamefont {A.}~\bibnamefont
  {Franceschini}}\ and\ \bibinfo {author} {\bibfnamefont {G.}~\bibnamefont
  {Rodighiero}},\ }\href@noop {} {\bibfield  {journal} {\bibinfo  {journal}
  {Astronomy \& Astrophysics}\ }\textbf {\bibinfo {volume} {603}},\ \bibinfo
  {pages} {A34} (\bibinfo {year} {2017})}\BibitemShut {NoStop}%
\bibitem [{\citenamefont {Xiao}\ \emph {et~al.}(2016)\citenamefont {Xiao},
  \citenamefont {Mészáros}, \citenamefont {Murase},\ and\ \citenamefont
  {Dai}}]{Xiao:2016rvd}%
  \BibitemOpen
  \bibfield  {author} {\bibinfo {author} {\bibfnamefont {D.}~\bibnamefont
  {Xiao}}, \bibinfo {author} {\bibfnamefont {P.}~\bibnamefont {Mészáros}},
  \bibinfo {author} {\bibfnamefont {K.}~\bibnamefont {Murase}}, \ and\ \bibinfo
  {author} {\bibfnamefont {Z.-g.}\ \bibnamefont {Dai}},\ }\href {\doibase
  10.3847/0004-637X/826/2/133} {\bibfield  {journal} {\bibinfo  {journal}
  {Astrophys. J.}\ }\textbf {\bibinfo {volume} {826}},\ \bibinfo {pages} {133}
  (\bibinfo {year} {2016})},\ \Eprint {http://arxiv.org/abs/1604.08131}
  {arXiv:1604.08131 [astro-ph.HE]} \BibitemShut {NoStop}%
\bibitem [{\citenamefont {Kun}\ \emph {et~al.}(2019)\citenamefont {Kun},
  \citenamefont {Biermann},\ and\ \citenamefont {Gergely}}]{kun2019very}%
  \BibitemOpen
  \bibfield  {author} {\bibinfo {author} {\bibfnamefont {E.}~\bibnamefont
  {Kun}}, \bibinfo {author} {\bibfnamefont {P.}~\bibnamefont {Biermann}}, \
  and\ \bibinfo {author} {\bibfnamefont {L.~{\'A}.}\ \bibnamefont {Gergely}},\
  }\href@noop {} {\bibfield  {journal} {\bibinfo  {journal} {Monthly Notices of
  the Royal Astronomical Society: Letters}\ }\textbf {\bibinfo {volume}
  {483}},\ \bibinfo {pages} {L42} (\bibinfo {year} {2019})}\BibitemShut
  {NoStop}%
\bibitem [{\citenamefont {Britzen}\ \emph {et~al.}(2019)\citenamefont
  {Britzen}, \citenamefont {Fendt}, \citenamefont {B{\"o}ttcher}, \citenamefont
  {Zaja{\v{c}}ek}, \citenamefont {Jaron}, \citenamefont {Pashchenko},
  \citenamefont {Araudo}, \citenamefont {Karas},\ and\ \citenamefont
  {Kurtanidze}}]{britzen2019cosmic}%
  \BibitemOpen
  \bibfield  {author} {\bibinfo {author} {\bibfnamefont {S.}~\bibnamefont
  {Britzen}}, \bibinfo {author} {\bibfnamefont {C.}~\bibnamefont {Fendt}},
  \bibinfo {author} {\bibfnamefont {M.}~\bibnamefont {B{\"o}ttcher}}, \bibinfo
  {author} {\bibfnamefont {M.}~\bibnamefont {Zaja{\v{c}}ek}}, \bibinfo {author}
  {\bibfnamefont {F.}~\bibnamefont {Jaron}}, \bibinfo {author} {\bibfnamefont
  {I.}~\bibnamefont {Pashchenko}}, \bibinfo {author} {\bibfnamefont
  {A.}~\bibnamefont {Araudo}}, \bibinfo {author} {\bibfnamefont
  {V.}~\bibnamefont {Karas}}, \ and\ \bibinfo {author} {\bibfnamefont
  {O.}~\bibnamefont {Kurtanidze}},\ }\href@noop {} {\bibfield  {journal}
  {\bibinfo  {journal} {Astronomy \& Astrophysics}\ }\textbf {\bibinfo {volume}
  {630}},\ \bibinfo {pages} {A103} (\bibinfo {year} {2019})}\BibitemShut
  {NoStop}%
\bibitem [{\citenamefont {de~Bruijn}\ \emph {et~al.}(2020)\citenamefont
  {de~Bruijn}, \citenamefont {Bartos}, \citenamefont {Biermann},\ and\
  \citenamefont {Tjus}}]{de2020recurrent}%
  \BibitemOpen
  \bibfield  {author} {\bibinfo {author} {\bibfnamefont {O.}~\bibnamefont
  {de~Bruijn}}, \bibinfo {author} {\bibfnamefont {I.}~\bibnamefont {Bartos}},
  \bibinfo {author} {\bibfnamefont {P.}~\bibnamefont {Biermann}}, \ and\
  \bibinfo {author} {\bibfnamefont {J.~B.}\ \bibnamefont {Tjus}},\ }\href@noop
  {} {\bibfield  {journal} {\bibinfo  {journal} {arXiv preprint
  arXiv:2006.11288}\ } (\bibinfo {year} {2020})}\BibitemShut {NoStop}%
\bibitem [{\citenamefont {Ros}\ \emph {et~al.}(2020)\citenamefont {Ros},
  \citenamefont {Kadler}, \citenamefont {Perucho}, \citenamefont {Boccardi},
  \citenamefont {Cao}, \citenamefont {Giroletti}, \citenamefont {Krauß},\ and\
  \citenamefont {Ojha}}]{Ros:2019bgo}%
  \BibitemOpen
  \bibfield  {author} {\bibinfo {author} {\bibfnamefont {E.}~\bibnamefont
  {Ros}}, \bibinfo {author} {\bibfnamefont {M.}~\bibnamefont {Kadler}},
  \bibinfo {author} {\bibfnamefont {M.}~\bibnamefont {Perucho}}, \bibinfo
  {author} {\bibfnamefont {B.}~\bibnamefont {Boccardi}}, \bibinfo {author}
  {\bibfnamefont {H.-M.}\ \bibnamefont {Cao}}, \bibinfo {author} {\bibfnamefont
  {M.}~\bibnamefont {Giroletti}}, \bibinfo {author} {\bibfnamefont
  {F.}~\bibnamefont {Krauß}}, \ and\ \bibinfo {author} {\bibfnamefont
  {R.}~\bibnamefont {Ojha}},\ }\href {\doibase 10.1051/0004-6361/201937206}
  {\bibfield  {journal} {\bibinfo  {journal} {Astron. Astrophys.}\ }\textbf
  {\bibinfo {volume} {633}},\ \bibinfo {pages} {L1} (\bibinfo {year} {2020})},\
  \Eprint {http://arxiv.org/abs/1912.01743} {arXiv:1912.01743 [astro-ph.GA]}
  \BibitemShut {NoStop}%
\end{thebibliography}%

\end{document}